\documentclass[aps,twocolumn,prb,reprint,superscriptaddress]{revtex4-2}
\usepackage{amsfonts}
\usepackage{amsmath}
\usepackage{amssymb}
\usepackage{amsthm}
\usepackage{mathrsfs}
\usepackage{graphicx}
\usepackage{color}
\usepackage{tensor}
\usepackage{upgreek}
\usepackage{soul, color, xcolor}
\usepackage{accents}
\usepackage[percent]{overpic}
\usepackage{times}
\usepackage{subfigure}
\usepackage{latexsym}
\usepackage{bm}
\usepackage{anyfontsize}
\usepackage{hyperref}
\hypersetup{colorlinks=true,linkcolor=blue,anchorcolor=blue,citecolor=blue,urlcolor=blue}
\usepackage[svgnames]{xcolor}
\usepackage{mathtools}
\usepackage{multirow}
\usepackage{booktabs}
\usepackage{makecell}
\usepackage{hhline}
\date{\today}
\begin{document}
\title{Multi-orbital Dirac superconductors and their realization of higher-order topology}
\author{Dao-He Ma}
\affiliation{National Laboratory of Solid State Microstructures, Department of Physics, Nanjing University, Nanjing 210093, China}
\author{Jin An}
\email{anjin@nju.edu.cn}
\affiliation{National Laboratory of Solid State Microstructures, Department of Physics, Nanjing University, Nanjing 210093, China}
\affiliation{Collaborative Innovation Center of Advanced Microstructures, Nanjing University, Nanjing 210093, China}
\begin{abstract}
Topological nodal superconductors (SCs) have attracted considerable interest due to their gapless bulk excitations and exotic surface states. In this paper, by establishing a general framework of the effective theory for multi-orbital SCs, we realize a class of three-dimensional (3D) time-reversal ($\mathcal{T}$)-invariant Dirac SCs, with their topologically protected gapless Dirac nodes being located at general positions in the Brillouin zone. By introducing $\mathcal{T}$-breaking pairing perturbations, we demonstrate the existence of Majorana hinge modes in these Dirac SCs as evidence of their realization of higher-order topology. We also propose a new kind of $\mathcal{T}$-breaking Dirac SCs, whose Dirac nodes possess nonzero even chiralities and so are characterized by surface Majorana arcs.  
\end{abstract}

\maketitle
\section{Introduction}
Topological SCs have gained tremendous attention in recent years due to their gapless Majorana boundary states \cite{TSC-2001-Kitaev, TSC_2008_prl_proximity, TSC_2010_RMP_Kane_review, TSC_2011_RMP_ZSC_QXL, TSC_2012_MF_sci, TSC_2012_nature, TSC_2012_RPP_MF, TSC_2012_SST_MF, TSC_2013_MF_3, TSC_2016_sato_review, TSC-2016-sci, TSC_2017_sato_review, TSC_2018_sci, TSC_2024_NC} which may have potential applications in quantum computation \cite{Q_computate_1_kitaev, Q_computate_2, Q_computate_3}. As a key category of topological SCs, the topological properties of $\mathcal{T}$-invariant odd-parity SCs are solely determined by the topology of their Fermi surfaces (FSs) \cite{Sato_Z2_2009, TSC_2010_prl_TRS_fu, Sato_Z2_2010}, irrespective of the SCs' detailed pairing symmetries. The emergence of higher-order topology \cite{HO_2017_prl_reflection, HO_2018_prb, HO_2018_prb_2, HO_2018_prb_inversion, HO_2018_prl_WZ, HO_2018_prl_2, HO_2019_prl_ZRX, HO_2019_prl_LCX, HO-2019-prl-yan, HO_2020_prl_2, HO-NSC-2020-prb, HO_2022_prx, HO-NSC-2023-prb, TSC_2024_NC, HO-2025-prb} has further extended the topological SCs to those with Majorana hinge or corner modes at lower-dimensional boundaries.

In contrast with fully gapped topological SCs, gapless nodal SCs \cite{NSC-2006-prb-Sato, NSC-2010-prl-Sato, NLSC_2011_prb, NSC_2012_prb_WSC, NSC_2012_prb_WSC_2, NSC_2013_prl, NSC_2013_njp, NLSC-2014-prb-Sato, NSC_2014_prl_DSCWSC, NSC_2014_cla, NSC-2015-prb-2, NSC_2015_JPCM, NSC_2015_prl_Sato, NSC-2015-prl-Sato-Weyl, NLSC-2016-prb-Sato, NSC_2016_prb_Sato, NSC-2016-prl-Zhao-classify, NSC-2016-prb-UPt, NLSC-2017-prl-ZSC, NLSC-2017-prl, NLSC-2018-prb-Sato, NSC-2018-prb-DSC, NSC-2018-prb-classify, NSC-2018-prl-HD, HO-NSC-2020-prb, NSC_2020_prb_ZRX, NSC-2022-prb-obs, NSC-2022-prl-Wan, HO-NSC-2023-prb, NLSC-2023-prb, NSC-2024-arxiv, NSC_2024_prb_DSC, NSC_2024_prb, HO_2024_china, NSC-HO-2025-arx, NSC_2025_nature, NSC-2025-prb, exp-NL-2019-prl, exp-NSC-2021-np, exp-2023-prl, exp-2024-prb, NLSC-2025-Ad} also show topologically nontrivial boundary characteristics, including Majorana surface arcs or flat bands, and Majorana drumhead surface states. Among these nodal SCs, Dirac SCs that host quadruple degenerate Dirac excitations have drawn significant interest. By analogy with the 3D Dirac semimetals, the Dirac points (DPs) of Dirac SCs generally lie on high-symmetry planes \cite{NSC_2014_prl_DSCWSC, NSC_2014_cla, NSC-2018-prb-DSC} or high-symmetry lines \cite{NSC_2015_prl_Sato, NSC_2016_prb_Sato, NSC_2020_prb_ZRX, NSC_2024_prb_DSC, NSC_2024_prb, HO_2024_china, NSC-HO-2025-arx} and topologically protected respectively by mirror (or glide mirror) or rotation symmetry. For the latter type of Dirac SCs which in most situations are superconducting doped Dirac semimetals, their nodes inherit those of the parent Dirac semimetals and realization of higher-order topology in these SCs has also been predicted \cite{NSC_2020_prb_ZRX, NSC_2024_prb_DSC, NSC_2024_prb, HO_2024_china, NSC-HO-2025-arx}. The questions naturally arise: Do 3D normal-metal based Dirac SCs exist, with their DPs being located at more general positions in the BZ? If so, can these nodes still be protected by symmetry and can higher-order topology be further realized in such system?

In this work, by developing the effective theory for the multi-orbital SCs, we provide confirmative answers to the above issues. Starting from multi-orbital normal metals with both $\mathcal{T}$ and space-inversion ($\mathcal{P}$) symmetries, we examine their possibility of realizing Dirac SCs. Based on our effective theory in which the effective and the original pairing potentials are found to share the same symmetries, we demonstrate by a two-orbital model as an example the realization of 3D $\mathcal{T}$-invariant Dirac SCs with the odd-parity helical $p$-wave pairing, with topologically protected DPs at general positions within the BZ. By mixing with a $\mathcal{T}$-breaking even-parity pairing potential, we further show that these SCs can exhibit higher-order topology, characterized by the existence of one-dimensional (1D) Majorana hinge modes. Additionally, we propose a new class of $\mathcal{T}$-breaking Dirac SCs, characterized by the nonzero even-chirality bulk DPs on the high-symmetry axes and surface Majorana arcs terminated by the projections of the DPs.

This paper is organized as follows. In Sec. \ref{sec2}, we introduce an effective theory for multi-orbital SCs and discuss the symmetry properties of the effective pairing potential. In Sec. \ref{sec3}, we present the construction of Dirac SCs with general-position DPs and the realization of their higher-order topology. In Sec. \ref{sec4}, we propose a new kind of $\mathcal{T}$-breaking Dirac SCs and discuss their surface Majorana arcs. In Sec. \ref{sec5}, we discuss the symmetry-breaking effect on the Dirac SCs and summarize our main results. Additional details regarding the effective theory, the real-space spin-orbit interactions, the relationship between the true and effective DPs, the edge theory for higher-order topology, as well as an odd-parity $s$-wave Dirac SC and its realization of higher-order topology, are provided in the Appendices \ref{app1}-\ref{app5}.
\section{Effective theory and symmetries for the EFFECTIVE PAIRING potential}
\label{sec2}
In this paper, we focus on multi-orbital SCs whose normal Hamiltonian $h(\bm{k})$ has both $\mathcal{T}$ and $\mathcal{P}$ symmetries (thus also respects $\mathcal{P}\mathcal{T}$ symmetry):
\begin{align}
    \mathcal{T}h(\bm{k})\mathcal{T}^{-1}&=h(-\bm{k}), \label{eq1}\\
    \mathcal{P} h(\bm{k}) \mathcal{P}^{-1}&=h(-\bm{k})\label{eq2},
\end{align}
where $\mathcal{T}= i\sigma_2 \mathcal{K}$, with $\mathcal{K}$ the complex conjugation and $\bm{\sigma}$ the spin Pauli matrices. In the weak-pairing limit (the order of magnitude of the pairing matrix is smaller compared with the Fermi energy $E_F$), the multi-orbital SCs can be simplified by invoking an effective theory without losing any essential physics of the SCs. In the following, we shall first introduce the framework of the effective theory and then we demonstrate that an effective pairing potential can always be defined that preserves the same symmetries, e.g., $\mathcal{T}$, $\mathcal{P}$, as those of the original one. 

\subsection{Effective pairing potential and its symmetries}
\label{sec2A}
Generally, a multi-orbital SC is described by the Bogoliubov-de Gennes (BdG) Hamiltonian 
\begin{align}
\label{eq3}
    H_{\mathrm{BdG}}(\bm{k})=\begin{pmatrix}
 h(\bm{k}) & \Delta(\bm{k})\\
 \Delta^{\dagger }(\bm{k}) &-h^T(-\bm{k})
\end{pmatrix}.
\end{align}
Due to $\mathcal{P}\mathcal{T}$ symmetry, $h(\bm{k})$ can be diagonalized into $N$ $\mathcal{P}\mathcal{T}$ pairs of bands:
\begin{align}
U^{\dagger}(\bm{k})h(\bm{k})U(\bm{k})&=\mathrm{diag}\left[\xi_1(\bm{k}),\xi_1(\bm{k}),\dots,\xi_{N}(\bm{k}),\xi_{N}(\bm{k})\right]\notag\\&\equiv\hat{h}(\bm{k}),
    \label{eq4}
\end{align}
where the degrees of freedom are assumed to include spin and $N$ others (such as orbitals and sublattices). Meanwhile, under the unitary transformation $\mathrm{diag}\left[U(\bm{k}),U^*(-\bm{k})\right]$, $H_{\mathrm{BdG}}(\bm{k})$ becomes:
\begin{align}
    H_{\mathrm{BdG}}(\bm{k})\longrightarrow H'_{\mathrm{BdG}}(\bm{k})=\begin{pmatrix}
        \hat{h}(\bm{k}) & \Delta'(\bm{k}) \\
        \Delta'^{\dagger}(\bm{k}) & -\hat{h}(-\bm{k})
    \end{pmatrix}\label{eq5},
\end{align}
with 
\begin{align}
\Delta'(\bm{k})=U^{\dagger}(\bm{k})\Delta(\bm{k})U^{*}(\boldsymbol{-k})\label{eq6}
\end{align}
the pairing matrix in band representation, where its $(mn)$ entry means the pairing between band $m$ and $n$. Since Cooper pairings in SCs only occur near the FSs, in the weak-pairing limit, only those $\mathcal{P}\mathcal{T}$ bands crossing $E_F$ should be retained. It can be shown that the $(mn)$ entry can be neglected as long as the retained bands $m, n$ belong to different $\mathcal{P}\mathcal{T}$ pairs (see Appendix \ref{app1}). As a result, $\Delta'(\bm{k})$ reduces to a block-diagonal matrix where the $j$th $2\times2$ block $\Delta'_j(\bm{k})$ corresponds to the $j$th $\mathcal{P}\mathcal{T}$ pair which contributes to the FSs. Because these $\mathcal{P}\mathcal{T}$ pairs are decoupled, one can suppose that only the first $\mathcal{P}\mathcal{T}$ bands are occupied without loss of generality. Consequently, Eq. (\ref{eq3}) of the multi-orbital SC reduces effectively to a $4\times4$ single-orbital SC $H^{r}_{\mathrm{BdG}}(\bm{k})$ with reduced normal Hamiltonian $h^{r}(\bm{k})=\xi_1(\bm{k})I$ and reduced $2\times2$ pairing matrix $\Delta^{r}(\bm{k})=\Delta'_{1}(\bm{k})$. Below we omit subscript $1$ for simplicity.

Now we examine the consequences of the symmetries of pairing potential $\Delta(\bm{k})$ in Eq. (\ref{eq3}). If the SC respects $\mathcal{P}$ symmetry, we have
\begin{align}
    \mathcal{P} \Delta(\bm{k}) \mathcal{P}^{T}&=\eta_\mathcal{P}\Delta(-\bm{k})\label{eq7},
\end{align}
where $\eta_\mathcal{P}=1$ ($-1$) stands for $\mathcal{P}$-even (-odd) parity. The corresponding $\Delta^{r}(\bm{k})$ generally lacks a similar equation. In the following, we shall demonstrate that a $2\times2$ unitary matrix $G(\bm{k})$ can always be found such that an effective pairing potential $\tilde{\Delta}(\bm{k})$ defined by 
\begin{align}
    \tilde{\Delta}(\bm{k})=G^{\dagger}(\bm{k})\Delta^{r}(\bm{k})G^{*}(-\bm{k})\label{eq8}
\end{align} obeys:
\begin{align}
     \tilde{\Delta}(\bm{k})&=\eta_{\mathcal{P}}\tilde{\Delta}(-\bm{k}).\label{eq9}
\end{align}
Similarly, we can prove that $\tilde{\Delta}(\bm{k})$ obeys: 
\begin{align}
    \tilde{\sigma}_2\tilde{\Delta}^{*}(\bm{k})\tilde{\sigma}_2&=\tilde{\Delta}(-\bm{k}),\label{eq10}
\end{align}
if the SC has $\mathcal{T}$ symmetry:
\begin{align}
    \sigma_2\Delta^{*}(\bm{k})\sigma_2&=\Delta(-\bm{k}),\label{eq11} 
\end{align} with $\tilde{\bm{\sigma}}$ the new pseudo-spin Pauli matrices. Eqs. (\ref{eq9})-(\ref{eq10}) can be viewed as the counterparts of Eqs. (\ref{eq7}) and (\ref{eq11}) for the effective pairing potential, where the effective $\mathcal{T}$ and $\mathcal{P}$ operators are defined in pseudo-spin basis of the occupied $\mathcal{PT}$ bands as $\tilde{\mathcal{T}}=i\tilde{\sigma}_2\mathcal{K}$ and $\tilde{\mathcal{P}}=\tilde{\sigma}_0$. These two equations are the central results of this section. The single-band pairing potential introduced for a two-orbital model in Refs. \cite{eff_theoty_Yip_tw,NSC_2016_prb_Sato} are analogous to $\Delta^{r}(\bm{k})$ here, and it directly satisfies Eqs. (\ref{eq9})-(\ref{eq10}) can be understood as its $G(\bm{k})$ being a unit matrix in Eq. (\ref{eq8}) due to a properly chosen unitary transformation of $U(\bm{k})$ for a two-orbital model.

In order to find $G(\bm{k})$ for the general situation, let us derive some preliminary formulas. Consider the relevant first $\mathcal{P}\mathcal{T}$ pair: $(u_{\mathrm{I}\bm{k}},u_{\mathrm{II}\bm{k}}$$\ =\ $$\mathcal{P}\mathcal{T}u_{\mathrm{I}\bm{k} })$, satisfying $h(\bm{k})u_{\mathrm{I}\bm{k}/\mathrm{II}\bm{k} } =\xi(\bm{k})u_{\mathrm{I}\bm{k}/\mathrm{II}\bm{k}}$. Accordingly, we have $h(-\bm{k})\mathcal{P}u_{\mathrm{I}\bm{k}}=\mathcal{P}h(\bm{k})u_{\mathrm{I}\bm{k} }=\xi(-\bm{k})\mathcal{P}u_{\mathrm{I}\bm{k}}$, where $\xi(\bm{k})=\xi(-\bm{k})$ due to $\mathcal{P}$ symmetry has been used. This indicates $\mathcal{P}u_{\mathrm{I} \bm{k}}$ is also an eigenvector of $h(-\bm{k})$ with eigenvalue $\xi(-\bm{k})$. Thus $\mathcal{P}u_{\mathrm{I}\bm{k} }$ can be expressed as a linear combination of $u_{\mathrm{I}-\bm{k} }$ and $u_{\mathrm{II} -\bm{k}}  $. Similar conclusion can be made for $\mathcal{P}u_{\mathrm{II}\bm{k} }$. Then we have:
\begin{align}
    \mathcal{P}(u_{\mathrm{I}\bm{k}} , u_{\mathrm{II}\bm{k} }   )=(u_{\mathrm{I} -\bm{k}},u_{\mathrm{II}-\bm{k}})p(-\bm{k}),\label{eq12}
\end{align}
where $p(\bm{k})$ is a $2\times2$ unitary matrix. Using  $\mathcal{P}^2=1$, we further have $(u_{\mathrm{I}\bm{k}},u_{\mathrm{II}\bm{k}}  )=\mathcal{P}(u_{\mathrm{I}-\bm{k} }  , u_{\mathrm{II}-\bm{k} })p(-\bm{k})=(u_{\mathrm{I}\bm{k}},u_{\mathrm{II}\bm{k}}) p(\bm{k})p(-\bm{k})$, implying $p^{\dagger}(\bm{k})=p(-\bm{k})$. Analogously, another $\mathcal{T}$-relevant $2\times2$ unitary matrix $w(\bm{k})$ can be defined by 
\begin{align}
    \mathcal{T}(u_{\mathrm{I}\bm{k}}, u_{\mathrm{II}\bm{k} })=(u_{\mathrm{I}-\bm{k} }, u_{\mathrm{II} -\bm{k}}) w(-\bm{k}),\label{eq13}
\end{align}
where $w(\bm{k})$ is actually the matrix introduced in Ref. \cite{TRS_2007_Fu_inversion} and satisfies $w^{T}(\bm{k})=-w(-\bm{k})$. Furthermore, a $\mathcal{P}\mathcal{T}$-relevant $2\times2$ unitary matrix $v(\bm{k})$ can be introduced by
\begin{align}
    \mathcal{P}\mathcal{T}(u_{\mathrm{I}\bm{k}},u_{\mathrm{II}\bm{k}} ) =(u_{\mathrm{I}\bm{k}},u_{\mathrm{II}\bm{k}})    v(\bm{k}).\label{eq14}
\end{align}
Combination of Eqs. (\ref{eq12})-(\ref{eq14}) gives rise to: 
\begin{align}
    p(\bm{k})w(-\bm{k})=v(\bm{k})=i\sigma'_2\label{eq15},
\end{align}
with a proper choice of phase, where $(\mathcal{P}\mathcal{T})^2=-1$ has been used and $\bm{\sigma}'$ are pseudo-spin Pauli matrices corresponding to the `old' basis $(u_{\mathrm{I}\bm{k}},u_{\mathrm{II}\bm{k}})$. So we have:
\begin{align}
    \sigma'_2p^*(\bm{k})\sigma'_2=p(\bm{k}).
    \label{eq16}
\end{align}

A general $2\times2$ unitary $p(\bm{k})$ satisfying Eq. (\ref{eq16}) takes the following form:
\begin{align}
    p(\bm{k})=a(\bm{k})+i\sqrt{1-a^2(\bm{k})} \bm{n}(\bm{k})\cdot \bm{\sigma}', \label{eq17}
\end{align}
where both $a(\bm{k})$ and unit vector $\bm{n}(\bm{k})$ are real. Utilizing $p^{\dagger}(\bm{k})=p(-\bm{k})$, we further have $a(\bm{k})=a(-\bm{k})$ and $\bm{n}(\bm{k})=-\bm{n}(-\bm{k})$. We now seek unitary matrices $G(\bm{k})$ satisfying $G^2(\bm{k})=p(\bm{k})$ and obtain two sets of solutions: 

\begin{align}
   G(\bm{k})=\begin{cases}
  \sqrt{\frac{1+a(\bm{k})}{2}}+i\sqrt{\frac{1-a(\bm{k})}{2}}\ \bm{n}(\bm{k})\cdot\bm{\sigma}',&\textrm{or}  \\
  i\sqrt{\frac{1-a(\bm{k})}{2}}+\sqrt{\frac{1+a(\bm{k})}{2}}\ \bm{n}(\bm{k})\cdot\bm{\sigma}'.& 
\end{cases}
\end{align}
This gives rise to:
\begin{align}
    G^{\dagger}(\bm{k})&=\pm G(-\bm{k})\label{eq19},\\
    \sigma'_2G^*(\bm{k})&=\pm G(\bm{k})\tilde{\sigma}_2\label{eq20},
\end{align}
where $+$ ($-$) corresponds to the 1st (2nd) solution of $G(\bm{k})$. Now we can define a new pseudo-spin basis within the same $\mathcal{PT}$ pair:
\begin{align}
    (\tilde{u}_{\Uparrow\bm{k}} , \tilde{u}_{\Downarrow \bm{k}}):=(u_{\mathrm{I}\bm{k}},u_{\mathrm{II}\bm{k}})G(\bm{k}).\label{eq21}
\end{align}
Combining Eqs. (\ref{eq19})-(\ref{eq20}), Eqs. (\ref{eq12})-(\ref{eq13}) become 
\begin{align}
    \mathcal{P}(\tilde{u}_{\Uparrow \bm{k}} , \tilde{u}_{\Downarrow \bm{k}}  )&=(\tilde{u}_{\Uparrow -\bm{k}},\tilde{u}_{\Downarrow-\bm{k}})(\pm \tilde{\sigma}_0)\label{eq22},\\
    \mathcal{T}(\tilde{u}_{\Uparrow \bm{k}} , \tilde{u}_{\Downarrow \bm{k}}    )&=(\tilde{u}_{\Uparrow -\bm{k}}  , \tilde{u}_{\Downarrow-\bm{k} })i\tilde{\sigma}_2,\label{eq23}
\end{align}
implying that the new pseudo-spin $\Uparrow\Downarrow$ behaves just like true spin, obeying $\left \langle \tilde{u}_{\Uparrow\bm{k}}|\bm{\sigma}  | \tilde{u}_{\Uparrow\bm{k}}  \right \rangle= \left \langle \tilde{u}_{\Uparrow-\bm{k}}|\bm{\sigma}  | \tilde{u}_{\Uparrow-\bm{k}}  \right \rangle=-\left \langle \tilde{u}_{\Downarrow \bm{k}}|\bm{\sigma}  | \tilde{u}_{\Downarrow \bm{k}}  \right \rangle=-\left \langle \tilde{u}_{\Downarrow -\bm{k}}|\bm{\sigma}  | \tilde{u}_{\Downarrow -\bm{k}}  \right \rangle$. This also gives a natural interpretation of why the effective $\mathcal{T}$ and $\mathcal{P}$ operators take the form of $\tilde{\mathcal{T}}=i\tilde{\sigma}_2\mathcal{K}$ and $\tilde{\mathcal{P}}=\pm \tilde{\sigma}_0$ in the new pseudo-spin basis. We further remark here that one can always invoke one solution of $G(\bm{k})$ to introduce the new pseudo-spin basis to keep it smooth and continuous in $\bm{k}$ space. But once one $G(\bm{k})$ is chosen, Eq. (\ref{eq22}) cannot be valid for the whole BZ since if so Eq. (\ref{eq22}) would mean the Bloch states at all $\mathcal{T}$-invariant $\bm{k}$ points share the same parity, which is evidently contradictory in most cases. Actually in most nontrivial situations, $G(\bm{k})$ always has some of the $\mathcal{T}$-invariant $\bm{k}$ points as its singularities, so that a $G(\bm{k})$ solution well-defined in the whole BZ can hardly be found. In the effective theory we focused on here, only FS and its adjacent area are concerned with other area in BZ irrelevant. As long as the FS is not passing through any $\mathcal{T}$-invariant $\bm{k}$ points, both equations are well defined and the effective theory works well. 

Next, we derive the relationship between $\Delta^{r}(\bm{k})$ and $p(\bm{k}), w(\bm{k})$. First of all, combination of Eqs. (\ref{eq6})-(\ref{eq7}) and (\ref{eq12}) leads to:
\begin{align}
    \Delta^{r}(-\bm{k})=\eta_\mathcal{P}\  p^{\dagger}(\bm{k})\Delta^{r}(\bm{k})p^{*}(-\bm{k})\label{eq24}.
\end{align}
By this equation, together with $p(\bm{k})=G^2(\bm{k})$, as well as Eq. (\ref{eq8}) and Eq. (\ref{eq19}), Eq. (\ref{eq9}) can be obtained. 

Analogously, using Eqs. (\ref{eq6}), (\ref{eq11}), (\ref{eq13}), one acquires:
\begin{align}
    \Delta^{r*}(-\bm{k})=w^{\dagger}(\bm{k})\Delta^{r}(\bm{k})w^{*}(-\bm{k})\label{eq25},
\end{align} 
so
\begin{align}
    \tilde{\sigma}_2\tilde{\Delta}^{*}(\bm{k})\tilde{\sigma}_2&=\tilde{\sigma}_2G^{T}(\bm{k})\Delta^{r*}(\bm{k})G(-\bm{k})\tilde{\sigma}_2\notag\\
    &=G^{\dagger}(\bm{k})\sigma'_2\Delta^{r*}(\bm{k})\sigma'_2G^{*}(-\bm{k})\notag\\
    &=G^{\dagger}(\bm{k})\sigma'_2w^{\dagger}(-\bm{k})\Delta^{r}(-\bm{k})w^{*}(\bm{k})\sigma'_2G^{*}(-\bm{k})\notag\\
    &=G^{\dagger}(\bm{k})(-i)p^{\dagger}(-\bm{k})\Delta^{r}(-\bm{k})ip^{*}(\bm{k})G^{*}(-\bm{k})\notag\\
    &=\eta_{\mathcal{P}}G^{\dagger}(\bm{k})\Delta^{r}(\bm{k})G^{*}(-\bm{k})\notag\\
    &=\eta_{\mathcal{P}}\tilde{\Delta}(\bm{k})=\tilde{\Delta}(-\bm{k})\label{eq26}
\end{align}
where in the above derivation Eqs. (\ref{eq8}), (\ref{eq20}), (\ref{eq25}), (\ref{eq15}), (\ref{eq24}), (\ref{eq8}), and (\ref{eq9}) have been used respectively in the equalities from the first to the last. This leads to Eq. (\ref{eq10}). So far, we have proven exactly that if the SC preserves either $\mathcal{P}$ or $\mathcal{T}$ symmetry, the effective pairing potential $\tilde{\Delta}(\bm{k})$ would obeys exactly similar constraint as that obeyed by the original pairing potential $\Delta(\bm{k})$. We notice that as we transform the pairing potential from $\Delta^r(\bm{k})$ to $\tilde{\Delta}(\bm{k})$, the BdG Hamiltonian would transform from the reduced $H^r_{\mathrm{BdG}}(\bm{k})$ to the effective $\tilde{H}_{\mathrm{BdG}}(\bm{k})$, but with the normal Hamiltonian $h^r(\bm{k})$ being always left unchanged since it is proportional to unit matrix.

The $2\times2$ $\tilde{\Delta}(\bm{k})$ can always be expressed as:
\begin{align}
    \tilde{\Delta}(\bm{k})=(\tilde{\psi}(\bm{k})+\tilde{\bm{d}}(\bm{k})\cdot\tilde{\bm{\sigma}})i\tilde{\sigma}_2 ,\label{eq27}
\end{align}
where $\tilde{\psi}(\bm{k})$ and $\tilde{\bm{d}}(\bm{k})$ correspond respectively to pseudo-spin singlet and triplet pairings. Fermionic statistics requires: $\tilde{\Delta}^{T}(\bm{k})=-\tilde{\Delta}(-\bm{k})$, indicating: $\tilde{\psi}(\bm{k})=\tilde{\psi}(-\bm{k})$ and $\tilde{\bm{d}}(\bm{k})=-\tilde{\bm{d}}(-\bm{k})$. 
Under the premise that the normal state of the SC has both $\mathcal{T}$ and $\mathcal{P}$ symmetries, the SC itself can break spontaneously either or both symmetries. If the SC has broken spontaneously both, the SC is generally described by a mixed pairing state of complex pseudo-spin singlet $\tilde{\psi}(\bm{k})$ and triplet $\tilde{\bm{d}}(\bm{k})$. In this situation, the energy excitation spectrum is:
\begin{align}    \tilde{E}_{\pm}=\sqrt{\xi^2+|\tilde{\psi}|^2+|\tilde{\bm{d}}|^2\pm\sqrt{|\tilde{\bm{d}}\times\tilde{\bm{d}}^*|^2+|\tilde{\psi}\tilde{\bm{d}}^{*}+\tilde{\psi}^{*}\tilde{\bm{d}}|^2}}\label{eq28}.
\end{align}
If the SC preserves $\mathcal{T}$ symmetry but have broken $\mathcal{P}$ symmetry, as discussed in Refs. \cite{symP-2004-prb, symP-2015-prl}, the SC is still generally a mixed pairing state, but by the real $\tilde{\psi}(\bm{k})$ and $\tilde{\bm{d}}(\bm{k})$ according to Eq. (\ref{eq10}), with its excitation spectrum given by: $\tilde{E}_{\pm}(\bm{k})=\sqrt{\xi^2(\bm{k})+(\tilde{\psi}(\bm{k})\pm|\tilde{\bm{d}}(\bm{k})|)^2}$. While if the SC preserves $\mathcal{P}$ and (but have broken) $\mathcal{T}$ symmetries, the SC is described by the real (complex) pseudo-spin singlet $\tilde{\psi}(\bm{k})$ for even parity with the degenerate excitation spectrum $\tilde{E}(\bm{k})=\sqrt{\xi^2(\bm{k})+|\tilde{\psi}(\bm{k})|^2}$, or by real (complex nonunitary) pseudo-spin triplet $\tilde{\bm{d}}(\bm{k})$ for odd parity with the excitation spectrum $\tilde{E}(\bm{k})=\sqrt{\xi^2(\bm{k})+\tilde{\bm{d}}^2(\bm{k})}$ ($\tilde{E}_{\pm}(\bm{k})=\sqrt{\xi^2(\bm{k})+|\tilde{\bm{d}}(\bm{k})|^2\pm|\tilde{\bm{d}}\times\tilde{\bm{d}}^*|}$), respectively. The superconducting states which break spontaneously $\mathcal{T}$ symmetry have been extensively studied in some superconducting materials \cite{symp-1993-prl, symp-1998-nat, symp-2009-prl, symp-2014-sci, symp-2021-jpcs, symp-2023-np, symp-2024-nm, symp-2025-sci}.   

\subsection{A two-orbital model}
To demonstrate the efficacy of the above framework, we apply it to a representative $N=2$ system, which is used in the next section. The normal Hamiltonian of this system is
\begin{align}
        h(\bm{k})=&\ \epsilon(\bm{k})I_4+\mathrm{Re} \lambda(\bm{k})\sigma_1\rho_3+\mathrm{Im} \lambda(\bm{k})\sigma_2\rho_3+\gamma(\bm{k})\sigma_3\rho_3\notag\\&+\mathrm{Re} g(\bm{k})\rho_1-\mathrm{Im} g(\bm{k})\rho_2,\label{eq29}
\end{align}
where $\bm{\rho}$ are Pauli matrices in orbital space. Items $\epsilon(\bm{k})$ and $\mathrm{Re}g(\bm{k})$ are even functions, while the rest are odd functions of $\bm{k}$. Thus, the system possesses both $\mathcal{T}$ and $\mathcal{P}$ symmetries with $\mathcal{P} =\rho_1$ and two $\mathcal{P}\mathcal{T}$ pairs of bands
\begin{align}
    \xi_{\pm}(\bm{k})=\epsilon(\bm{k})\pm q(\bm{k}),
\end{align}
where $q(\bm{k})=\sqrt{\gamma(\bm{k})^2+|g(\bm{k})|^2+|\lambda(\bm{k})|^2}$. If the lower $\mathcal{P}\mathcal{T}$ pair of bands is occupied, it is enough to consider the eigenvectors of $\xi_{-}(\bm{k})$ in the basis of $(\uparrow1,\uparrow 2, \downarrow1,\downarrow2)$:
\begin{align}
   u_{\mathrm{I}\bm{k} }  &=\frac{1}{\mathcal{A}(\bm{k})}[-g(\bm{k}),\gamma(\bm{k})+q(\bm{k}),0,\lambda (\bm{k})]^{T}, \notag\\
u_{\mathrm{II}\bm{k} }  &=
\frac{1}{\mathcal{A}(\bm{k})} [\lambda^*(\bm{k}),0,-\gamma(\bm{k})-q(\bm{k}),g^*(\bm{k})]^{T}
 ,\label{eq31}
\end{align}
where $\mathcal{A}(\bm{k})=\sqrt{2q(\bm{k})\left [ q(\bm{k})+\gamma(\bm{k})\right ]}$. Then, using $\mathcal{P}=\rho_1$, we can get $p(\bm{k})=\frac{1}{\mathcal{Q}(\bm{k})}\begin{pmatrix}
 -g^*(\bm{k}) & -\lambda^*(\bm{k})\\
  \lambda(\bm{k})&-g(\bm{k})
\end{pmatrix}$ and 
\begin{align}
   G(\bm{k})=\begin{cases}
  \frac{1}{\mathcal{L}_{-}(\bm{k})}\begin{pmatrix}
 \mathcal{Q}(\bm{k})-g^*(\bm{k}) & -\lambda^*(\bm{k})\\
  \lambda(\bm{k})&\mathcal{Q}(\bm{k})-g(\bm{k})
\end{pmatrix},&\textrm{or}  \\
  \frac{i}{\mathcal{L}_{+}(\bm{k})}\begin{pmatrix}
 \mathcal{Q}(\bm{k})+g^*(\bm{k}) & \lambda^*(\bm{k})\\
  -\lambda(\bm{k})&\mathcal{Q}(\bm{k})+g(\bm{k})
\end{pmatrix},& 
\end{cases}
\end{align}
where $\mathcal{L}_{\pm}(\bm{k})=\sqrt{2\mathcal{Q}(\bm{k})[\mathcal{Q}(\bm{k})\pm\mathrm{Re}g(\bm{k})]}$ with $\mathcal{Q}(\bm{k})=\sqrt{|g(\bm{k})|^2+|\lambda(\bm{k})|^2}$.
We consider a $\mathcal{T}$-invariant odd-parity pairing potential, 
\begin{align}
    \Delta(\bm{k})=\begin{pmatrix}
        iS(\bm{k})&0\\0&iS^*(\bm{k})
    \end{pmatrix}\rho_0, \label{eq33}
\end{align}
where $S(\bm{k})$ is a complex odd function. This means this SC can be described effectively by a real pseudo-spin triplet $\tilde{\bm{d}}$ whose components can be derived as (the 1st solution of $G(\bm{k})$ is chosen here): 
\begin{align}
    \tilde{d}_x(\bm{k})&=\frac{1}{q(\bm{k})}\left \{ \mathcal{Q}(\bm{k})\mathrm{Im}S(\bm{k})+\frac{\mathrm{Im}\lambda(\bm{k})\mathrm{Re}\left [ S(\bm{k})\lambda(\bm{k}) \right ] }{\mathcal{Q}(\bm{k})-\mathrm{Re}g(\bm{k})}\right \} ,\notag\\
    \tilde{d}_y(\bm{k})&=\frac{1}{q(\bm{k})}\left \{\mathcal{Q}(\bm{k})\mathrm{Re}S(\bm{k})- \frac{\mathrm{Re}\lambda(\bm{k})\mathrm{Re}\left [ S(\bm{k})\lambda(\bm{k}) \right ]}{\mathcal{Q}(\bm{k})-\mathrm{Re}g(\bm{k})}\right \} ,\notag\\
    \tilde{d}_z(\bm{k})&=\frac{1}{q(\bm{k})}\frac{\mathrm{Im}g(\bm{k})\mathrm{Re}\left [ S(\bm{k})\lambda(\bm{k}) \right ]}{\mathcal{Q}(\bm{k})-\mathrm{Re}g(\bm{k})}\label{eq34}.
\end{align}
After some simplifications, we obtain:
\begin{align}
    \tilde{\bm{d}}^2(\bm{k})=\left\{|g(\bm{k})S(\bm{k})|^2+\mathrm{Im}^2\left [S(\bm{k})\lambda(\bm{k}) \right ]\right\}/q^2(\bm{k})\label{eq35}.
\end{align}
If another $\mathcal{P}\mathcal{T}$ pair of bands $\xi_{+}(\bm{k})$ is also occupied, one can follow the same process to obtain the corresponding $\tilde{\bm{d}}^2(\bm{k})$, which is found to share the exactly same expression to Eq. (\ref{eq35}), although its value is actually different since it is defined on the FS of $\xi_{+}(\bm{k})=0$. For later convenience, this and other representative pairing potentials and their effective excitation gaps are listed in Table \ref{table1}.

\begin{table}[t]
\caption{\label{table1} Typical pairing potentials, corresponding $\mathcal{P}$ parities and effective excitation gaps, where $\mathcal{P}=\rho_1$ and the former takes the general form: $\Delta(\bm{k})=(\psi_{a}(\bm{k})+\bm{d}_{a}(\bm{k})\cdot \bm{\sigma})i\sigma_{2}\rho_{a}, a=0,1,2,3$. In the 1st column, $\bm{d}_{a}(\bm{k})=(\mathrm{Im}S(\bm{k}),\mathrm{Re}S(\bm{k}),0)$  with odd (even) function $S(\bm{k})$ and $\psi_a(\bm{k})$ is even (odd) function for $a=0,1,3$ ($a=2$).
}
\begin{ruledtabular}
\begin{tabular}{cccc}
\textrm{$\Delta$}& \textrm{parity}&
\textrm{$|\tilde{\psi}|^2$ or $|\tilde{\bm{d}}|^2 $ }\\
\colrule
$\psi_0$ & even &$|\tilde{\psi}|^2= |\psi_0|^2$ \\
$\psi_1$& even &$|\tilde{\psi}|^2= |\psi_1\mathrm{Re}g|^2/q^2$ \\
$\psi_2$& even &$|\tilde{\psi}|^2= |\psi_2\mathrm{Im}g|^2/q^2$ \\
$\psi_3$& odd &$ |\tilde{\bm{d}}|^2=|\psi_3|^2(\gamma^2+|\lambda|^2)/q^2$ \\
$\bm{d}_{0}$& odd &$ |\tilde{\bm{d}}|^2=\left [ |gS|^2+\mathrm{Im}^2(S \lambda)\right ] /q^2$ \\
$\bm{d}_{1}$& odd &$ |\tilde{\bm{d}}|^2=\left [ (\gamma^2+\mathrm{Re}^2g)|S|^2+\mathrm{Re}^2(S \lambda)\right ] /q^2$ \\$\bm{d}_{2}$& odd &$ |\tilde{\bm{d}}|^2=\left [ (\gamma^2+\mathrm{Im}^2g)|S|^2+\mathrm{Re}^2(S \lambda)\right ] /q^2$ \\
$\bm{d}_{3}$& even &$|\tilde{\psi}|^2= \mathrm{Im}^2(S \lambda)/q^2$ \\
\end{tabular}
\end{ruledtabular}
\end{table}
\section{DIRAC SCs with DPs at general positions and their realization of higher-order topology}
\label{sec3}
\begin{figure*}[t]
	\centering

	\includegraphics[width=\linewidth]{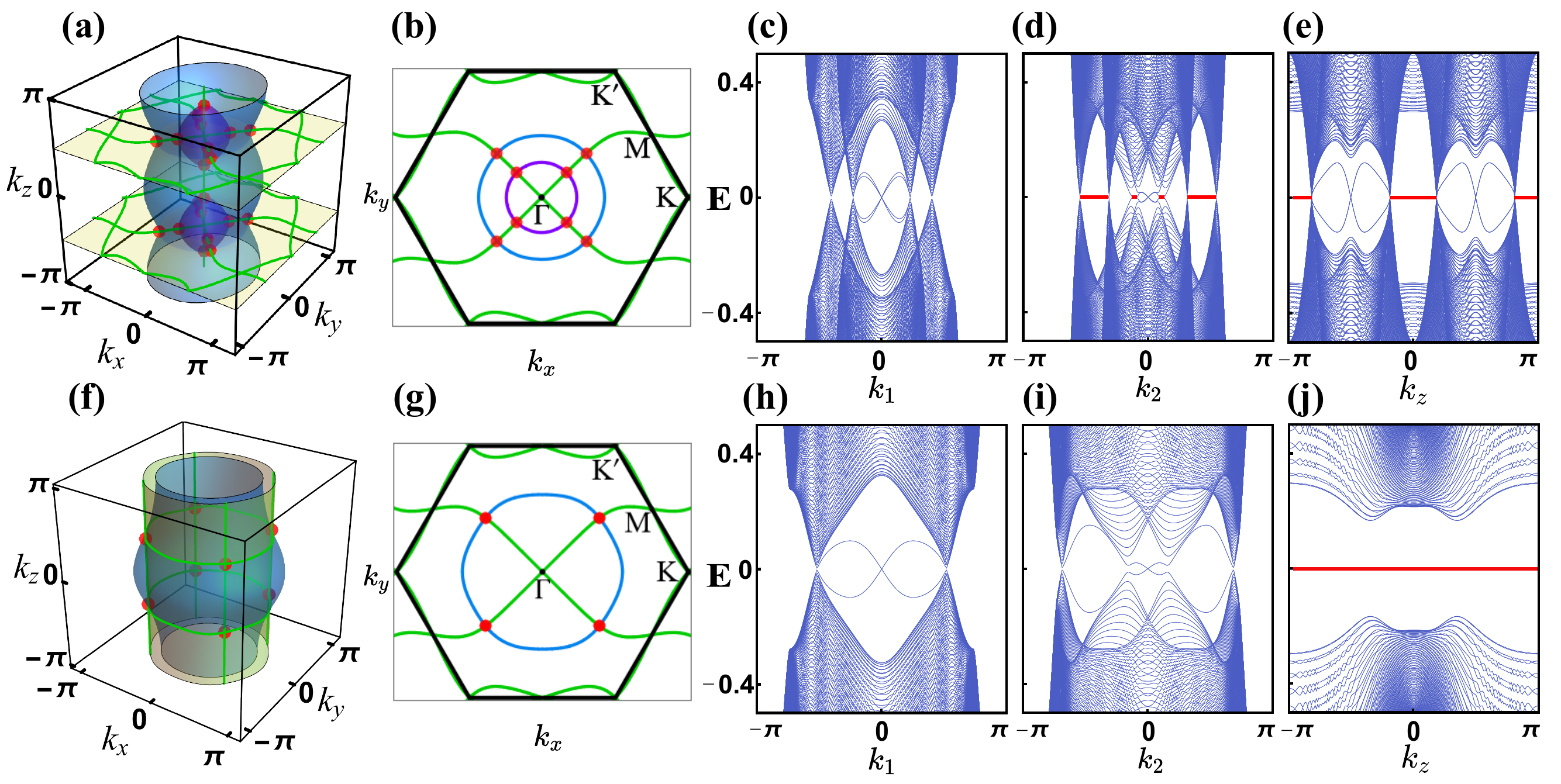}
    \caption{Odd-parity Dirac SCs with DPs at general positions or on high-symmetry lines. The upper (lower) panels correspond to the 1st (2nd) scheme. (a), (f) FSs (blue and purple surfaces) and the nodal lines (green) of the effective $\bm{d}$-vector $\tilde{\bm{d}}(\bm{k})$ located on the surface of $g=0$ (light yellow), their intersections giving the DPs (denoted by red solid dots). (b), (g) Horizontal plane at $k_z=k_0$ containing general-position DPs in the upper half of BZ, where the black hexagon is the 2D BZ boundary and the closed loops (blue and purple) represent the 1D FS cross-sections. (c), (h) [(d), (i)] Energy spectra of $k_z=k_0$ subsystem with boundary along $\bm{a}_1$ ($\bm{a}_2$) direction. (e), (j) Energy spectra of $k_2=0$ subsystem with boundary along $\bm{a}_z$ direction. Parameters: $(t_z,t_2,t_z^{'},\gamma_{so},\lambda_{so},k_0,\mu,\Delta_0)=(0,0,1.5,0.4,0.4,\pi/2,-3.5,0.1)$ and $(t_z,t_2,t_{1g},t_{2g},\gamma_{so},\lambda_{so},\mu,\Delta_0)=(0.8,0,1,0.2,0.4,0.4,-2.5,0.1)$ for the upper and lower panels, respectively.}
    \label{fig1}
\end{figure*}
Based on the above framework and the two-orbital model, in this section, we shall construct SCs with bulk DPs. First, we exhibit two schemes for constructing Dirac SCs with DPs at general positions in BZ. Then we display how to realize higher-order topology in both schemes. Here, we adopt a triangular lattice stacked along $\bm{z}$ direction, where each lattice site possesses two orbitals: $s\pm p_z$ and the lattice vectors are denoted as $\boldsymbol{a}_{1}=(1,0,0)$, $\boldsymbol{a}_{2}=(-1/2,\sqrt{3}/2,0)$ and $\boldsymbol{a}_{z}=(0,0,1)$, with $\boldsymbol{a}_{3}=-\boldsymbol{a}_{1}-\boldsymbol{a}_{2}$ being introduced for convenience. Each layer of the triangular lattice is so decorated by other atoms that it breaks $\mathcal{C}_6$ rotation symmetry to $\mathcal{C}_3$ (See Appendix \ref{app2}). Any lattice site can be chosen as the inversion center, and two orbitals are interchanged under $\mathcal{P}$. The expressions $\epsilon(\bm{k})$, $\gamma(\bm{k})$, $\lambda(\bm{k})$ in Eq. (\ref{eq29}) are given by:
\begin{align}
    \epsilon(\bm{k})&=-\mu-2(\cos{k_1}+\cos{k_2}+\cos{k_3})-2t_z \cos{k_z}\notag\\&-2t_2[\cos{(k_{1}-k_{3})}+\cos{(k_{2}-k_{1})}+\cos{(k_{3}-k_{2})}],\notag\\    \gamma(\bm{k})&=2\gamma_{so}(\sin{k_1}+\sin{k_2}+\sin{k_3}),\label{eq36}\\     \lambda(\bm{k})&=-2i\lambda_{so}(\sin{k_{1}}+\omega\sin{k_{2}}+\omega^{2}\sin{k_{3}}),\notag
\end{align}
where $k_{j}=\bm{k}\cdot\boldsymbol{a}_{j},\ j=1,2,3$, and $\epsilon(\bm{k})$ includes the intralayer nearest-neighbor (NN), second NN hoppings and interlayer NN hopping along $\bm{z}$ direction, with $\mu$ the chemical potential. $\lambda(\bm{k})$ and $\gamma(\bm{k})$ stand for the intrinsic SOCs with and without spin-flipping, preserving both $\mathcal{P}$ and $\mathcal{T}$ symmetries, where $\omega=e^{i2\pi/3}$ (see Appendix \ref{app2}). The pairing function in Eq. (\ref{eq33}) is chosen as a NN $p+ip$ wave:
\begin{align}
    S(\bm{k})&=2\Delta_0(\sin{k_{1}}+\omega\sin{k_{2}}+\omega^{2}\sin{k_{3}}).\label{eq37}
\end{align}
So the system we shall start with is an odd-parity helical $p$-wave SC.

\subsection{Construction of Dirac SCs with general-position DPs}
\label{sec3A}
According to the SC's degenerate excitation spectrum, whether it possesses gapless DPs depends on whether there exist nodes of $\tilde{\bm{d}}(\bm{k})$ on the FS. It is obvious that any nodes of $S(\bm{k})$ in BZ would also be those of $\tilde{\Delta}(\bm{k})$, or $\tilde{\bm{d}}(\bm{k})$. In this section we mainly concentrate on the nontrivial nodes of $\tilde{\Delta}(\bm{k})$ where $S(\bm{k})\ne0$, which would coincide with the common zeros of the two items: $g(\bm{k})$ and $\mathrm{Im}\left [S(\bm{k}) \lambda(\bm{k}) \right ]$. Their common zeros would be expected to form several lines, namely, the nontrivial nodal lines of $\tilde{\bm{d}}(\bm{k})$. Below, by constructing two different forms of the inter-orbital hopping term $g(\bm{k})$, we shall demonstrate as examples two schemes of realization of nodal lines of $\tilde{\bm{d}}(\bm{k})$, whose intersections with the FS give rise to Dirac-like excitations located at general positions.  

In the first scheme, we set $t_z=t_2=0$, and $g(\bm{k})=-2t_z^{'} (\cos{k_z}-\cos{k_0})$ whose nodal surfaces are the two planes at $k_z=\pm k_0$. On the other hand, since both $S(\bm{k})$ and $\lambda(\bm{k})$ are independent of $k_z$, $\mathrm{Im}\left [S(\bm{k}) \lambda(\bm{k}) \right ]$ vanishes on a cylinder with its generatrix being along $k_z$ direction. The intersections of the two surfaces generically yield several nodal lines within planes $k_z=\pm k_0$, as shown in Figs. \ref{fig1}(a)-(b). Now we show that these nodal lines of $\tilde{\bm{d}}(\bm{k})$ must pass through some high-symmetry points. Let $\bm{k}=(\bm{k}_\perp,\pm k_0)$, we have $\lambda(\bm{k})=0$ if $\bm{k}_{\perp}=K/K'$, $\Gamma$ and $M$, so the nodal lines of $\tilde{\bm{d}}(\bm{k})$ are always passing through these points [see Fig. \ref{fig1}(a)]. This is because $\lambda(\bm{k})$ is an eigenfunction of $\mathcal{C}_3$ around $k_z$ axis and so $\lambda(\mathcal{C}_3\bm{k})=\omega\lambda(\bm{k})=\lambda(\bm{k})$ when $\bm{k}_{\perp}=K/K'$, which means $\lambda(\bm{k})=0$ since $\omega\ne 1$. While $\lambda(\bm{k})=0$ when $\bm{k}_{\perp}=\Gamma$ and $M$ is because $\Gamma$ and $M$ are $\mathcal{T}$-invariant points and $\lambda(\bm{k})$ is an odd function of $\bm{k}_{\perp}$. Consequently, provided that the 1D FS sections at $k_z=\pm k_0$ enclose these points, the intersections between the FSs and $\mathrm{Im}\left [S(\bm{k}) \lambda(\bm{k}) \right ]=0$, namely, Dirac nodes of the SCs, would inevitably emerge. Although their locations may change due to parameter variations, the existence of these general-position DPs does not rely on the detailed form of $\lambda(\bm{k})$, as long as the SOC $\lambda(\bm{k})$ is an odd function of $\bm{k}$, which is naturally constrained by $\mathcal{P}$ or $\mathcal{T}$ symmetry. Besides these general-position DPs, there also exist four DPs located on $k_z$ axis, which is provided by the trivial nodal lines of $\tilde{\bm{d}}(\bm{k})$, or equivalently, by $S(\bm{k})=0$. 

To exclude the DPs on $k_z$ axis, in the second scheme, we set $t_2=0$ and a $k_z$ independent $g(\bm{k})$: $g(\bm{k})=-2t_{1g}(\cos{k_1}+\cos{k_2}+\cos{k_3})-2 t_{2g}[\cos{(k_{1}-k_{3})}+\cos{(k_{2}-k_{1})}+\cos{(k_{3}-k_{2})}]$, whose nodal surface is also a cylinder parallel to $k_z$. This results in four vertical lines as nodal lines of $\tilde{\bm{d}}(\bm{k})$, as shown in Fig. \ref{fig1}(f). Now we construct the FS as follows: By tuning the parameter $t_z$, the FS takes such a structure that its section at $k_z=0$ ($k_z=\pi$) plane encloses (excludes) the four nodal lines. Thus, the continuity of the FS would guarantee that the occurrence of the eight points of intersections (DPs) of the FS and the four nodal lines becomes inevitable. This gives rise to general-position DPs. 

To illustrate the topological nature of these DPs of the SC, we exhibit the bulk-edge correspondence in Figs. \ref{fig1}(c)-\ref{fig1}(e), \ref{fig1}(h)-\ref{fig1}(j), where we present their energy spectra with open surfaces normal to $\boldsymbol{a}_{2}\times\boldsymbol{a}_{z}$ or $\boldsymbol{a}_{1}\times\boldsymbol{a}_{z}$ for both schemes. For the first scheme, there exist Majorana arcs between the projections of the general-position DPs [Fig. \ref{fig1}(d)], or between the projections of the DPs on $k_z$ axis [Fig. \ref{fig1}(e)]. For the second scheme, there exist no Majorana arcs between the projections of the DPs, but the whole line of $k_2=0$ serves as the Majorana surface states [Fig. \ref{fig1}(j)]. The latter is because the FS is tabular, so each $k_z$-fixed two-dimensional (2D) subsystem has the FS cross-section as its FS which always encloses only one $\mathcal{T}$-invariant point $\Gamma$ so is a $\mathbb{Z}_2$ nontrivial topological SC, according to the criterion in Refs. \cite{Sato_Z2_2009, TSC_2010_prl_TRS_fu, Sato_Z2_2010}.  

We remark here that the general-position DPs derived from the effective theory are located exactly on the FS, and their existence is meaningful only under the condition of the weak-pairing limit. Actually, the corresponding true DPs of the BdG Hamiltonians do exist but their locations are slightly deviated from the FS. The shifted distances from the corresponding effective ones are found to be proportional to $\Delta^2_0/E_F$, which is negligible in the weak-pairing limit, implying that both kinds of DPs are equivalent in determining the essential physics of the nodal SCs discussed here. See Appendix \ref{app3} for details.

\subsection{Realization of higher-order Dirac SCs}
\label{sec3B}
Now we consider a superconducting state that spontaneously breaks both $\mathcal{P}$ and $\mathcal{T}$ symmetries but preserves $\mathcal{PT}$ symmetry, which means a mixed pairing state of even-parity and odd-parity states with complex but proper superposition coefficients. This kind of mixed-parity pairing states has been discussed in Refs. \cite{symp-2017-prl, symp-2022-cn}. Based on the $\mathcal{T}$-invariant odd-parity SCs with DPs introduced above, by further introducing a $\mathcal{T}$-breaking even-parity pairing term, without changing the positions of the DPs, we show in this section that for both schemes the SCs can realize higher-order topology: The SCs have no 2D Majorana surface states but host 1D Majorana hinge modes.

\begin{figure}[t]
	\centering
	\includegraphics[width=\linewidth]{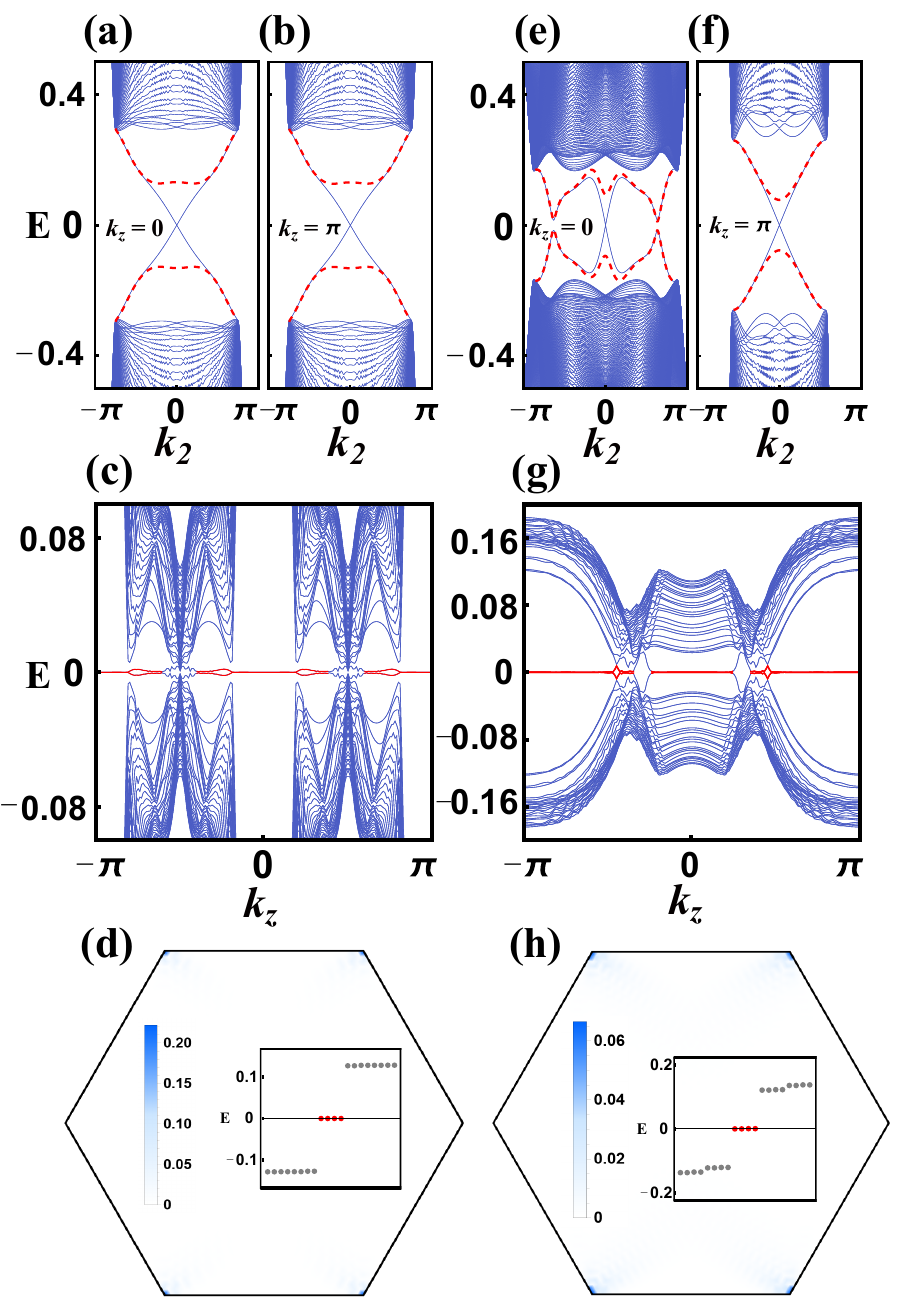}
    \caption{Higher-order topology of the $\mathcal{T}$-breaking Dirac SCs with mixed parity, where the odd-parity Dirac SCs with DPs in the 1st (left panels) and 2nd scheme (right panels) in the last section, are perturbed by a $\mathcal{T}$-breaking even-parity pairing term. (a), (e) [(b), (f)] Energy spectra for $k_z=0$ ($k_z=\pi$) planes with boundary along $\bm{a}_2$ direction. The blue solid (red dashed) lines are the edge states without (with) the $\mathcal{T}$-breaking term. (c), (g) Energy spectra for an infinitely long hexagonal prism with side lengths of 30 and 40, respectively, where the red segments denote the Majorana hinge modes. (d), (h) Real-space distribution of the Majorana corner states at $k_z=0$ and $k_z=\pi$ in (c) and (g), respectively, where the insets show energy eigenvalues near zero. Parameters: $(\Delta_0,\Delta_1)=(0.1,0.1)$ and $(0.1,0.25)$ for (a)-(f) and (g)-(h), respectively.}
    \label{fig2}
\end{figure}

Both $k_z=0$ and $k_z=\pi$ planes of the SCs in either scheme of the last section are $\mathcal{T}$-invariant planes. These two 2D systems possess nontrivial $\mathbb{Z}_2$ invariants and host gapless helical Majorana edge states, which can be viewed in Figs. \ref{fig1}(e), \ref{fig1}(j) and Figs. \ref{fig2}(a)-\ref{fig2}(b), \ref{fig2}(e)-\ref{fig2}(f). To realize higher-order topology, a relatively small $\mathcal{T}$-breaking pairing term is introduced to open gaps in these Majorana edge states [see the dashed lines in Figs. \ref{fig2}(a)-\ref{fig2}(b), \ref{fig2}(e)-\ref{fig2}(f)], with their bulk gaps always being kept open during this process. 
If the perturbed pairing term is so constructed that the signs of the edge gaps opened for some adjacent boundaries are opposite, these sign alternations would create zero-energy domain walls: the Majorana corner states \cite{corner_1976_prd,corner_2007_prl}. The horizontal planes in the vicinity of $k_z=0$ or $\pi$ can be viewed as being continuously connected to one of the latter two, so their bulk and edge gaps still keep open and the sign changes of the edge gaps for adjacent boundaries still keep unchanged. This indicates they belong to the same topological class as those at $k_z=0$ or $k_z=\pi$ and thereby possess the Majorana corner states at the same corners, giving rise to the Majorana hinge modes for the corresponding 3D system.

Specifically, we add to the pairing potential in Eq. (\ref{eq33}) an even-parity $\mathcal{T}$-breaking one $\Delta_e(\bm{k})=i\times\psi_1(\bm{k})i\sigma_2\rho_1$ with $\psi_1(\bm{k})=2\Delta_1(\delta_1\cos{k_1}+\delta_2\cos{k_2}+\delta_3\cos{k_3})$, which corresponds to an effective pseudo-spin singlet pairing component $\tilde{\psi}(\bm{k})=i \mathrm{Re}g(\bm{k})\psi_1(\bm{k})/q(\bm{k})$ (see Table \ref{table1}), being purely imaginary. This term breaks both $\mathcal{T}$ and $\mathcal{P}$ symmetries of the SCs, but preserves $\mathcal{P}\mathcal{T}$ symmetry. This mixed-parity $\mathcal{P}\mathcal{T}$-symmetric pairing state is quite similar to those discussed in Refs. \cite{symp-2017-prl, symp-2022-cn}. Because $\tilde{\bm{d}}(\bm{k})$ is a real function and the phase difference between $\tilde{\bm{d}}(\bm{k})$ and $\tilde{\psi}(\bm{k})$ is $\pi/2$, the last two items of Eq. (\ref{eq28}) are both zero, so the effective excitation spectrum becomes $\tilde{E}(\bm{k})=\sqrt{\xi^2(\bm{k})+\tilde{\psi}^2(\bm{k})+\tilde{\bm{d}}^2(\bm{k})}$ with $\tilde{\psi}^2(\bm{k})+\tilde{\bm{d}}^2(\bm{k})=\left [g^2(\bm{k})(\psi_1^2(\bm{k})+|S(\bm{k})|^2) +\mathrm{Im}^2(S(\bm{k}) \lambda(\bm{k}))\right ]/q^2(\bm{k}) $. Since the general-position DPs are all located at the nodal surfaces of $g(\bm{k})$, the addition of $\Delta_e(\bm{k})$ does not remove or even shift these DPs. The DPs on the $k_z$ axis for the first scheme will also remain if $\delta_1+\delta_2+\delta_3=0$ is chosen so that $\psi_1(\bm{k})=0$ there. According to the edge theory (see Appendix \ref{app4}), the effective gaps for the edges along $\boldsymbol{a}_{1}$, $\boldsymbol{a}_{2}$, and $\boldsymbol{a}_{3}$ directions are respectively proportional to $\delta_2+\delta_3, \delta_1+\delta_3 $ and $ \delta_1+\delta_2$. Therefore, by properly selecting these three values, opposite-sign edge gaps can be constructed in some adjacent boundaries, thereby achieving a higher-order topological state. Here, we set $(\delta_1, \delta_2, \delta_3)=(1,-1/2,-1/2)$. In Figs. \ref{fig2}(c) and \ref{fig2}(g) we show the energy spectra for an infinitely long hexagonal prism along $\bm{z}$ direction, where the Majorana hinge modes appear near $k_z=0$ and $k_z=\pi$. The 2D real-space distributions of the Majorana hinge modes at $k_z=0$ and $k_z=\pi$ are shown in Figs. \ref{fig2}(d) and \ref{fig2}(h), exhibiting the nature of being localized at the four corners between edges along $\bm{a}_1$ and those along $\bm{a}_2$ or $\bm{a}_3$. We also remark here that when the perturbed pairing term is introduced, although the true general-position DPs of the BdG Hamiltonians can not be removed, their locations are actually slightly changed, further confirming the validity and simplification of the effective DPs (see Appendix \ref{app3}). 

By examining various possible pairing potentials within the effective theory, we find that it is possible to even start with conventional $s$-wave pairing and achieve a Dirac SC characterized by nontrivial surface Majorana arcs. The SC can further realize higher-order topology. See Appendix \ref{app5} for more details.

\section{$\mathcal{T}$-breaking Dirac SCs}
\label{sec4}
In this section, we construct a new kind of Dirac SCs, which possess DPs at high-symmetry axes but are $\mathcal{T}$-breaking, in contrast to all the previously studied Dirac SCs. For some $k_z$-fixed 2D planes between some of the DPs, they have non-zero Chern number. So analogous to Weyl semimetals, this kind of Dirac SCs host nontrivial surface states, characterized by Majorana arcs that terminate at the projections of the DPs.
\begin{figure}[t]
	\centering
	\includegraphics[width=\linewidth]{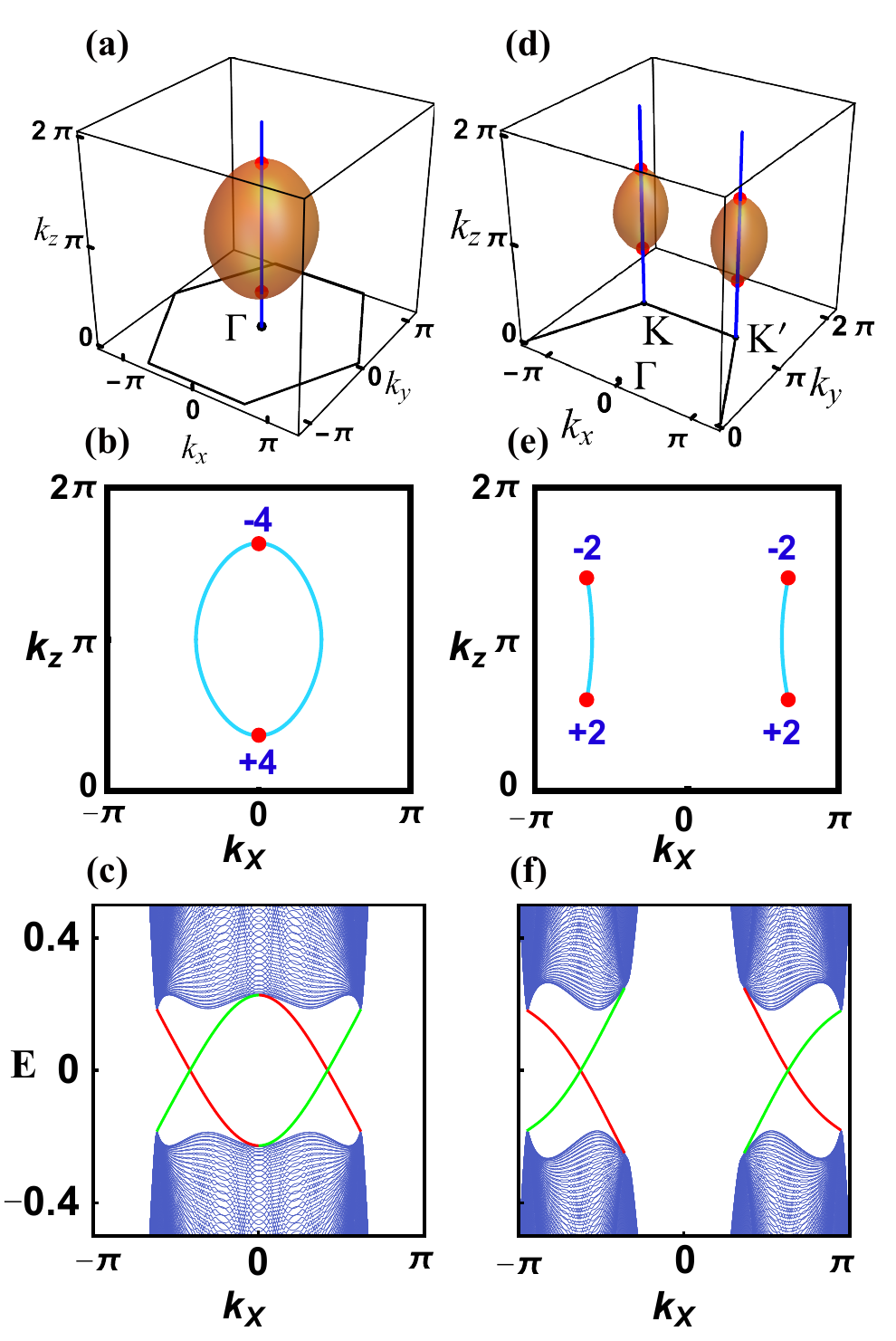}
    \caption{$\mathcal{T}$-breaking Dirac SCs with DPs on high-symmetry lines, where the left (right) panels correspond to $\mu=-6$ (7). (a), (d) FSs centered at $\bm{k}=(0,0,\pi)$ and $(\pm 2\pi/3, 2\pi/\sqrt{3}, \pi)$, respectively, with DPs (red solid dots) located on high symmetry lines. (b), (e) Surface Majorana arcs at ($010$) surface BZ, where the numbers represent the chiralities of DPs. (c), (f) Energy spectra of $k_z=\pi$ plane with open boundary along $\bm{a}_1$ direction, where green (red) solid lines correspond to the Majorana edge bands localized at the right (left) boundary. Parameters: $(t_z,t_2,t_z^{'},\gamma_{so},\lambda_{so},k_0,\Delta_2)=(0,-0.2,1,0.2,0.2,0,0.1)$.}
 \label{fig3}
\end{figure}
The normal Hamiltonian is identical to that of the first scheme in the last section. Then we choose as an example a $\mathcal{T}$-breaking even-parity pairing potential $\Delta(\bm{k})=\psi_0(\bm{k}) i\sigma_2\rho_0$ with $\psi_0(\bm{k})$ the NN $d+id$-wave $\psi_0(\bm{k})=2\Delta_2(\cos{k_1}+\omega\cos{k_2}+\omega^2\cos{k_3})$, which corresponds to a pseudo-spin singlet effective pairing identical to $\psi_0(\bm{k})$: $\tilde{\psi}(\bm{k})=\psi_0(\bm{k})$. The DPs are thus given by the intersections of the FS and nodal lines of $\psi_0(\bm{k})$. The FSs and their DPs at two typical chemical potentials are shown in Figs. \ref{fig3}(a), \ref{fig3}(d). The gapless Majorana edge states of the topologically nontrivial horizontal planes between the upper and lower DPs are connected to form Majorana arcs, as shown in Figs. \ref{fig3}(b), \ref{fig3}(e). Figs. \ref{fig3}(c), \ref{fig3}(f) give the energy spectra of $k_z=\pi$ plane with boundary along $\bm{a}_1$ direction. The edge bands with different colors are located at different boundaries. Since at each boundary there are four edge states (each edge band is accidentally doubly degenerate) propagating along the same direction, this implies that for either case the Chern number is $4$. 

This can also be understood by the effective pairing potential as follows. Consider a 2D horizontal plane near the upper DP (DPs) on the $k_z$ axis ($K$ and $K'$ axes) for the case of $\mu=-6$ ($\mu=7$), which is topologically equivalent to $k_z=\pi$ plane and thus share the same Chern number. This 2D SC has a small Fermi circle encircling $\Gamma$ ($K$ and $K'$), along which $\psi_0(\bm{k})$ takes the form $\psi_0(\bm{k})\sim(k_x-ik_y)^2$ ($\sim \delta k_x-i\delta k_y$ around both $K$ and $K'$). Therefore, for both chemical potentials, the total phase winding of $\psi_0(\bm{k})$ is $-2$. By taking into account the FS is doubly degenerate for the $\mathcal{PT}$ pair, the final total phase winding of the pairing potential is $-4$, explaining the Chern number $4$ of the 2D SCs. In fact, this Chen number can be associated with the chiralities of the DPs which can be defined by analogy with the Weyl points in 3D Weyl semimetals \cite{WSM-2011-prb-Wan, WSM-2015-sci, WSM-2018-RMP}. Consider a mapping from a sufficiently small 2D sphere $S^2$ encircling a DP to a unit vector: $\bm{k}\mapsto (\mathrm{Re}\tilde{\psi}(\bm{k}),\mathrm{Im}\tilde{\psi}(\bm{k}),\xi(\bm{k}))/\tilde{E}(\bm{k})$. The number of wrappings of this mapping on $S^2$ gives the chirality of the DP. The chirality of the upper DP on the $k_z$ axis ($K$ or $K'$ axis) for the case of $\mu = -6$ ($\mu = 7$) is found to be $-4$ ($-2$), consistent with the Chern number of the corresponding 2D SCs.

\section{Further discussions and conclusions}
\label{sec5}
Now under the premise that the normal state always has both $\mathcal{T}$ and $\mathcal{P}$ symmetries, we briefly examine the effects of breaking $\mathcal{C}_3$, $\mathcal{P}$ or $\mathcal{T}$ symmetries on Dirac SCs described in Secs. \ref{sec3A} and \ref{sec4}, and then we summarize our main results. 

Let us first examine the stability of the DPs of the Dirac SCs under $\mathcal{C}_3$-breaking perturbations. As long as $\mathcal{P}$ or $\mathcal{T}$ symmetry is preserved, $\lambda(\bm{k})$ always retains its odd-function property, so the FSs around the $\mathcal{T}$-invariant points must intersect the nodal lines of the effective $\tilde{\bm{d}}(\bm{k})$, and the general-position DPs in the first scheme still exist. Since the DPs in the second scheme exhibit no significant dependence on $\mathcal{C}_3$, the absence of $\mathcal{C}_3$ typically only slightly shifts the positions of the DPs. As for the $\mathcal{T}$-breaking Dirac SCs in Sec. \ref{sec4}, a simple $\mathcal{C}_3$-breaking process is a small deviation of $\psi_0(\bm{k})$ from the NN $d+id$-wave. This would generally split the nodal line of $k_z$ axis into a pair of symmetrically positioned nodal lines, thus resulting in each DP being replaced by a pair of symmetrically positioned DPs. While the nodal lines at $K$ and $K'$ axes will only deviate slightly from their original positions in a symmetrical manner due to $\mathcal{P}:K\leftrightarrow K'$, causing the shift of DPs.

Then to break $\mathcal{P}$ symmetry, for Dirac SCs of both schemes in Sec. \ref{sec3A}, we further introduce a perturbative real even-parity pairing potential, such as $\psi_0(\bm{k}), \bm{d}_3(\bm{k})$, which corresponds to an effective pseudo-spin singlet pairing $\tilde{\psi}(\bm{k})$. The effective excitation spectrum near the original DPs is $\tilde{E}_{\pm}(\bm{k})=\sqrt{\xi^2(\bm{k})+(\tilde{\psi}(\bm{k})\pm|\tilde{\bm{d}}(\bm{k})|)^2}$, where the original doubly degenerate bands split, with one branch being gapped while the other gapless branch being changed from the node into a small nodal loop encircling it. Therefore, by breaking $\mathcal{P}$ symmetry, the Dirac SCs discussed here would become Weyl-loop SCs. For $\mathcal{T}$-breaking process, a similar conclusion can also be drawn. 

In summary, within the framework of the effective theory, we have shown the effective pairing potential maintains the same symmetry as the original pairing potential. Based on this we have demonstrated realization of 3D Dirac SCs, where their effective Dirac nodes are located at general positions in BZ and are protected by symmetry. We further show they can exhibit higher-order topology, manifested by the existence of Majorana hinge modes. A new kind of 3D $\mathcal{T}$-breaking Dirac SCs possessing surface Majorana arcs is also proposed.

\section{Acknowledgment}
D.H.M. thanks J.J. Fu, S.T. Guan, and J.P. Xiao for helpful discussions. This work was supported by NSFC under Grant No.~11874202, and Innovation Program for Quantum Science and Technology under Grant No.2024ZD0300101.
\appendix
\renewcommand{\appendixname}{APPENDIX}
\section{EFFECTIVE THEORY}\label{app1}
In Sec. \ref{sec2A}, we have introduced the framework of the effective theory to deal with the multi-orbital SCs whose normal state Hamiltonian $h(\bm{k})$ has both $\mathcal{T}$ and $\mathcal{P}$ symmetries. In this appendix, we shall explain more rigorously why the pairing potential $\Delta'_{mn}(\bm{k})$ between different $\mathcal{P}\mathcal{T}$ bands contributing to the different FSs but with large band gap can be directly ignored without qualitatively affecting the underlying physics.

We now make a transformation of $\Delta'(\bm{k})$ by a unitary transformation $U'(\bm{k})$. For SCs in the weak-pairing limit, $\Delta'(\bm{k})$ in $H'_{\mathrm{BdG}}(\bm{k})$ is a small-quantity matrix, so we can set $U'(\bm{k})=I+\delta U(\bm{k})$ with $\delta U(\bm{k})$ also a small-quantity matrix. Using $U'^{\dagger}(\bm{k})U'(\bm{k})=I$, we have $\delta U^{\dagger}(\bm{k})=-\delta U(\bm{k})$. Then, 
\begin{align}
    &\quad\  U'^{\dagger}(\bm{k})H'_{\mathrm{BdG}}(\bm{k})U'(\bm{k})\notag\\&=(I+\delta U^{\dagger}(\bm{k}))H'_{\mathrm{BdG}}(\bm{k})(I+\delta U(\bm{k}))\notag\\&\simeq H'_{\mathrm{BdG}}(\bm{k})+\delta U^{\dagger}(\bm{k})H'_{\mathrm{BdG}}(\bm{k})+H'_{\mathrm{BdG}}(\bm{k})\delta U(\bm{k})\notag\\
    &=H'_{\mathrm{BdG}}(\bm{k})+[H'_{\mathrm{BdG}}(\bm{k}), \delta U(\bm{k})]\notag
    \\& \simeq H'_{\mathrm{BdG}}(\bm{k})+[\check{h} (\bm{k}), \delta U(\bm{k})]\label{eqA1}
\end{align}
with $\check{h} (\bm{k})=\mathrm{diag}( \hat{h}(\bm{k}),-\hat{h}(-\bm{k}))$, where we have neglected terms of order $\Delta^2_0$, with $\Delta_0$ presenting the order of magnitude of $\Delta'(\bm{k})$. To eliminate $\Delta'_{mn}(\bm{k})$ with $m$ and $n$ belonging to different $\mathcal{P}\mathcal{T}$ bands contributing to the FSs, Eq. (\ref{eqA1}) requires $\Delta'_{mn}(\bm{k})+\check{h}(\bm{k})_{mm}\delta U_{mn}(\bm{k})-\delta U_{mn}(\bm{k})\check{h}(\bm{k})_{nn}=0$, i.e., 
\begin{align}
    \delta U_{mn}(\bm{k})=-\Delta'_{mn}(\bm{k})/(\check{h}_{mm}(\bm{k})-\check{h}_{nn}(\bm{k}))\label{eqA2}.
\end{align} 
If all $\mathcal{P}\mathcal{T}$ bands contributing to FSs are sufficiently separated with each other, one can choose $\delta U_{mn}(\bm{k})$ according to the above equation to achieve the elimination without any change of the diagonal blocks of $\Delta'(\bm{k})$. This transformation also does not affect the diagonal entries of $H'_{\mathrm{BdG}}(\bm{k})$ (the normal dispersion of the corresponding $\mathcal{P}\mathcal{T}$ bands) due to the block-off-diagonal structure of $\delta U(\bm{k})$ and Eq. (\ref{eqA1}).

From Eq. (\ref{eqA2}), $\delta U(\bm{k})$ is of order $\Delta_0/E_F$. Therefore, the final effective pairing potential in the main text is accurate only within the limit of $(\Delta_0/E_F)^2\ll1$, which is well guaranteed by the weak-pairing limit.

\section{LATTICE STRUCTURE AND SPIN-ORBIT INTERACTIONS IN REAL SPACE}
\label{app2}
\begin{figure}[t]
	\centering
	\includegraphics[width=\linewidth]{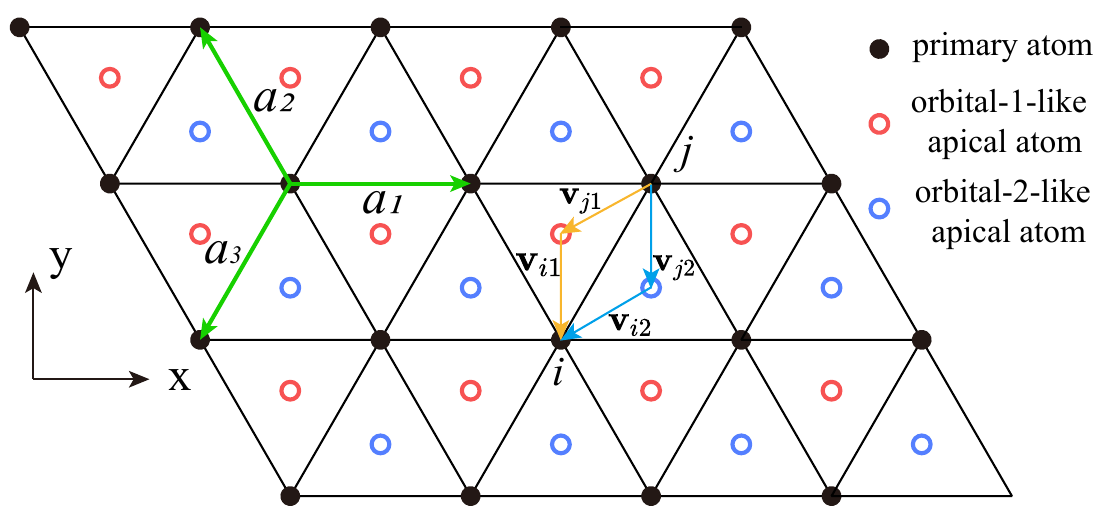}
    \caption{Schematic diagram of a single layer of the layered model adopted in the main text, where the black solid dots represent the primary atoms, possessing two orbitals ($1$ and $2$) per site and forming a triangular lattice, while the red (blue) hollow dots denote the decorated apical atoms with orbital-$1$ ($2$)-like orbitals. }
    \label{fig4}
\end{figure}

Here we provide in detail the lattice structure of our model in the main text, and then we give the real-space representations of the intrinsic SOCs $\gamma(\bm{k})$ and $\lambda(\bm{k})$. Each layer is shown schematically in Fig. \ref{fig4}, which is a triangular lattice, with each lattice site (black solid dots) possessing two orbitals: $s\pm p_z$, denoted by orbital-1 and -2 respectively. The lattice is decorated by two kinds of apical atoms, denoted by the red and blue hollow dots in Fig. \ref{fig4}. Since the six adjacent apical atoms centered on a lattice site are arranged in an alternating pattern, the $\mathcal{C}_6$ rotation symmetry is broken to $\mathcal{C}_3$. $\mathcal{P}$ symmetry is still preserved, if the two kinds of apical atoms can be viewed to have respectively orbital-1-like and orbital-2-like orbital geometry, and they are interchanged under $\mathcal{P}$. These apical atoms are only symmetry relevant and do not contribute to the kinetic terms in the Hamiltonian.

The real-space representation of $\gamma(\bm{k})$ is analogous to the intrinsic SOC in the K-M model \cite{KMmodel_2005_prl}, and is given by $i\gamma_{so}\sum_{\left \langle ij \right \rangle ,\eta}\nu_{ij,\eta}c_{i\eta}^{\dagger}\sigma_zc_{j\eta}$, where the sum is over the NN sites $\left \langle ij \right \rangle$ and orbital index $\eta=1,2$. Here $\nu_{ij,\eta}=\bm{a}_z\cdot(\mathbf{v}_{j\eta}\times\mathbf{v}_{i\eta})/|\mathbf{v}_{j\eta}\times\mathbf{v}_{i\eta}|=\pm1$, where $\mathbf{v}_{j\eta}$ represents the displacement from site $j$ to the orbital-$\eta$-like apical atom nearest to site $i$ and $j$ (see Fig. \ref{fig4}). The real-space expression of $\lambda(\bm{k})$ is given by $i \lambda_{so} \sum_{\left \langle ij \right \rangle ,\eta}l_{\eta}c_{i\eta}^{\dagger}(\bm{\sigma}\times\bm{d}_{ij})_z c_{j\eta}$ with $\bm{d}_{ij}$ the displacement between NN sites $i$ and $j$ within the same layer. In a single-orbital system, this term represents a Rashba SOC that breaks $\mathcal{P}$ symmetry. By choosing $l_1=1$ and $l_2=-1$, this term maintains $\mathcal{P}$ symmetry, and can thus be regarded as an intrinsic SOC of the crystal structure we discuss here. 

\section{RELATION BETWEEN TRUE DPS AND THOSE DERIVED FROM THE EFFECTIVE THEORY}
\label{app3}
The DPs realized in the SCs discussed in Sec. \ref{sec3A} should be regarded as the effective nodes whose locations are derived from the effective theory in the weak-pairing limit. In this Appendix, by examining the rigorous energy excitations of the SCs, we shall show that the true general-position DPs of $H_{\mathrm{BdG}}(\bm{k})$ do exist and then we shall discuss the relationship between them and the corresponding effective DPs.

According to Eqs. (\ref{eq29}), (\ref{eq33}), the exact excitation spectrum of $H_{\mathrm{BdG}}(\bm{k})$ can be given by:
\begin{align}
    E_{\pm}=\sqrt{\epsilon^2+q^2+|S|^2\pm 2\sqrt{(\epsilon q)^2+(\gamma|S|)^2+\mathrm{Re}^2(S \lambda)}}.
\end{align}
where $q^2=|g|^2+|\lambda|^2+\gamma^2$. Setting $E_{-}=0$, the equation determining the zero-energy excitations of the SC can be derived to be:
\begin{align}
    (\epsilon^2-q^2+|S|^2)^2+4|gS|^2+4\mathrm{Im}^2(S \lambda)=0.
\end{align}
Since the expression on the left side of this equation is positive-definite, its zeros should be jointly determined by those of each of the three terms. The zeros of the latter two terms actually give the nodal lines of $\tilde{\bm{d}}(\bm{k})$ in Eq. (\ref{eq35}). The nodal surface determined by zeros of the first term, namely, $\epsilon^2-q^2+|S|^2=\xi_{+}\xi_{-}+|S|^2=0$ replaces here the FS in discussing the effective DPs in the main text. Since $S$ is of order $\Delta_0$, the energy difference between the nodal surface and the FS should be of order $\Delta_0^2/E_F$, which is rather small in the weak-pairing limit. So the Dirac SCs discussed in the main text indeed host true general-position DPs, but their locations just slightly deviate from those of the corresponding effective DPs by a value of order $\delta k/ k_F\sim (\Delta_0/E_F)^2$, with $k_F$ the Fermi wave vector.

The strict zero-energy equation of the SC with perturbation term $\Delta_e(\bm{k})$ is $(\epsilon^2-q^2+|S|^2+\psi_1^2)^2+4|g|^2(|S|^2+\psi_1^2)+4\mathrm{Im}^2(S \lambda)=0$. It can be seen that without changing the nontrivial nodal lines of $\tilde{\bm{d}}(\bm{k})$ by this equation, the nodal surface defined by the zeros of the first term slightly changes compared to that without perturbation, indicating that although the locations of the effective general-position DPs are left unchanged, those of the true DPs do slightly vary.

\section{EDGE THEORY FOR HIGHER-ORDER TOPOLOGY}
\label{app4}
In Sec. \ref{sec3B}, we directly provide the induced edge gaps $\delta_2+\delta_3, \delta_1+\delta_3 $ and $ \delta_1+\delta_2$ for boundaries along $\boldsymbol{a}_{1}$, $\boldsymbol{a}_{2}$, and $\boldsymbol{a}_{3}$ directions, respectively, from the additionally introduced paring potential $\Delta_e(\bm{k})=i\times\psi_1(\bm{k})i\sigma_2\rho_1$. In this appendix, we shall give a rigorous proof by using the edge theory, which shares similar basic spirit in method to that of Refs. \cite{HO_2018_prl_WZ, HO_2020_prl_2, HO_2024_china}. 

Take the 2D $k_z=0$ subsystem in the first scheme as an example, where only the $\xi_{-}(\bm{k})$ band is occupied. Since the gapless edge Kramers states lie at the $\mathcal{T}$-invariant point $\bm{k}_{\perp}=(0,0)$, without losing any essential physics, one can only focus on the region of $\bm{k}_{\perp}\rightarrow0$, where one can make the expansions: $\epsilon(\bm{k})\sim 3(k_x^2+k_y^2)/2-6$, $g(\bm{k})\sim -2t'_z$, $\lambda(\bm{k})\sim o(k)$, $\gamma(\bm{k})\sim o(k^3)$ and $S(\bm{k})\sim 3\Delta_0(k_x+ik_y)$. Then, retaining the reduced pairing potential $\Delta^{r}(\bm{k})$ [obtained from Eq. (\ref{eq33})] to first order in $k$ leads to
\begin{align}
    \Delta^{r}(\bm{k})\sim i\begin{pmatrix}
 S(\bm{k}) & \\
  &S^*(\bm{k})
\end{pmatrix}=3\Delta_0(ik_x-k_y\sigma'_3).
\end{align} Define the continuous BdG Hamiltonian $H^{\mathrm{con}}_{\mathrm{BdG}}(k_x,k_y)$ as $H^{r}_{\mathrm{BdG}}(k_x,k_y,k_z=0)$ in the limit of $\bm{k}_{\perp}\rightarrow0$, we have  $H^{\mathrm{con}}_{\mathrm{BdG}}(k_x,k_y)= [b(k_x^2+k_y^2)-c ] \tau_3-3\Delta_0(k_x\tau_2+k_y\tau_1\sigma'_3)$, where $\bm{\tau}$ are particle-hole Pauli matrices and $\bm{\sigma}'$ are pseudo-spin Pauli matrices introduced in Sec. \ref{sec2A}, $b=3/2-9\lambda_{so}^2/4t'_z$ and $c=6+\mu+2t'_z$. 

We consider a 2D semi-infinite system whose boundary is inclined at angle $\theta$ to the horizontal, where the coordinates along the positive and normal directions of the boundary are denoted as $x_{\parallel}$ and $x_{\perp}$ respectively, with their corresponding momenta being denoted as $k$ and $k_{\perp}$. Notice that $k_x+ik_y=e^{i\theta}(k+ik_{\perp})$ and let $k_{\perp}\rightarrow -i\partial$, we have $H^{\mathrm{con}}_{\mathrm{BdG}}(k_x,k_y)\longrightarrow H^{(0)}_{\mathrm{BdG}}(-i\partial,\theta)+H^{(1)}_{\mathrm{BdG}}(k,\theta)$ with the $0$th-order Hamiltonian $H^{(0)}_{\mathrm{BdG}}(-i\partial,\theta)=-(b\partial^2+c) \tau_3+3i\Delta_0(-\sin{\theta}\tau_2+\cos{\theta}\tau_1\sigma'_3)\partial$ and the $1$st-order Hamiltonian $H^{(1)}_{\mathrm{BdG}}(k,\theta)=-3\Delta_0(\cos{\theta}\tau_2+\sin{\theta}\tau_1\sigma'_3)k$.
\begin{figure}[t]   
	\centering    
	\includegraphics[width=\linewidth]{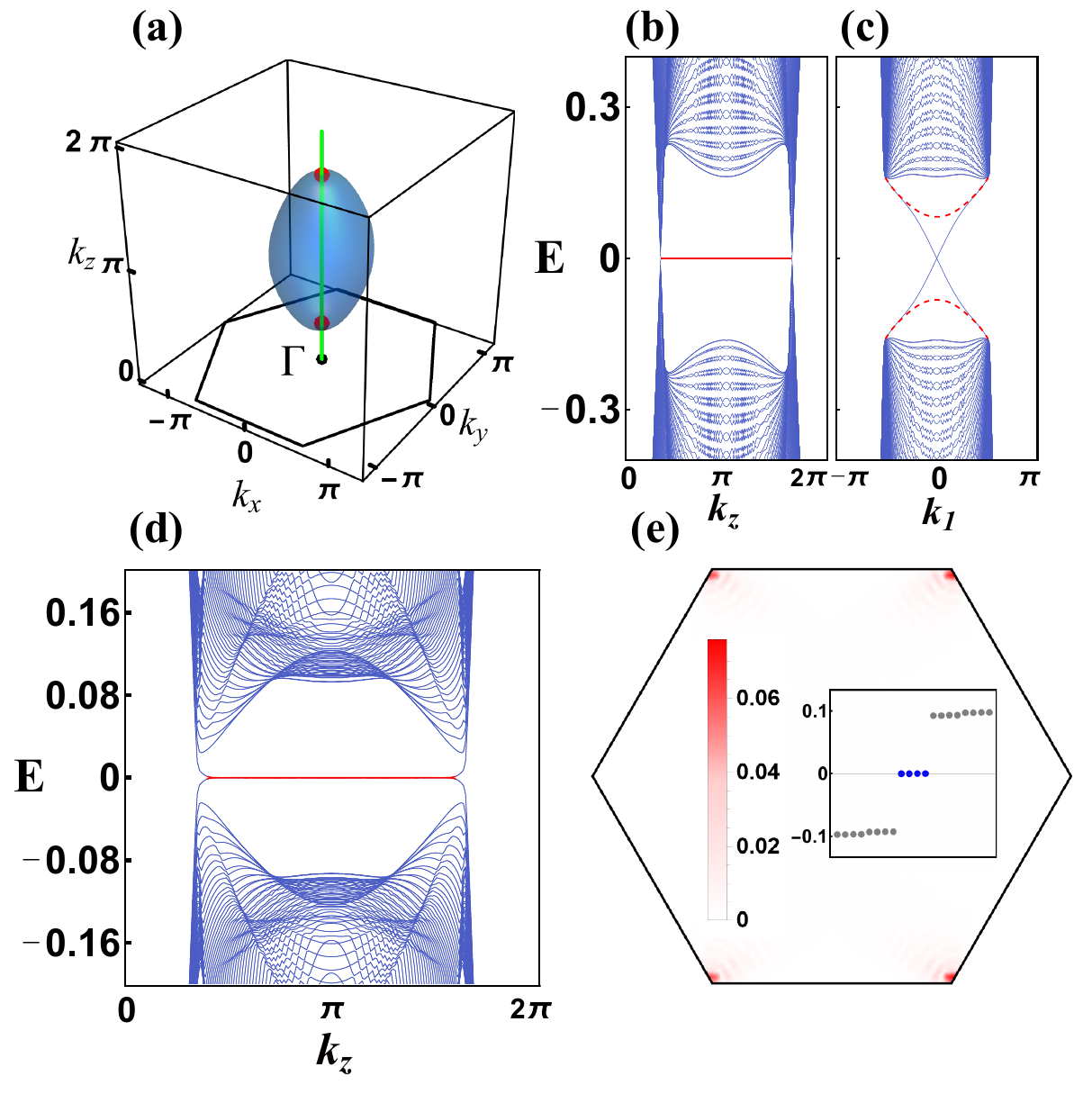}
    \caption{Odd-parity $s$-wave Dirac SC and its realization of higher-order topology. (a) FS (blue surface) and the effective DPs (red solid dots). (b) [(c)] Energy spectra of the $k_1=0$ ($k_z=\pi$) subsystem with boundary along $\bm{a}_z$ ($\bm{a}_1$) direction. The solid blue (dashed red) lines in (c) represent the edge states of $k_z=\pi$ before (after) the inclusion of the additional pairing potential $\Delta_e(\bm{k})$. (d) Energy spectra for an infinitely long hexagonal prism with side lengths of 40, where the red segments denote the Majorana hinge modes. (e) Real-space distribution of the Majorana corner states at $k_z = \pi$ in (d), where the insets show energy eigenvalues near zero. Parameters: $(t_z,t_2,t_z^{'},\gamma_{so},\lambda_{so},k_0,\mu,\Delta_3)=(0,0,1,0.5,0.5,0,-7,0.2)$, $\Delta_4=0.05$ and $0.15$ for (c) and (d)-(e), respectively.}\label{fig5}
\end{figure}

The edge states should satisfy the boundary condition $|\Phi(\theta)\rangle=0$ at $x_{\perp}=0$. By solving $H^{(0)}_{\mathrm{BdG}}(-i\partial,\theta)| \Phi(\theta)\rangle=0$, we reveal two zero-energy solutions: 
\begin{align}
    |\Phi_1(\theta)\rangle=\phi(x_{\parallel},x_{\perp})|\chi_1(\theta)\rangle,\notag\\|\Phi_2(\theta)\rangle=\phi(x_{\parallel},x_{\perp})|\chi_2(\theta)\rangle,
\end{align} 

where $\phi(x_{\parallel},x_{\perp})=\mathcal{N}e^{-\alpha x_{\perp}}\sin{(\beta x_\perp)}e^{i kx_{\parallel}}$, normalization constant $\mathcal{N}^2=4\alpha(\alpha^2+\beta^2)/\beta^2$, $\alpha=3\Delta_0/2b$ and $\beta=\sqrt{4bc-9\Delta_0^2}/2b$. In addition, spinors $|\chi_{1,2}(\theta)\rangle$ are eigenvectors of the matrix $\sin{\theta}\tau_1+\cos{\theta}\tau_2\sigma'_3$ with eigenvalue 1:
\begin{align}
    |\chi_{1}(\theta)\rangle&=\frac{1}{\sqrt{2}}\begin{pmatrix}
 ie^{-i\theta/2}\\
e^{i\theta/2}
\end{pmatrix}\otimes\begin{pmatrix}
 0\\1
\end{pmatrix},\notag\\|\chi_{2}(\theta)\rangle&=\frac{1}{\sqrt{2}}\begin{pmatrix}
 -ie^{i\theta/2}\\
e^{-i\theta/2}
\end{pmatrix}\otimes\begin{pmatrix}
 1\\0
\end{pmatrix}.
\end{align} Then, by projection of $H^{(1)}_{\mathrm{BdG}}(k,\theta)$ into the subspace of the above two zero-energy states, the low-energy edge Hamiltonian can be obtained: $H^{\mathrm{edge}}(k)=3\Delta_0 k\hat{\sigma}_3$, with $\hat{\bm{\sigma}}$ the Pauli matrices in the two-state subspace, indicating the existence of a pair of Majorana helical states at a generic boundary [see the edge states (blue solid lines) in Fig. \ref{fig2}(a)].

Next, we examine the effect of $\Delta_{e}(\bm{k})$ on the above pair of helical states. First of all, retaining $\Delta^{r}_{e}(\bm{k})$ to the lowest order in $k$ leads to $\psi_1(\bm{k})\sigma'_2$ which takes the form of $\psi_1(\bm{k})\tau_1\sigma'_2$ in the particle-hole representation. Then, the edge Hamiltonian $\Delta^{\mathrm{edge}}(\bm{k},\theta)$ is obtained by calculating $\Delta^{\mathrm{edge}}_{mn}(\bm{k},\theta)=\langle\Phi_{m}(\theta)|\psi_1(\bm{k})\tau_1\sigma'_2|\Phi_{n}(\theta)\rangle$ with $m, n=1,2$.
For the matrix part, we have
\begin{align}
    \begin{pmatrix}
 \langle\chi_1(\theta)| \tau_1\sigma'_2|\chi_1(\theta)\rangle& \langle\chi_1(\theta)|\tau_1\sigma'_2|\chi_2(\theta)\rangle\\
\langle\chi_2(\theta)| \tau_1\sigma'_2|\chi_1(\theta)\rangle &\langle\chi_2(\theta)|\tau_1\sigma'_2|\chi_2(\theta)\rangle
\end{pmatrix}=\hat{\sigma}_1.\label{eqD4}
\end{align}
In order to obtain other parts, we explicitly write out the expansion of $\psi_1(\bm{k})$ at $\bm{k}_{\perp}=(0,0)$: $\psi_1(k_x,k_y)\sim-\frac{\Delta_1}{4}[(4\delta_1+\delta_2+\delta_3)k_x^2+3(\delta_2+\delta_3)k_y^2+2\sqrt{3}(\delta_3-\delta_2)k_{x}k_{y}]$, where we have neglected constant term. Then, transforming $\psi_1(k_x,k_y)$ to $\psi_1(k, k_{\perp})$, we have $\psi_1(0, k_{\perp})=-\frac{\Delta_1k_{\perp}^2}{4}[2(\delta_1+\delta_2+\delta_3)+(\delta_2+\delta_3-2\delta_1)\cos{2\theta}+\sqrt{3}(\delta_2-\delta_3)\sin{2\theta}]$. Using $\int_{0}^{\infty}\phi^*(x_{\parallel},x_{\perp})\partial^{2}\phi(x_{\parallel},x_{\perp})\mathrm{d}x_{\perp}=-(\alpha^2+\beta^2)=-b/c$ and Eq. (\ref{eqD4}), we obtain $\Delta^{\mathrm{edge}}(0,\theta)=-\frac{c\Delta_1}{4b}[2(\delta_1+\delta_2+\delta_3)+(\delta_2+\delta_3-2\delta_1)\cos{2\theta}+\sqrt{3}(\delta_2-\delta_3)\sin{2\theta}]\hat{\sigma}_1$. When $\theta$ takes the values $0,-\frac{\pi}{3},\frac{\pi}{3}$, by neglecting unimportant proportionality coefficient, the effective edge gaps corresponding to $\boldsymbol{a}_{1}$, $\boldsymbol{a}_{2}$, and $\boldsymbol{a}_{3}$ directions are $\delta_2+\delta_3, \delta_1+\delta_3 $ and $\delta_1+\delta_2$, respectively. When we set $(\delta_1, \delta_2, \delta_3)=(1,-1/2,-1/2)$, Majorana corner modes are expected to emerge at the corners between the boundary along $\bm{a}_1$ direction and those along $\bm{a}_2$ or $\bm{a}_3$ directions [see Fig. \ref{fig2} (d) (h)]. So far, we have provided a rigorous proof of the conclusion in Sec. \ref{sec3B} by using the edge theory.

\section{ODD-PARITY $S$-WAVE DIRAC SC AND ITS REALIZATION OF HIGHER-ORDER TOPOLOGY}
\label{app5}
In this appendix, we start from a conventional $s$-wave pairing to realize the Dirac SC with its nodes located on the $k_z$ axis, and display higher-order topology within this SC.

The normal Hamiltonian is identical to that of the first scheme in Sec. \ref{sec3A}. Then we choose an odd-parity $s$-wave $\mathcal{T}$-invariant pairing potential $\Delta(\bm{k})=2\Delta_3i\sigma_2\rho_3$, which corresponds to a pseudo-spin triplet effective pairing $\tilde{\bm{d}}$ satisfying $|\tilde{\bm{d}}|^2=4\Delta_3^2(\gamma^2(\bm{k})+|\lambda(\bm{k})|^2)$ (see Table \ref{table1}). Thus, the effective DPs appear at the intersections of the FS and the $k_z$ axis—the zeros of $(\gamma^2(\bm{k})+|\lambda(\bm{k})|^2)$ [see Fig. \ref{fig5}(a)]. The energy spectra of the $k_1=0$ plane with boundary along $\bm{a}_z$ [Fig. \ref{fig5}(b)] show that surface Majorana arcs exist between the projections of the DPs on the $k_z$ axis. 

By introducing additionally an even-parity $\mathcal{T}$-breaking term $\Delta_e(\bm{k})=i\times\psi_0(\bm{k})i\sigma_2\rho_0$ with $\psi_0(\bm{k})$ the NN $d$-wave pairing: $\psi_0(\bm{k})=2\Delta_4(\cos{k_1}-\frac{1}{2}\cos{k_2}-\frac{1}{2}\cos{k_3})$, corresponding to an effective pseudo-spin singlet pairing potential $\tilde{\psi}(\bm{k})=i\psi_0(\bm{k})$, we can achieve higher-order topology in this SC with similar reasons in Sec. \ref{sec3B}. The relevant results are shown in Fig. \ref{fig5}.

\bibliography{ref} 

\begin{thebibliography}{94}%
\makeatletter
\providecommand \@ifxundefined [1]{%
 \@ifx{#1\undefined}
}%
\providecommand \@ifnum [1]{%
 \ifnum #1\expandafter \@firstoftwo
 \else \expandafter \@secondoftwo
 \fi
}%
\providecommand \@ifx [1]{%
 \ifx #1\expandafter \@firstoftwo
 \else \expandafter \@secondoftwo
 \fi
}%
\providecommand \natexlab [1]{#1}%
\providecommand \enquote  [1]{``#1''}%
\providecommand \bibnamefont  [1]{#1}%
\providecommand \bibfnamefont [1]{#1}%
\providecommand \citenamefont [1]{#1}%
\providecommand \href@noop [0]{\@secondoftwo}%
\providecommand \href [0]{\begingroup \@sanitize@url \@href}%
\providecommand \@href[1]{\@@startlink{#1}\@@href}%
\providecommand \@@href[1]{\endgroup#1\@@endlink}%
\providecommand \@sanitize@url [0]{\catcode `\\12\catcode `\$12\catcode
  `\&12\catcode `\#12\catcode `\^12\catcode `\_12\catcode `\%12\relax}%
\providecommand \@@startlink[1]{}%
\providecommand \@@endlink[0]{}%
\providecommand \url  [0]{\begingroup\@sanitize@url \@url }%
\providecommand \@url [1]{\endgroup\@href {#1}{\urlprefix }}%
\providecommand \urlprefix  [0]{URL }%
\providecommand \Eprint [0]{\href }%
\providecommand \doibase [0]{https://doi.org/}%
\providecommand \selectlanguage [0]{\@gobble}%
\providecommand \bibinfo  [0]{\@secondoftwo}%
\providecommand \bibfield  [0]{\@secondoftwo}%
\providecommand \translation [1]{[#1]}%
\providecommand \BibitemOpen [0]{}%
\providecommand \bibitemStop [0]{}%
\providecommand \bibitemNoStop [0]{.\EOS\space}%
\providecommand \EOS [0]{\spacefactor3000\relax}%
\providecommand \BibitemShut  [1]{\csname bibitem#1\endcsname}%
\let\auto@bib@innerbib\@empty
%</preamble>
\bibitem [{\citenamefont {Kitaev}(2001)}]{TSC-2001-Kitaev}%
  \BibitemOpen
  \bibfield  {author} {\bibinfo {author} {\bibfnamefont {A.~Y.}\ \bibnamefont
  {Kitaev}},\ }\bibfield  {title} {\bibinfo {title} {{Unpaired Majorana
  fermions in quantum wires}},\ }\href
  {https://doi.org/10.1070/1063-7869/44/10S/S29} {\bibfield  {journal}
  {\bibinfo  {journal} {Phys. Usp.}\ }\textbf {\bibinfo {volume} {44}},\
  \bibinfo {pages} {131} (\bibinfo {year} {2001})}\BibitemShut {NoStop}%
\bibitem [{\citenamefont {Fu}\ and\ \citenamefont
  {Kane}(2008)}]{TSC_2008_prl_proximity}%
  \BibitemOpen
  \bibfield  {author} {\bibinfo {author} {\bibfnamefont {L.}~\bibnamefont
  {Fu}}\ and\ \bibinfo {author} {\bibfnamefont {C.~L.}\ \bibnamefont {Kane}},\
  }\bibfield  {title} {\bibinfo {title} {{Superconducting Proximity Effect and
  Majorana Fermions at the Surface of a Topological Insulator}},\ }\href
  {https://doi.org/10.1103/PhysRevLett.100.096407} {\bibfield  {journal}
  {\bibinfo  {journal} {Phys. Rev. Lett.}\ }\textbf {\bibinfo {volume} {100}},\
  \bibinfo {pages} {096407} (\bibinfo {year} {2008})}\BibitemShut {NoStop}%
\bibitem [{\citenamefont {Hasan}\ and\ \citenamefont
  {Kane}(2010)}]{TSC_2010_RMP_Kane_review}%
  \BibitemOpen
  \bibfield  {author} {\bibinfo {author} {\bibfnamefont {M.~Z.}\ \bibnamefont
  {Hasan}}\ and\ \bibinfo {author} {\bibfnamefont {C.~L.}\ \bibnamefont
  {Kane}},\ }\bibfield  {title} {\bibinfo {title} {{Colloquium: Topological
  insulators}},\ }\href {https://doi.org/10.1103/RevModPhys.82.3045} {\bibfield
   {journal} {\bibinfo  {journal} {Rev. Mod. Phys.}\ }\textbf {\bibinfo
  {volume} {82}},\ \bibinfo {pages} {3045} (\bibinfo {year}
  {2010})}\BibitemShut {NoStop}%
\bibitem [{\citenamefont {Qi}\ and\ \citenamefont
  {Zhang}(2011)}]{TSC_2011_RMP_ZSC_QXL}%
  \BibitemOpen
  \bibfield  {author} {\bibinfo {author} {\bibfnamefont {X.-L.}\ \bibnamefont
  {Qi}}\ and\ \bibinfo {author} {\bibfnamefont {S.-C.}\ \bibnamefont {Zhang}},\
  }\bibfield  {title} {\bibinfo {title} {Topological insulators and
  superconductors},\ }\href {https://doi.org/10.1103/RevModPhys.83.1057}
  {\bibfield  {journal} {\bibinfo  {journal} {Rev. Mod. Phys.}\ }\textbf
  {\bibinfo {volume} {83}},\ \bibinfo {pages} {1057} (\bibinfo {year}
  {2011})}\BibitemShut {NoStop}%
\bibitem [{\citenamefont {Mourik}\ \emph {et~al.}(2012)\citenamefont {Mourik},
  \citenamefont {Zuo}, \citenamefont {Frolov}, \citenamefont {Plissard},
  \citenamefont {Bakkers},\ and\ \citenamefont
  {Kouwenhoven}}]{TSC_2012_MF_sci}%
  \BibitemOpen
  \bibfield  {author} {\bibinfo {author} {\bibfnamefont {V.}~\bibnamefont
  {Mourik}}, \bibinfo {author} {\bibfnamefont {K.}~\bibnamefont {Zuo}},
  \bibinfo {author} {\bibfnamefont {S.~M.}\ \bibnamefont {Frolov}}, \bibinfo
  {author} {\bibfnamefont {S.~R.}\ \bibnamefont {Plissard}}, \bibinfo {author}
  {\bibfnamefont {E.~P. A.~M.}\ \bibnamefont {Bakkers}},\ and\ \bibinfo
  {author} {\bibfnamefont {L.~P.}\ \bibnamefont {Kouwenhoven}},\ }\bibfield
  {title} {\bibinfo {title} {{Signatures of Majorana Fermions in Hybrid
  Superconductor-Semiconductor Nanowire Devices}},\ }\href
  {https://doi.org/10.1126/science.1222360} {\bibfield  {journal} {\bibinfo
  {journal} {Science}\ }\textbf {\bibinfo {volume} {336}},\ \bibinfo {pages}
  {1003} (\bibinfo {year} {2012})}\BibitemShut {NoStop}%
\bibitem [{\citenamefont {Das}\ \emph {et~al.}(2012)\citenamefont {Das},
  \citenamefont {Ronen}, \citenamefont {Most}, \citenamefont {Oreg},
  \citenamefont {Heiblum},\ and\ \citenamefont {Shtrikman}}]{TSC_2012_nature}%
  \BibitemOpen
  \bibfield  {author} {\bibinfo {author} {\bibfnamefont {A.}~\bibnamefont
  {Das}}, \bibinfo {author} {\bibfnamefont {Y.}~\bibnamefont {Ronen}}, \bibinfo
  {author} {\bibfnamefont {Y.}~\bibnamefont {Most}}, \bibinfo {author}
  {\bibfnamefont {Y.}~\bibnamefont {Oreg}}, \bibinfo {author} {\bibfnamefont
  {M.}~\bibnamefont {Heiblum}},\ and\ \bibinfo {author} {\bibfnamefont
  {H.}~\bibnamefont {Shtrikman}},\ }\bibfield  {title} {\bibinfo {title}
  {Zero-bias peaks and splitting in an al–{InAs} nanowire topological
  superconductor as a signature of majorana fermions},\ }\href
  {https://doi.org/10.1038/nphys2479} {\bibfield  {journal} {\bibinfo
  {journal} {Nat. Phys.}\ }\textbf {\bibinfo {volume} {8}},\ \bibinfo {pages}
  {887} (\bibinfo {year} {2012})}\BibitemShut {NoStop}%
\bibitem [{\citenamefont {Alicea}(2012)}]{TSC_2012_RPP_MF}%
  \BibitemOpen
  \bibfield  {author} {\bibinfo {author} {\bibfnamefont {J.}~\bibnamefont
  {Alicea}},\ }\bibfield  {title} {\bibinfo {title} {New directions in the
  pursuit of majorana fermions in solid state systems},\ }\href
  {https://doi.org/10.1088/0034-4885/75/7/076501} {\bibfield  {journal}
  {\bibinfo  {journal} {Rep. Prog. Phys.}\ }\textbf {\bibinfo {volume} {75}},\
  \bibinfo {pages} {076501} (\bibinfo {year} {2012})}\BibitemShut {NoStop}%
\bibitem [{\citenamefont {Leijnse}\ and\ \citenamefont
  {Flensberg}(2012)}]{TSC_2012_SST_MF}%
  \BibitemOpen
  \bibfield  {author} {\bibinfo {author} {\bibfnamefont {M.}~\bibnamefont
  {Leijnse}}\ and\ \bibinfo {author} {\bibfnamefont {K.}~\bibnamefont
  {Flensberg}},\ }\bibfield  {title} {\bibinfo {title} {{Introduction to
  topological superconductivity and Majorana fermions}},\ }\href
  {https://doi.org/10.1088/0268-1242/27/12/124003} {\bibfield  {journal}
  {\bibinfo  {journal} {Semicond. Sci. Technol.}\ }\textbf {\bibinfo {volume}
  {27}},\ \bibinfo {pages} {124003} (\bibinfo {year} {2012})}\BibitemShut
  {NoStop}%
\bibitem [{\citenamefont {Beenakker}(2013)}]{TSC_2013_MF_3}%
  \BibitemOpen
  \bibfield  {author} {\bibinfo {author} {\bibfnamefont {C.}~\bibnamefont
  {Beenakker}},\ }\bibfield  {title} {\bibinfo {title} {{Search for Majorana
  Fermions in Superconductors}},\ }\href
  {https://doi.org/10.1146/annurev-conmatphys-030212-184337} {\bibfield
  {journal} {\bibinfo  {journal} {Annu. Rev. Condens. Matter Phys.}\ }\textbf
  {\bibinfo {volume} {4}},\ \bibinfo {pages} {113} (\bibinfo {year}
  {2013})}\BibitemShut {NoStop}%
\bibitem [{\citenamefont {Sato}\ and\ \citenamefont
  {Fujimoto}(2016)}]{TSC_2016_sato_review}%
  \BibitemOpen
  \bibfield  {author} {\bibinfo {author} {\bibfnamefont {M.}~\bibnamefont
  {Sato}}\ and\ \bibinfo {author} {\bibfnamefont {S.}~\bibnamefont
  {Fujimoto}},\ }\bibfield  {title} {\bibinfo {title} {{Majorana Fermions and
  Topology in Superconductors}},\ }\href
  {https://doi.org/10.7566/jpsj.85.072001} {\bibfield  {journal} {\bibinfo
  {journal} {J. Phys. Soc. Jpn.}\ }\textbf {\bibinfo {volume} {85}},\ \bibinfo
  {pages} {072001} (\bibinfo {year} {2016})}\BibitemShut {NoStop}%
\bibitem [{\citenamefont {Kozii}\ \emph {et~al.}(2016)\citenamefont {Kozii},
  \citenamefont {Venderbos},\ and\ \citenamefont {Fu}}]{TSC-2016-sci}%
  \BibitemOpen
  \bibfield  {author} {\bibinfo {author} {\bibfnamefont {V.}~\bibnamefont
  {Kozii}}, \bibinfo {author} {\bibfnamefont {J.~W.~F.}\ \bibnamefont
  {Venderbos}},\ and\ \bibinfo {author} {\bibfnamefont {L.}~\bibnamefont
  {Fu}},\ }\bibfield  {title} {\bibinfo {title} {{Three-dimensional Majorana
  fermions in chiral superconductors}},\ }\href
  {https://doi.org/10.1126/sciadv.1601835} {\bibfield  {journal} {\bibinfo
  {journal} {Sci. Adv.}\ }\textbf {\bibinfo {volume} {2}},\ \bibinfo {pages}
  {e1601835} (\bibinfo {year} {2016})}\BibitemShut {NoStop}%
\bibitem [{\citenamefont {Sato}\ and\ \citenamefont
  {Ando}(2017)}]{TSC_2017_sato_review}%
  \BibitemOpen
  \bibfield  {author} {\bibinfo {author} {\bibfnamefont {M.}~\bibnamefont
  {Sato}}\ and\ \bibinfo {author} {\bibfnamefont {Y.}~\bibnamefont {Ando}},\
  }\bibfield  {title} {\bibinfo {title} {Topological superconductors: a
  review},\ }\href {https://doi.org/10.1088/1361-6633/aa6ac7} {\bibfield
  {journal} {\bibinfo  {journal} {Rep. Prog. Phys.}\ }\textbf {\bibinfo
  {volume} {80}},\ \bibinfo {pages} {076501} (\bibinfo {year}
  {2017})}\BibitemShut {NoStop}%
\bibitem [{\citenamefont {Zhang}\ \emph {et~al.}(2018)\citenamefont {Zhang},
  \citenamefont {Yaji}, \citenamefont {Hashimoto}, \citenamefont {Ota},
  \citenamefont {Kondo}, \citenamefont {Okazaki}, \citenamefont {Wang},
  \citenamefont {Wen}, \citenamefont {Gu}, \citenamefont {Ding},\ and\
  \citenamefont {Shin}}]{TSC_2018_sci}%
  \BibitemOpen
  \bibfield  {author} {\bibinfo {author} {\bibfnamefont {P.}~\bibnamefont
  {Zhang}}, \bibinfo {author} {\bibfnamefont {K.}~\bibnamefont {Yaji}},
  \bibinfo {author} {\bibfnamefont {T.}~\bibnamefont {Hashimoto}}, \bibinfo
  {author} {\bibfnamefont {Y.}~\bibnamefont {Ota}}, \bibinfo {author}
  {\bibfnamefont {T.}~\bibnamefont {Kondo}}, \bibinfo {author} {\bibfnamefont
  {K.}~\bibnamefont {Okazaki}}, \bibinfo {author} {\bibfnamefont
  {Z.}~\bibnamefont {Wang}}, \bibinfo {author} {\bibfnamefont {J.}~\bibnamefont
  {Wen}}, \bibinfo {author} {\bibfnamefont {G.~D.}\ \bibnamefont {Gu}},
  \bibinfo {author} {\bibfnamefont {H.}~\bibnamefont {Ding}},\ and\ \bibinfo
  {author} {\bibfnamefont {S.}~\bibnamefont {Shin}},\ }\bibfield  {title}
  {\bibinfo {title} {Observation of topological superconductivity on the
  surface of an iron-based superconductor},\ }\href
  {https://doi.org/10.1126/science.aan4596} {\bibfield  {journal} {\bibinfo
  {journal} {Science}\ }\textbf {\bibinfo {volume} {360}},\ \bibinfo {pages}
  {182} (\bibinfo {year} {2018})}\BibitemShut {NoStop}%
\bibitem [{\citenamefont {Zhang}\ \emph {et~al.}(2024)\citenamefont {Zhang},
  \citenamefont {Wu}, \citenamefont {Fang}, \citenamefont {Zhang},
  \citenamefont {Hu}, \citenamefont {Wang},\ and\ \citenamefont
  {Qin}}]{TSC_2024_NC}%
  \BibitemOpen
  \bibfield  {author} {\bibinfo {author} {\bibfnamefont {Z.}~\bibnamefont
  {Zhang}}, \bibinfo {author} {\bibfnamefont {Z.}~\bibnamefont {Wu}}, \bibinfo
  {author} {\bibfnamefont {C.}~\bibnamefont {Fang}}, \bibinfo {author}
  {\bibfnamefont {F.-c.}\ \bibnamefont {Zhang}}, \bibinfo {author}
  {\bibfnamefont {J.}~\bibnamefont {Hu}}, \bibinfo {author} {\bibfnamefont
  {Y.}~\bibnamefont {Wang}},\ and\ \bibinfo {author} {\bibfnamefont
  {S.}~\bibnamefont {Qin}},\ }\bibfield  {title} {\bibinfo {title} {Topological
  superconductivity from unconventional band degeneracy with conventional
  pairing},\ }\href {https://doi.org/10.1038/s41467-024-52156-1} {\bibfield
  {journal} {\bibinfo  {journal} {Nat. Commun.}\ }\textbf {\bibinfo {volume}
  {15}},\ \bibinfo {pages} {7971} (\bibinfo {year} {2024})}\BibitemShut
  {NoStop}%
\bibitem [{\citenamefont {Kitaev}(2003)}]{Q_computate_1_kitaev}%
  \BibitemOpen
  \bibfield  {author} {\bibinfo {author} {\bibfnamefont {A.~Y.}\ \bibnamefont
  {Kitaev}},\ }\bibfield  {title} {\bibinfo {title} {Fault-tolerant quantum
  computation by anyons},\ }\href
  {https://doi.org/https://doi.org/10.1016/S0003-4916(02)00018-0} {\bibfield
  {journal} {\bibinfo  {journal} {Ann. Phys.}\ }\textbf {\bibinfo {volume}
  {303}},\ \bibinfo {pages} {2} (\bibinfo {year} {2003})}\BibitemShut {NoStop}%
\bibitem [{\citenamefont {Nayak}\ \emph {et~al.}(2008)\citenamefont {Nayak},
  \citenamefont {Simon}, \citenamefont {Stern}, \citenamefont {Freedman},\ and\
  \citenamefont {Das~Sarma}}]{Q_computate_2}%
  \BibitemOpen
  \bibfield  {author} {\bibinfo {author} {\bibfnamefont {C.}~\bibnamefont
  {Nayak}}, \bibinfo {author} {\bibfnamefont {S.~H.}\ \bibnamefont {Simon}},
  \bibinfo {author} {\bibfnamefont {A.}~\bibnamefont {Stern}}, \bibinfo
  {author} {\bibfnamefont {M.}~\bibnamefont {Freedman}},\ and\ \bibinfo
  {author} {\bibfnamefont {S.}~\bibnamefont {Das~Sarma}},\ }\bibfield  {title}
  {\bibinfo {title} {{Non-Abelian anyons and topological quantum
  computation}},\ }\href {https://doi.org/10.1103/RevModPhys.80.1083}
  {\bibfield  {journal} {\bibinfo  {journal} {Rev. Mod. Phys.}\ }\textbf
  {\bibinfo {volume} {80}},\ \bibinfo {pages} {1083} (\bibinfo {year}
  {2008})}\BibitemShut {NoStop}%
\bibitem [{\citenamefont {Lian}\ \emph {et~al.}(2018)\citenamefont {Lian},
  \citenamefont {Sun}, \citenamefont {Vaezi}, \citenamefont {Qi},\ and\
  \citenamefont {Zhang}}]{Q_computate_3}%
  \BibitemOpen
  \bibfield  {author} {\bibinfo {author} {\bibfnamefont {B.}~\bibnamefont
  {Lian}}, \bibinfo {author} {\bibfnamefont {X.-Q.}\ \bibnamefont {Sun}},
  \bibinfo {author} {\bibfnamefont {A.}~\bibnamefont {Vaezi}}, \bibinfo
  {author} {\bibfnamefont {X.-L.}\ \bibnamefont {Qi}},\ and\ \bibinfo {author}
  {\bibfnamefont {S.-C.}\ \bibnamefont {Zhang}},\ }\bibfield  {title} {\bibinfo
  {title} {{Topological quantum computation based on chiral Majorana
  fermions}},\ }\href {https://doi.org/10.1073/pnas.1810003115} {\bibfield
  {journal} {\bibinfo  {journal} {Proc. Natl. Acad. Sci. USA}\ }\textbf
  {\bibinfo {volume} {115}},\ \bibinfo {pages} {10938} (\bibinfo {year}
  {2018})}\BibitemShut {NoStop}%
\bibitem [{\citenamefont {Sato}(2009)}]{Sato_Z2_2009}%
  \BibitemOpen
  \bibfield  {author} {\bibinfo {author} {\bibfnamefont {M.}~\bibnamefont
  {Sato}},\ }\bibfield  {title} {\bibinfo {title} {{Topological properties of
  spin-triplet superconductors and Fermi surface topology in the normal
  state}},\ }\href {https://doi.org/10.1103/PhysRevB.79.214526} {\bibfield
  {journal} {\bibinfo  {journal} {Phys. Rev. B}\ }\textbf {\bibinfo {volume}
  {79}},\ \bibinfo {pages} {214526} (\bibinfo {year} {2009})}\BibitemShut
  {NoStop}%
\bibitem [{\citenamefont {Fu}\ and\ \citenamefont
  {Berg}(2010)}]{TSC_2010_prl_TRS_fu}%
  \BibitemOpen
  \bibfield  {author} {\bibinfo {author} {\bibfnamefont {L.}~\bibnamefont
  {Fu}}\ and\ \bibinfo {author} {\bibfnamefont {E.}~\bibnamefont {Berg}},\
  }\bibfield  {title} {\bibinfo {title} {{Odd-Parity Topological
  Superconductors: Theory and Application to
  ${\mathrm{Cu}}_{x}{\mathrm{Bi}}_{2}{\mathrm{Se}}_{3}$}},\ }\href
  {https://doi.org/10.1103/PhysRevLett.105.097001} {\bibfield  {journal}
  {\bibinfo  {journal} {Phys. Rev. Lett.}\ }\textbf {\bibinfo {volume} {105}},\
  \bibinfo {pages} {097001} (\bibinfo {year} {2010})}\BibitemShut {NoStop}%
\bibitem [{\citenamefont {Sato}(2010)}]{Sato_Z2_2010}%
  \BibitemOpen
  \bibfield  {author} {\bibinfo {author} {\bibfnamefont {M.}~\bibnamefont
  {Sato}},\ }\bibfield  {title} {\bibinfo {title} {{Topological odd-parity
  superconductors}},\ }\href {https://doi.org/10.1103/PhysRevB.81.220504}
  {\bibfield  {journal} {\bibinfo  {journal} {Phys. Rev. B}\ }\textbf {\bibinfo
  {volume} {81}},\ \bibinfo {pages} {220504} (\bibinfo {year}
  {2010})}\BibitemShut {NoStop}%
\bibitem [{\citenamefont {Langbehn}\ \emph {et~al.}(2017)\citenamefont
  {Langbehn}, \citenamefont {Peng}, \citenamefont {Trifunovic}, \citenamefont
  {von Oppen},\ and\ \citenamefont {Brouwer}}]{HO_2017_prl_reflection}%
  \BibitemOpen
  \bibfield  {author} {\bibinfo {author} {\bibfnamefont {J.}~\bibnamefont
  {Langbehn}}, \bibinfo {author} {\bibfnamefont {Y.}~\bibnamefont {Peng}},
  \bibinfo {author} {\bibfnamefont {L.}~\bibnamefont {Trifunovic}}, \bibinfo
  {author} {\bibfnamefont {F.}~\bibnamefont {von Oppen}},\ and\ \bibinfo
  {author} {\bibfnamefont {P.~W.}\ \bibnamefont {Brouwer}},\ }\bibfield
  {title} {\bibinfo {title} {{Reflection-Symmetric Second-Order Topological
  Insulators and Superconductors}},\ }\href
  {https://doi.org/10.1103/PhysRevLett.119.246401} {\bibfield  {journal}
  {\bibinfo  {journal} {Phys. Rev. Lett.}\ }\textbf {\bibinfo {volume} {119}},\
  \bibinfo {pages} {246401} (\bibinfo {year} {2017})}\BibitemShut {NoStop}%
\bibitem [{\citenamefont {Zhu}(2018)}]{HO_2018_prb}%
  \BibitemOpen
  \bibfield  {author} {\bibinfo {author} {\bibfnamefont {X.}~\bibnamefont
  {Zhu}},\ }\bibfield  {title} {\bibinfo {title} {{Tunable Majorana corner
  states in a two-dimensional second-order topological superconductor induced
  by magnetic fields}},\ }\href {https://doi.org/10.1103/PhysRevB.97.205134}
  {\bibfield  {journal} {\bibinfo  {journal} {Phys. Rev. B}\ }\textbf {\bibinfo
  {volume} {97}},\ \bibinfo {pages} {205134} (\bibinfo {year}
  {2018})}\BibitemShut {NoStop}%
\bibitem [{\citenamefont {Wang}\ \emph
  {et~al.}(2018{\natexlab{a}})\citenamefont {Wang}, \citenamefont {Lin},\ and\
  \citenamefont {Hughes}}]{HO_2018_prb_2}%
  \BibitemOpen
  \bibfield  {author} {\bibinfo {author} {\bibfnamefont {Y.}~\bibnamefont
  {Wang}}, \bibinfo {author} {\bibfnamefont {M.}~\bibnamefont {Lin}},\ and\
  \bibinfo {author} {\bibfnamefont {T.~L.}\ \bibnamefont {Hughes}},\ }\bibfield
   {title} {\bibinfo {title} {{Weak-pairing higher order topological
  superconductors}},\ }\href {https://doi.org/10.1103/PhysRevB.98.165144}
  {\bibfield  {journal} {\bibinfo  {journal} {Phys. Rev. B}\ }\textbf {\bibinfo
  {volume} {98}},\ \bibinfo {pages} {165144} (\bibinfo {year}
  {2018}{\natexlab{a}})}\BibitemShut {NoStop}%
\bibitem [{\citenamefont {Khalaf}(2018)}]{HO_2018_prb_inversion}%
  \BibitemOpen
  \bibfield  {author} {\bibinfo {author} {\bibfnamefont {E.}~\bibnamefont
  {Khalaf}},\ }\bibfield  {title} {\bibinfo {title} {Higher-order topological
  insulators and superconductors protected by inversion symmetry},\ }\href
  {https://doi.org/10.1103/PhysRevB.97.205136} {\bibfield  {journal} {\bibinfo
  {journal} {Phys. Rev. B}\ }\textbf {\bibinfo {volume} {97}},\ \bibinfo
  {pages} {205136} (\bibinfo {year} {2018})}\BibitemShut {NoStop}%
\bibitem [{\citenamefont {Yan}\ \emph {et~al.}(2018)\citenamefont {Yan},
  \citenamefont {Song},\ and\ \citenamefont {Wang}}]{HO_2018_prl_WZ}%
  \BibitemOpen
  \bibfield  {author} {\bibinfo {author} {\bibfnamefont {Z.}~\bibnamefont
  {Yan}}, \bibinfo {author} {\bibfnamefont {F.}~\bibnamefont {Song}},\ and\
  \bibinfo {author} {\bibfnamefont {Z.}~\bibnamefont {Wang}},\ }\bibfield
  {title} {\bibinfo {title} {{Majorana Corner Modes in a High-Temperature
  Platform}},\ }\href {https://doi.org/10.1103/PhysRevLett.121.096803}
  {\bibfield  {journal} {\bibinfo  {journal} {Phys. Rev. Lett.}\ }\textbf
  {\bibinfo {volume} {121}},\ \bibinfo {pages} {096803} (\bibinfo {year}
  {2018})}\BibitemShut {NoStop}%
\bibitem [{\citenamefont {Wang}\ \emph
  {et~al.}(2018{\natexlab{b}})\citenamefont {Wang}, \citenamefont {Liu},
  \citenamefont {Lu},\ and\ \citenamefont {Zhang}}]{HO_2018_prl_2}%
  \BibitemOpen
  \bibfield  {author} {\bibinfo {author} {\bibfnamefont {Q.}~\bibnamefont
  {Wang}}, \bibinfo {author} {\bibfnamefont {C.-C.}\ \bibnamefont {Liu}},
  \bibinfo {author} {\bibfnamefont {Y.-M.}\ \bibnamefont {Lu}},\ and\ \bibinfo
  {author} {\bibfnamefont {F.}~\bibnamefont {Zhang}},\ }\bibfield  {title}
  {\bibinfo {title} {{High-Temperature Majorana Corner States}},\ }\href
  {https://doi.org/10.1103/PhysRevLett.121.186801} {\bibfield  {journal}
  {\bibinfo  {journal} {Phys. Rev. Lett.}\ }\textbf {\bibinfo {volume} {121}},\
  \bibinfo {pages} {186801} (\bibinfo {year} {2018}{\natexlab{b}})}\BibitemShut
  {NoStop}%
\bibitem [{\citenamefont {Zhang}\ \emph {et~al.}(2019)\citenamefont {Zhang},
  \citenamefont {Cole}, \citenamefont {Wu},\ and\ \citenamefont
  {Das~Sarma}}]{HO_2019_prl_ZRX}%
  \BibitemOpen
  \bibfield  {author} {\bibinfo {author} {\bibfnamefont {R.-X.}\ \bibnamefont
  {Zhang}}, \bibinfo {author} {\bibfnamefont {W.~S.}\ \bibnamefont {Cole}},
  \bibinfo {author} {\bibfnamefont {X.}~\bibnamefont {Wu}},\ and\ \bibinfo
  {author} {\bibfnamefont {S.}~\bibnamefont {Das~Sarma}},\ }\bibfield  {title}
  {\bibinfo {title} {{Higher-Order Topology and Nodal Topological
  Superconductivity in Fe(Se,Te) Heterostructures}},\ }\href
  {https://doi.org/10.1103/PhysRevLett.123.167001} {\bibfield  {journal}
  {\bibinfo  {journal} {Phys. Rev. Lett.}\ }\textbf {\bibinfo {volume} {123}},\
  \bibinfo {pages} {167001} (\bibinfo {year} {2019})}\BibitemShut {NoStop}%
\bibitem [{\citenamefont {Pan}\ \emph {et~al.}(2019)\citenamefont {Pan},
  \citenamefont {Yang}, \citenamefont {Chen}, \citenamefont {Xu}, \citenamefont
  {Liu},\ and\ \citenamefont {Liu}}]{HO_2019_prl_LCX}%
  \BibitemOpen
  \bibfield  {author} {\bibinfo {author} {\bibfnamefont {X.-H.}\ \bibnamefont
  {Pan}}, \bibinfo {author} {\bibfnamefont {K.-J.}\ \bibnamefont {Yang}},
  \bibinfo {author} {\bibfnamefont {L.}~\bibnamefont {Chen}}, \bibinfo {author}
  {\bibfnamefont {G.}~\bibnamefont {Xu}}, \bibinfo {author} {\bibfnamefont
  {C.-X.}\ \bibnamefont {Liu}},\ and\ \bibinfo {author} {\bibfnamefont
  {X.}~\bibnamefont {Liu}},\ }\bibfield  {title} {\bibinfo {title}
  {{Lattice-Symmetry-Assisted Second-Order Topological Superconductors and
  Majorana Patterns}},\ }\href {https://doi.org/10.1103/PhysRevLett.123.156801}
  {\bibfield  {journal} {\bibinfo  {journal} {Phys. Rev. Lett.}\ }\textbf
  {\bibinfo {volume} {123}},\ \bibinfo {pages} {156801} (\bibinfo {year}
  {2019})}\BibitemShut {NoStop}%
\bibitem [{\citenamefont {Yan}(2019)}]{HO-2019-prl-yan}%
  \BibitemOpen
  \bibfield  {author} {\bibinfo {author} {\bibfnamefont {Z.}~\bibnamefont
  {Yan}},\ }\bibfield  {title} {\bibinfo {title} {{Higher-Order Topological
  Odd-Parity Superconductors}},\ }\href
  {https://doi.org/10.1103/PhysRevLett.123.177001} {\bibfield  {journal}
  {\bibinfo  {journal} {Phys. Rev. Lett.}\ }\textbf {\bibinfo {volume} {123}},\
  \bibinfo {pages} {177001} (\bibinfo {year} {2019})}\BibitemShut {NoStop}%
\bibitem [{\citenamefont {Wu}\ \emph {et~al.}(2020)\citenamefont {Wu},
  \citenamefont {Hou}, \citenamefont {Li}, \citenamefont {Luo}, \citenamefont
  {Shi},\ and\ \citenamefont {Zhang}}]{HO_2020_prl_2}%
  \BibitemOpen
  \bibfield  {author} {\bibinfo {author} {\bibfnamefont {Y.-J.}\ \bibnamefont
  {Wu}}, \bibinfo {author} {\bibfnamefont {J.}~\bibnamefont {Hou}}, \bibinfo
  {author} {\bibfnamefont {Y.-M.}\ \bibnamefont {Li}}, \bibinfo {author}
  {\bibfnamefont {X.-W.}\ \bibnamefont {Luo}}, \bibinfo {author} {\bibfnamefont
  {X.}~\bibnamefont {Shi}},\ and\ \bibinfo {author} {\bibfnamefont
  {C.}~\bibnamefont {Zhang}},\ }\bibfield  {title} {\bibinfo {title} {{In-Plane
  Zeeman-Field-Induced Majorana Corner and Hinge Modes in an $s$-Wave
  Superconductor Heterostructure}},\ }\href
  {https://doi.org/10.1103/PhysRevLett.124.227001} {\bibfield  {journal}
  {\bibinfo  {journal} {Phys. Rev. Lett.}\ }\textbf {\bibinfo {volume} {124}},\
  \bibinfo {pages} {227001} (\bibinfo {year} {2020})}\BibitemShut {NoStop}%
\bibitem [{\citenamefont {Ahn}\ and\ \citenamefont
  {Yang}(2020)}]{HO-NSC-2020-prb}%
  \BibitemOpen
  \bibfield  {author} {\bibinfo {author} {\bibfnamefont {J.}~\bibnamefont
  {Ahn}}\ and\ \bibinfo {author} {\bibfnamefont {B.-J.}\ \bibnamefont {Yang}},\
  }\bibfield  {title} {\bibinfo {title} {{Higher-order topological
  superconductivity of spin-polarized fermions}},\ }\href
  {https://doi.org/10.1103/PhysRevResearch.2.012060} {\bibfield  {journal}
  {\bibinfo  {journal} {Phys. Rev. Res.}\ }\textbf {\bibinfo {volume} {2}},\
  \bibinfo {pages} {012060} (\bibinfo {year} {2020})}\BibitemShut {NoStop}%
\bibitem [{\citenamefont {Qin}\ \emph {et~al.}(2022)\citenamefont {Qin},
  \citenamefont {Fang}, \citenamefont {Zhang},\ and\ \citenamefont
  {Hu}}]{HO_2022_prx}%
  \BibitemOpen
  \bibfield  {author} {\bibinfo {author} {\bibfnamefont {S.}~\bibnamefont
  {Qin}}, \bibinfo {author} {\bibfnamefont {C.}~\bibnamefont {Fang}}, \bibinfo
  {author} {\bibfnamefont {F.-C.}\ \bibnamefont {Zhang}},\ and\ \bibinfo
  {author} {\bibfnamefont {J.}~\bibnamefont {Hu}},\ }\bibfield  {title}
  {\bibinfo {title} {Topological superconductivity in an extended $s$-wave
  superconductor and its implication to iron-based superconductors},\ }\href
  {https://doi.org/10.1103/PhysRevX.12.011030} {\bibfield  {journal} {\bibinfo
  {journal} {Phys. Rev. X}\ }\textbf {\bibinfo {volume} {12}},\ \bibinfo
  {pages} {011030} (\bibinfo {year} {2022})}\BibitemShut {NoStop}%
\bibitem [{\citenamefont {Nakai}\ and\ \citenamefont
  {Nomura}(2023)}]{HO-NSC-2023-prb}%
  \BibitemOpen
  \bibfield  {author} {\bibinfo {author} {\bibfnamefont {R.}~\bibnamefont
  {Nakai}}\ and\ \bibinfo {author} {\bibfnamefont {K.}~\bibnamefont {Nomura}},\
  }\bibfield  {title} {\bibinfo {title} {{Higher-order topological
  superconductor phases in a multilayer system}},\ }\href
  {https://doi.org/10.1103/PhysRevB.108.184517} {\bibfield  {journal} {\bibinfo
   {journal} {Phys. Rev. B}\ }\textbf {\bibinfo {volume} {108}},\ \bibinfo
  {pages} {184517} (\bibinfo {year} {2023})}\BibitemShut {NoStop}%
\bibitem [{\citenamefont {Luo}\ \emph {et~al.}(2025)\citenamefont {Luo},
  \citenamefont {Li}, \citenamefont {Xiao},\ and\ \citenamefont
  {Wu}}]{HO-2025-prb}%
  \BibitemOpen
  \bibfield  {author} {\bibinfo {author} {\bibfnamefont {X.-J.}\ \bibnamefont
  {Luo}}, \bibinfo {author} {\bibfnamefont {J.-Z.}\ \bibnamefont {Li}},
  \bibinfo {author} {\bibfnamefont {M.}~\bibnamefont {Xiao}},\ and\ \bibinfo
  {author} {\bibfnamefont {F.}~\bibnamefont {Wu}},\ }\bibfield  {title}
  {\bibinfo {title} {Characterization of higher-order topological
  superconductors using bott indices},\ }\href
  {https://doi.org/10.1103/PhysRevB.111.184516} {\bibfield  {journal} {\bibinfo
   {journal} {Phys. Rev. B}\ }\textbf {\bibinfo {volume} {111}},\ \bibinfo
  {pages} {184516} (\bibinfo {year} {2025})}\BibitemShut {NoStop}%
\bibitem [{\citenamefont {Sato}(2006)}]{NSC-2006-prb-Sato}%
  \BibitemOpen
  \bibfield  {author} {\bibinfo {author} {\bibfnamefont {M.}~\bibnamefont
  {Sato}},\ }\bibfield  {title} {\bibinfo {title} {{Nodal structure of
  superconductors with time-reversal invariance and $Z_{2}$ topological
  number}},\ }\href {https://doi.org/10.1103/PhysRevB.73.214502} {\bibfield
  {journal} {\bibinfo  {journal} {Phys. Rev. B}\ }\textbf {\bibinfo {volume}
  {73}},\ \bibinfo {pages} {214502} (\bibinfo {year} {2006})}\BibitemShut
  {NoStop}%
\bibitem [{\citenamefont {Sato}\ and\ \citenamefont
  {Fujimoto}(2010)}]{NSC-2010-prl-Sato}%
  \BibitemOpen
  \bibfield  {author} {\bibinfo {author} {\bibfnamefont {M.}~\bibnamefont
  {Sato}}\ and\ \bibinfo {author} {\bibfnamefont {S.}~\bibnamefont
  {Fujimoto}},\ }\bibfield  {title} {\bibinfo {title} {{Existence of Majorana
  Fermions and Topological Order in Nodal Superconductors with Spin-Orbit
  Interactions in External Magnetic Fields}},\ }\href
  {https://doi.org/10.1103/PhysRevLett.105.217001} {\bibfield  {journal}
  {\bibinfo  {journal} {Phys. Rev. Lett.}\ }\textbf {\bibinfo {volume} {105}},\
  \bibinfo {pages} {217001} (\bibinfo {year} {2010})}\BibitemShut {NoStop}%
\bibitem [{\citenamefont {Schnyder}\ and\ \citenamefont
  {Ryu}(2011)}]{NLSC_2011_prb}%
  \BibitemOpen
  \bibfield  {author} {\bibinfo {author} {\bibfnamefont {A.~P.}\ \bibnamefont
  {Schnyder}}\ and\ \bibinfo {author} {\bibfnamefont {S.}~\bibnamefont {Ryu}},\
  }\bibfield  {title} {\bibinfo {title} {Topological phases and surface flat
  bands in superconductors without inversion symmetry},\ }\href
  {https://doi.org/10.1103/PhysRevB.84.060504} {\bibfield  {journal} {\bibinfo
  {journal} {Phys. Rev. B}\ }\textbf {\bibinfo {volume} {84}},\ \bibinfo
  {pages} {060504} (\bibinfo {year} {2011})}\BibitemShut {NoStop}%
\bibitem [{\citenamefont {Meng}\ and\ \citenamefont
  {Balents}(2012)}]{NSC_2012_prb_WSC}%
  \BibitemOpen
  \bibfield  {author} {\bibinfo {author} {\bibfnamefont {T.}~\bibnamefont
  {Meng}}\ and\ \bibinfo {author} {\bibfnamefont {L.}~\bibnamefont {Balents}},\
  }\bibfield  {title} {\bibinfo {title} {Weyl superconductors},\ }\href
  {https://doi.org/10.1103/PhysRevB.86.054504} {\bibfield  {journal} {\bibinfo
  {journal} {Phys. Rev. B}\ }\textbf {\bibinfo {volume} {86}},\ \bibinfo
  {pages} {054504} (\bibinfo {year} {2012})}\BibitemShut {NoStop}%
\bibitem [{\citenamefont {Sau}\ and\ \citenamefont
  {Tewari}(2012)}]{NSC_2012_prb_WSC_2}%
  \BibitemOpen
  \bibfield  {author} {\bibinfo {author} {\bibfnamefont {J.~D.}\ \bibnamefont
  {Sau}}\ and\ \bibinfo {author} {\bibfnamefont {S.}~\bibnamefont {Tewari}},\
  }\bibfield  {title} {\bibinfo {title} {{Topologically protected surface
  Majorana arcs and bulk Weyl fermions in ferromagnetic superconductors}},\
  }\href {https://doi.org/10.1103/PhysRevB.86.104509} {\bibfield  {journal}
  {\bibinfo  {journal} {Phys. Rev. B}\ }\textbf {\bibinfo {volume} {86}},\
  \bibinfo {pages} {104509} (\bibinfo {year} {2012})}\BibitemShut {NoStop}%
\bibitem [{\citenamefont {Zhao}\ and\ \citenamefont
  {Wang}(2013)}]{NSC_2013_prl}%
  \BibitemOpen
  \bibfield  {author} {\bibinfo {author} {\bibfnamefont {Y.~X.}\ \bibnamefont
  {Zhao}}\ and\ \bibinfo {author} {\bibfnamefont {Z.~D.}\ \bibnamefont
  {Wang}},\ }\bibfield  {title} {\bibinfo {title} {{Topological Classification
  and Stability of Fermi Surfaces}},\ }\href
  {https://doi.org/10.1103/PhysRevLett.110.240404} {\bibfield  {journal}
  {\bibinfo  {journal} {Phys. Rev. Lett.}\ }\textbf {\bibinfo {volume} {110}},\
  \bibinfo {pages} {240404} (\bibinfo {year} {2013})}\BibitemShut {NoStop}%
\bibitem [{\citenamefont {Matsuura}\ \emph {et~al.}(2013)\citenamefont
  {Matsuura}, \citenamefont {Chang}, \citenamefont {Schnyder},\ and\
  \citenamefont {Ryu}}]{NSC_2013_njp}%
  \BibitemOpen
  \bibfield  {author} {\bibinfo {author} {\bibfnamefont {S.}~\bibnamefont
  {Matsuura}}, \bibinfo {author} {\bibfnamefont {P.-Y.}\ \bibnamefont {Chang}},
  \bibinfo {author} {\bibfnamefont {A.~P.}\ \bibnamefont {Schnyder}},\ and\
  \bibinfo {author} {\bibfnamefont {S.}~\bibnamefont {Ryu}},\ }\bibfield
  {title} {\bibinfo {title} {Protected boundary states in gapless topological
  phases},\ }\href {https://doi.org/10.1088/1367-2630/15/6/065001} {\bibfield
  {journal} {\bibinfo  {journal} {New J. Phys.}\ }\textbf {\bibinfo {volume}
  {15}},\ \bibinfo {pages} {065001} (\bibinfo {year} {2013})}\BibitemShut
  {NoStop}%
\bibitem [{\citenamefont {Kobayashi}\ \emph {et~al.}(2014)\citenamefont
  {Kobayashi}, \citenamefont {Shiozaki}, \citenamefont {Tanaka},\ and\
  \citenamefont {Sato}}]{NLSC-2014-prb-Sato}%
  \BibitemOpen
  \bibfield  {author} {\bibinfo {author} {\bibfnamefont {S.}~\bibnamefont
  {Kobayashi}}, \bibinfo {author} {\bibfnamefont {K.}~\bibnamefont {Shiozaki}},
  \bibinfo {author} {\bibfnamefont {Y.}~\bibnamefont {Tanaka}},\ and\ \bibinfo
  {author} {\bibfnamefont {M.}~\bibnamefont {Sato}},\ }\bibfield  {title}
  {\bibinfo {title} {{Topological Blount's theorem of odd-parity
  superconductors}},\ }\href {https://doi.org/10.1103/PhysRevB.90.024516}
  {\bibfield  {journal} {\bibinfo  {journal} {Phys. Rev. B}\ }\textbf {\bibinfo
  {volume} {90}},\ \bibinfo {pages} {024516} (\bibinfo {year}
  {2014})}\BibitemShut {NoStop}%
\bibitem [{\citenamefont {Yang}\ \emph {et~al.}(2014)\citenamefont {Yang},
  \citenamefont {Pan},\ and\ \citenamefont {Zhang}}]{NSC_2014_prl_DSCWSC}%
  \BibitemOpen
  \bibfield  {author} {\bibinfo {author} {\bibfnamefont {S.~A.}\ \bibnamefont
  {Yang}}, \bibinfo {author} {\bibfnamefont {H.}~\bibnamefont {Pan}},\ and\
  \bibinfo {author} {\bibfnamefont {F.}~\bibnamefont {Zhang}},\ }\bibfield
  {title} {\bibinfo {title} {{Dirac and Weyl Superconductors in Three
  Dimensions}},\ }\href {https://doi.org/10.1103/PhysRevLett.113.046401}
  {\bibfield  {journal} {\bibinfo  {journal} {Phys. Rev. Lett.}\ }\textbf
  {\bibinfo {volume} {113}},\ \bibinfo {pages} {046401} (\bibinfo {year}
  {2014})}\BibitemShut {NoStop}%
\bibitem [{\citenamefont {Chiu}\ and\ \citenamefont
  {Schnyder}(2014)}]{NSC_2014_cla}%
  \BibitemOpen
  \bibfield  {author} {\bibinfo {author} {\bibfnamefont {C.-K.}\ \bibnamefont
  {Chiu}}\ and\ \bibinfo {author} {\bibfnamefont {A.~P.}\ \bibnamefont
  {Schnyder}},\ }\bibfield  {title} {\bibinfo {title} {Classification of
  reflection-symmetry-protected topological semimetals and nodal
  superconductors},\ }\href {https://doi.org/10.1103/PhysRevB.90.205136}
  {\bibfield  {journal} {\bibinfo  {journal} {Phys. Rev. B}\ }\textbf {\bibinfo
  {volume} {90}},\ \bibinfo {pages} {205136} (\bibinfo {year}
  {2014})}\BibitemShut {NoStop}%
\bibitem [{\citenamefont {Kao}\ \emph {et~al.}(2015)\citenamefont {Kao},
  \citenamefont {Huang}, \citenamefont {Mou},\ and\ \citenamefont
  {Tsuei}}]{NSC-2015-prb-2}%
  \BibitemOpen
  \bibfield  {author} {\bibinfo {author} {\bibfnamefont {J.-T.}\ \bibnamefont
  {Kao}}, \bibinfo {author} {\bibfnamefont {S.-M.}\ \bibnamefont {Huang}},
  \bibinfo {author} {\bibfnamefont {C.-Y.}\ \bibnamefont {Mou}},\ and\ \bibinfo
  {author} {\bibfnamefont {C.~C.}\ \bibnamefont {Tsuei}},\ }\bibfield  {title}
  {\bibinfo {title} {{Tunneling spectroscopy and Majorana modes emergent from
  topological gapless phases in high-${T}_{c}$ cuprate superconductors}},\
  }\href {https://doi.org/10.1103/PhysRevB.91.134501} {\bibfield  {journal}
  {\bibinfo  {journal} {Phys. Rev. B}\ }\textbf {\bibinfo {volume} {91}},\
  \bibinfo {pages} {134501} (\bibinfo {year} {2015})}\BibitemShut {NoStop}%
\bibitem [{\citenamefont {Schnyder}\ and\ \citenamefont
  {Brydon}(2015)}]{NSC_2015_JPCM}%
  \BibitemOpen
  \bibfield  {author} {\bibinfo {author} {\bibfnamefont {A.~P.}\ \bibnamefont
  {Schnyder}}\ and\ \bibinfo {author} {\bibfnamefont {P.~M.~R.}\ \bibnamefont
  {Brydon}},\ }\bibfield  {title} {\bibinfo {title} {Topological surface states
  in nodal superconductors},\ }\href
  {https://doi.org/10.1088/0953-8984/27/24/243201} {\bibfield  {journal}
  {\bibinfo  {journal} {J. Phys.: Condens. Matter}\ }\textbf {\bibinfo {volume}
  {27}},\ \bibinfo {pages} {243201} (\bibinfo {year} {2015})}\BibitemShut
  {NoStop}%
\bibitem [{\citenamefont {Kobayashi}\ and\ \citenamefont
  {Sato}(2015)}]{NSC_2015_prl_Sato}%
  \BibitemOpen
  \bibfield  {author} {\bibinfo {author} {\bibfnamefont {S.}~\bibnamefont
  {Kobayashi}}\ and\ \bibinfo {author} {\bibfnamefont {M.}~\bibnamefont
  {Sato}},\ }\bibfield  {title} {\bibinfo {title} {{Topological
  Superconductivity in Dirac Semimetals}},\ }\href
  {https://doi.org/10.1103/PhysRevLett.115.187001} {\bibfield  {journal}
  {\bibinfo  {journal} {Phys. Rev. Lett.}\ }\textbf {\bibinfo {volume} {115}},\
  \bibinfo {pages} {187001} (\bibinfo {year} {2015})}\BibitemShut {NoStop}%
\bibitem [{\citenamefont {Lu}\ \emph {et~al.}(2015)\citenamefont {Lu},
  \citenamefont {Yada}, \citenamefont {Sato},\ and\ \citenamefont
  {Tanaka}}]{NSC-2015-prl-Sato-Weyl}%
  \BibitemOpen
  \bibfield  {author} {\bibinfo {author} {\bibfnamefont {B.}~\bibnamefont
  {Lu}}, \bibinfo {author} {\bibfnamefont {K.}~\bibnamefont {Yada}}, \bibinfo
  {author} {\bibfnamefont {M.}~\bibnamefont {Sato}},\ and\ \bibinfo {author}
  {\bibfnamefont {Y.}~\bibnamefont {Tanaka}},\ }\bibfield  {title} {\bibinfo
  {title} {{Crossed Surface Flat Bands of Weyl Semimetal Superconductors}},\
  }\href {https://doi.org/10.1103/PhysRevLett.114.096804} {\bibfield  {journal}
  {\bibinfo  {journal} {Phys. Rev. Lett.}\ }\textbf {\bibinfo {volume} {114}},\
  \bibinfo {pages} {096804} (\bibinfo {year} {2015})}\BibitemShut {NoStop}%
\bibitem [{\citenamefont {Kobayashi}\ \emph {et~al.}(2016)\citenamefont
  {Kobayashi}, \citenamefont {Yanase},\ and\ \citenamefont
  {Sato}}]{NLSC-2016-prb-Sato}%
  \BibitemOpen
  \bibfield  {author} {\bibinfo {author} {\bibfnamefont {S.}~\bibnamefont
  {Kobayashi}}, \bibinfo {author} {\bibfnamefont {Y.}~\bibnamefont {Yanase}},\
  and\ \bibinfo {author} {\bibfnamefont {M.}~\bibnamefont {Sato}},\ }\bibfield
  {title} {\bibinfo {title} {{Topologically stable gapless phases in
  nonsymmorphic superconductors}},\ }\href
  {https://doi.org/10.1103/PhysRevB.94.134512} {\bibfield  {journal} {\bibinfo
  {journal} {Phys. Rev. B}\ }\textbf {\bibinfo {volume} {94}},\ \bibinfo
  {pages} {134512} (\bibinfo {year} {2016})}\BibitemShut {NoStop}%
\bibitem [{\citenamefont {Hashimoto}\ \emph {et~al.}(2016)\citenamefont
  {Hashimoto}, \citenamefont {Kobayashi}, \citenamefont {Tanaka},\ and\
  \citenamefont {Sato}}]{NSC_2016_prb_Sato}%
  \BibitemOpen
  \bibfield  {author} {\bibinfo {author} {\bibfnamefont {T.}~\bibnamefont
  {Hashimoto}}, \bibinfo {author} {\bibfnamefont {S.}~\bibnamefont
  {Kobayashi}}, \bibinfo {author} {\bibfnamefont {Y.}~\bibnamefont {Tanaka}},\
  and\ \bibinfo {author} {\bibfnamefont {M.}~\bibnamefont {Sato}},\ }\bibfield
  {title} {\bibinfo {title} {{Superconductivity in doped Dirac semimetals}},\
  }\href {https://doi.org/10.1103/PhysRevB.94.014510} {\bibfield  {journal}
  {\bibinfo  {journal} {Phys. Rev. B}\ }\textbf {\bibinfo {volume} {94}},\
  \bibinfo {pages} {014510} (\bibinfo {year} {2016})}\BibitemShut {NoStop}%
\bibitem [{\citenamefont {Zhao}\ \emph {et~al.}(2016)\citenamefont {Zhao},
  \citenamefont {Schnyder},\ and\ \citenamefont
  {Wang}}]{NSC-2016-prl-Zhao-classify}%
  \BibitemOpen
  \bibfield  {author} {\bibinfo {author} {\bibfnamefont {Y.~X.}\ \bibnamefont
  {Zhao}}, \bibinfo {author} {\bibfnamefont {A.~P.}\ \bibnamefont {Schnyder}},\
  and\ \bibinfo {author} {\bibfnamefont {Z.~D.}\ \bibnamefont {Wang}},\
  }\bibfield  {title} {\bibinfo {title} {{Unified Theory of $PT$ and $CP$
  Invariant Topological Metals and Nodal Superconductors}},\ }\href
  {https://doi.org/10.1103/PhysRevLett.116.156402} {\bibfield  {journal}
  {\bibinfo  {journal} {Phys. Rev. Lett.}\ }\textbf {\bibinfo {volume} {116}},\
  \bibinfo {pages} {156402} (\bibinfo {year} {2016})}\BibitemShut {NoStop}%
\bibitem [{\citenamefont {Yanase}(2016)}]{NSC-2016-prb-UPt}%
  \BibitemOpen
  \bibfield  {author} {\bibinfo {author} {\bibfnamefont {Y.}~\bibnamefont
  {Yanase}},\ }\bibfield  {title} {\bibinfo {title} {{Nonsymmorphic Weyl
  superconductivity in ${\mathrm{UPt}}_{3}$ based on ${E}_{2u}$
  representation}},\ }\href {https://doi.org/10.1103/PhysRevB.94.174502}
  {\bibfield  {journal} {\bibinfo  {journal} {Phys. Rev. B}\ }\textbf {\bibinfo
  {volume} {94}},\ \bibinfo {pages} {174502} (\bibinfo {year}
  {2016})}\BibitemShut {NoStop}%
\bibitem [{\citenamefont {Sun}\ \emph {et~al.}(2017)\citenamefont {Sun},
  \citenamefont {Lian},\ and\ \citenamefont {Zhang}}]{NLSC-2017-prl-ZSC}%
  \BibitemOpen
  \bibfield  {author} {\bibinfo {author} {\bibfnamefont {X.-Q.}\ \bibnamefont
  {Sun}}, \bibinfo {author} {\bibfnamefont {B.}~\bibnamefont {Lian}},\ and\
  \bibinfo {author} {\bibfnamefont {S.-C.}\ \bibnamefont {Zhang}},\ }\bibfield
  {title} {\bibinfo {title} {{Double Helix Nodal Line Superconductor}},\ }\href
  {https://doi.org/10.1103/PhysRevLett.119.147001} {\bibfield  {journal}
  {\bibinfo  {journal} {Phys. Rev. Lett.}\ }\textbf {\bibinfo {volume} {119}},\
  \bibinfo {pages} {147001} (\bibinfo {year} {2017})}\BibitemShut {NoStop}%
\bibitem [{\citenamefont {Micklitz}\ and\ \citenamefont
  {Norman}(2017)}]{NLSC-2017-prl}%
  \BibitemOpen
  \bibfield  {author} {\bibinfo {author} {\bibfnamefont {T.}~\bibnamefont
  {Micklitz}}\ and\ \bibinfo {author} {\bibfnamefont {M.~R.}\ \bibnamefont
  {Norman}},\ }\bibfield  {title} {\bibinfo {title} {{Symmetry-Enforced Line
  Nodes in Unconventional Superconductors}},\ }\href
  {https://doi.org/10.1103/PhysRevLett.118.207001} {\bibfield  {journal}
  {\bibinfo  {journal} {Phys. Rev. Lett.}\ }\textbf {\bibinfo {volume} {118}},\
  \bibinfo {pages} {207001} (\bibinfo {year} {2017})}\BibitemShut {NoStop}%
\bibitem [{\citenamefont {Kobayashi}\ \emph {et~al.}(2018)\citenamefont
  {Kobayashi}, \citenamefont {Sumita}, \citenamefont {Yanase},\ and\
  \citenamefont {Sato}}]{NLSC-2018-prb-Sato}%
  \BibitemOpen
  \bibfield  {author} {\bibinfo {author} {\bibfnamefont {S.}~\bibnamefont
  {Kobayashi}}, \bibinfo {author} {\bibfnamefont {S.}~\bibnamefont {Sumita}},
  \bibinfo {author} {\bibfnamefont {Y.}~\bibnamefont {Yanase}},\ and\ \bibinfo
  {author} {\bibfnamefont {M.}~\bibnamefont {Sato}},\ }\bibfield  {title}
  {\bibinfo {title} {{Symmetry-protected line nodes and Majorana flat bands in
  nodal crystalline superconductors}},\ }\href
  {https://doi.org/10.1103/PhysRevB.97.180504} {\bibfield  {journal} {\bibinfo
  {journal} {Phys. Rev. B}\ }\textbf {\bibinfo {volume} {97}},\ \bibinfo
  {pages} {180504} (\bibinfo {year} {2018})}\BibitemShut {NoStop}%
\bibitem [{\citenamefont {Park}\ \emph {et~al.}(2018)\citenamefont {Park},
  \citenamefont {Raza}, \citenamefont {Gilbert},\ and\ \citenamefont
  {Teo}}]{NSC-2018-prb-DSC}%
  \BibitemOpen
  \bibfield  {author} {\bibinfo {author} {\bibfnamefont {M.~J.}\ \bibnamefont
  {Park}}, \bibinfo {author} {\bibfnamefont {S.}~\bibnamefont {Raza}}, \bibinfo
  {author} {\bibfnamefont {M.~J.}\ \bibnamefont {Gilbert}},\ and\ \bibinfo
  {author} {\bibfnamefont {J.~C.~Y.}\ \bibnamefont {Teo}},\ }\bibfield  {title}
  {\bibinfo {title} {{Coupled wire models of interacting Dirac nodal
  superconductors}},\ }\href {https://doi.org/10.1103/PhysRevB.98.184514}
  {\bibfield  {journal} {\bibinfo  {journal} {Phys. Rev. B}\ }\textbf {\bibinfo
  {volume} {98}},\ \bibinfo {pages} {184514} (\bibinfo {year}
  {2018})}\BibitemShut {NoStop}%
\bibitem [{\citenamefont {Sumita}\ and\ \citenamefont
  {Yanase}(2018)}]{NSC-2018-prb-classify}%
  \BibitemOpen
  \bibfield  {author} {\bibinfo {author} {\bibfnamefont {S.}~\bibnamefont
  {Sumita}}\ and\ \bibinfo {author} {\bibfnamefont {Y.}~\bibnamefont
  {Yanase}},\ }\bibfield  {title} {\bibinfo {title} {{Unconventional
  superconducting gap structure protected by space group symmetry}},\ }\href
  {https://doi.org/10.1103/PhysRevB.97.134512} {\bibfield  {journal} {\bibinfo
  {journal} {Phys. Rev. B}\ }\textbf {\bibinfo {volume} {97}},\ \bibinfo
  {pages} {134512} (\bibinfo {year} {2018})}\BibitemShut {NoStop}%
\bibitem [{\citenamefont {Li}\ and\ \citenamefont
  {Haldane}(2018)}]{NSC-2018-prl-HD}%
  \BibitemOpen
  \bibfield  {author} {\bibinfo {author} {\bibfnamefont {Y.}~\bibnamefont
  {Li}}\ and\ \bibinfo {author} {\bibfnamefont {F.~D.~M.}\ \bibnamefont
  {Haldane}},\ }\bibfield  {title} {\bibinfo {title} {{Topological Nodal Cooper
  Pairing in Doped Weyl Metals}},\ }\href
  {https://doi.org/10.1103/PhysRevLett.120.067003} {\bibfield  {journal}
  {\bibinfo  {journal} {Phys. Rev. Lett.}\ }\textbf {\bibinfo {volume} {120}},\
  \bibinfo {pages} {067003} (\bibinfo {year} {2018})}\BibitemShut {NoStop}%
\bibitem [{\citenamefont {Zhang}\ \emph {et~al.}(2020)\citenamefont {Zhang},
  \citenamefont {Hsu},\ and\ \citenamefont {Das~Sarma}}]{NSC_2020_prb_ZRX}%
  \BibitemOpen
  \bibfield  {author} {\bibinfo {author} {\bibfnamefont {R.-X.}\ \bibnamefont
  {Zhang}}, \bibinfo {author} {\bibfnamefont {Y.-T.}\ \bibnamefont {Hsu}},\
  and\ \bibinfo {author} {\bibfnamefont {S.}~\bibnamefont {Das~Sarma}},\
  }\bibfield  {title} {\bibinfo {title} {{Higher-order topological Dirac
  superconductors}},\ }\href {https://doi.org/10.1103/PhysRevB.102.094503}
  {\bibfield  {journal} {\bibinfo  {journal} {Phys. Rev. B}\ }\textbf {\bibinfo
  {volume} {102}},\ \bibinfo {pages} {094503} (\bibinfo {year}
  {2020})}\BibitemShut {NoStop}%
\bibitem [{\citenamefont {Yu}\ \emph {et~al.}(2022)\citenamefont {Yu},
  \citenamefont {Chen},\ and\ \citenamefont {Das~Sarma}}]{NSC-2022-prb-obs}%
  \BibitemOpen
  \bibfield  {author} {\bibinfo {author} {\bibfnamefont {J.}~\bibnamefont
  {Yu}}, \bibinfo {author} {\bibfnamefont {Y.-A.}\ \bibnamefont {Chen}},\ and\
  \bibinfo {author} {\bibfnamefont {S.}~\bibnamefont {Das~Sarma}},\ }\bibfield
  {title} {\bibinfo {title} {{Euler-obstructed Cooper pairing: Nodal
  superconductivity and hinge Majorana zero modes}},\ }\href
  {https://doi.org/10.1103/PhysRevB.105.104515} {\bibfield  {journal} {\bibinfo
   {journal} {Phys. Rev. B}\ }\textbf {\bibinfo {volume} {105}},\ \bibinfo
  {pages} {104515} (\bibinfo {year} {2022})}\BibitemShut {NoStop}%
\bibitem [{\citenamefont {Tang}\ \emph {et~al.}(2022)\citenamefont {Tang},
  \citenamefont {Ono}, \citenamefont {Wan},\ and\ \citenamefont
  {Watanabe}}]{NSC-2022-prl-Wan}%
  \BibitemOpen
  \bibfield  {author} {\bibinfo {author} {\bibfnamefont {F.}~\bibnamefont
  {Tang}}, \bibinfo {author} {\bibfnamefont {S.}~\bibnamefont {Ono}}, \bibinfo
  {author} {\bibfnamefont {X.}~\bibnamefont {Wan}},\ and\ \bibinfo {author}
  {\bibfnamefont {H.}~\bibnamefont {Watanabe}},\ }\bibfield  {title} {\bibinfo
  {title} {{High-Throughput Investigations of Topological and Nodal
  Superconductors}},\ }\href {https://doi.org/10.1103/PhysRevLett.129.027001}
  {\bibfield  {journal} {\bibinfo  {journal} {Phys. Rev. Lett.}\ }\textbf
  {\bibinfo {volume} {129}},\ \bibinfo {pages} {027001} (\bibinfo {year}
  {2022})}\BibitemShut {NoStop}%
\bibitem [{\citenamefont {Wu}\ and\ \citenamefont
  {Wang}(2023)}]{NLSC-2023-prb}%
  \BibitemOpen
  \bibfield  {author} {\bibinfo {author} {\bibfnamefont {Z.}~\bibnamefont
  {Wu}}\ and\ \bibinfo {author} {\bibfnamefont {Y.}~\bibnamefont {Wang}},\
  }\bibfield  {title} {\bibinfo {title} {{Nodal topological superconductivity
  in nodal-line semimetals}},\ }\href
  {https://doi.org/10.1103/PhysRevB.108.224503} {\bibfield  {journal} {\bibinfo
   {journal} {Phys. Rev. B}\ }\textbf {\bibinfo {volume} {108}},\ \bibinfo
  {pages} {224503} (\bibinfo {year} {2023})}\BibitemShut {NoStop}%
\bibitem [{\citenamefont {Bobrow}\ and\ \citenamefont {Li}()}]{NSC-2024-arxiv}%
  \BibitemOpen
  \bibfield  {author} {\bibinfo {author} {\bibfnamefont {E.}~\bibnamefont
  {Bobrow}}\ and\ \bibinfo {author} {\bibfnamefont {Y.}~\bibnamefont {Li}},\
  }\bibfield  {title} {\bibinfo {title} {{Monopole Superconductivity in
  Magnetically Doped Cd$_3$As$_2$}},\ }\href {http://arxiv.org/abs/2204.04249}
  {\bibinfo  {journal} {arXiv.2204.04249}\ }\BibitemShut {NoStop}%
\bibitem [{\citenamefont {Wu}\ and\ \citenamefont
  {Wang}(2022)}]{NSC_2024_prb_DSC}%
  \BibitemOpen
\bibfield  {journal} {  }\bibfield  {author} {\bibinfo {author} {\bibfnamefont
  {Z.}~\bibnamefont {Wu}}\ and\ \bibinfo {author} {\bibfnamefont
  {Y.}~\bibnamefont {Wang}},\ }\bibfield  {title} {\bibinfo {title} {{Nodal
  higher-order topological superconductivity from a
  ${\mathcal{C}}_{4}$-symmetric Dirac semimetal}},\ }\href
  {https://doi.org/10.1103/PhysRevB.106.214510} {\bibfield  {journal} {\bibinfo
   {journal} {Phys. Rev. B}\ }\textbf {\bibinfo {volume} {106}},\ \bibinfo
  {pages} {214510} (\bibinfo {year} {2022})}\BibitemShut {NoStop}%
\bibitem [{\citenamefont {Feng}(2024)}]{NSC_2024_prb}%
  \BibitemOpen
  \bibfield  {author} {\bibinfo {author} {\bibfnamefont {G.-H.}\ \bibnamefont
  {Feng}},\ }\bibfield  {title} {\bibinfo {title} {{Nodal higher-order
  topological superconductivity from ${C}_{6}$-symmetric Dirac semimetals}},\
  }\href {https://doi.org/10.1103/PhysRevB.110.174513} {\bibfield  {journal}
  {\bibinfo  {journal} {Phys. Rev. B}\ }\textbf {\bibinfo {volume} {110}},\
  \bibinfo {pages} {174513} (\bibinfo {year} {2024})}\BibitemShut {NoStop}%
\bibitem [{\citenamefont {Xie}\ \emph {et~al.}(2024)\citenamefont {Xie},
  \citenamefont {Wu}, \citenamefont {Fang},\ and\ \citenamefont
  {Wang}}]{HO_2024_china}%
  \BibitemOpen
  \bibfield  {author} {\bibinfo {author} {\bibfnamefont {Y.}~\bibnamefont
  {Xie}}, \bibinfo {author} {\bibfnamefont {X.}~\bibnamefont {Wu}}, \bibinfo
  {author} {\bibfnamefont {Z.}~\bibnamefont {Fang}},\ and\ \bibinfo {author}
  {\bibfnamefont {Z.}~\bibnamefont {Wang}},\ }\bibfield  {title} {\bibinfo
  {title} {{Hinge Majorana flat band in type-{II} Dirac semimetals}},\ }\href
  {https://doi.org/10.1007/s11433-024-2483-8} {\bibfield  {journal} {\bibinfo
  {journal} {Sci. China-Phys. Mech. Astron.}\ }\textbf {\bibinfo {volume}
  {67}},\ \bibinfo {pages} {127011} (\bibinfo {year} {2024})}\BibitemShut
  {NoStop}%
\bibitem [{\citenamefont {Zhang}\ \emph {et~al.}()\citenamefont {Zhang},
  \citenamefont {Fang}, \citenamefont {Qin}, \citenamefont {Zhang},
  \citenamefont {Po},\ and\ \citenamefont {Wu}}]{NSC-HO-2025-arx}%
  \BibitemOpen
  \bibfield  {author} {\bibinfo {author} {\bibfnamefont {Z.}~\bibnamefont
  {Zhang}}, \bibinfo {author} {\bibfnamefont {Z.}~\bibnamefont {Fang}},
  \bibinfo {author} {\bibfnamefont {S.}~\bibnamefont {Qin}}, \bibinfo {author}
  {\bibfnamefont {P.}~\bibnamefont {Zhang}}, \bibinfo {author} {\bibfnamefont
  {H.~C.}\ \bibnamefont {Po}},\ and\ \bibinfo {author} {\bibfnamefont
  {X.}~\bibnamefont {Wu}},\ }\bibfield  {title} {\bibinfo {title} {{Double
  Majorana Vortex Flat Bands in the Topological Dirac Superconductor}},\ }\href
  {http://arxiv.org/abs/2501.05317} {\bibinfo  {journal} {arXiv.2501.05317}\
  }\BibitemShut {NoStop}%
\bibitem [{\citenamefont {Zeng}\ \emph {et~al.}(2025)\citenamefont {Zeng},
  \citenamefont {He}, \citenamefont {Ning}, \citenamefont {Xu},\ and\
  \citenamefont {Wang}}]{NSC_2025_nature}%
  \BibitemOpen
\bibfield  {journal} {  }\bibfield  {author} {\bibinfo {author} {\bibfnamefont
  {J.}~\bibnamefont {Zeng}}, \bibinfo {author} {\bibfnamefont {J.~J.}\
  \bibnamefont {He}}, \bibinfo {author} {\bibfnamefont {Z.}~\bibnamefont
  {Ning}}, \bibinfo {author} {\bibfnamefont {D.-H.}\ \bibnamefont {Xu}},\ and\
  \bibinfo {author} {\bibfnamefont {R.}~\bibnamefont {Wang}},\ }\bibfield
  {title} {\bibinfo {title} {Spin signature of majorana fermions in topological
  nodal-point superconductors},\ }\href
  {https://doi.org/10.1038/s41535-025-00768-1} {\bibfield  {journal} {\bibinfo
  {journal} {npj Quantum Mater.}\ }\textbf {\bibinfo {volume} {10}},\ \bibinfo
  {pages} {48} (\bibinfo {year} {2025})}\BibitemShut {NoStop}%
\bibitem [{\citenamefont {Zhang}\ \emph {et~al.}(2025)\citenamefont {Zhang},
  \citenamefont {Shi},\ and\ \citenamefont {Zhao}}]{NSC-2025-prb}%
  \BibitemOpen
  \bibfield  {author} {\bibinfo {author} {\bibfnamefont {X.}~\bibnamefont
  {Zhang}}, \bibinfo {author} {\bibfnamefont {X.}~\bibnamefont {Shi}},\ and\
  \bibinfo {author} {\bibfnamefont {M.}~\bibnamefont {Zhao}},\ }\bibfield
  {title} {\bibinfo {title} {{Effective nodal topological superconductivity
  driven by $s$-wave pairing and the ferromagnetic proximity effect in
  materials with persistent spin textures}},\ }\href
  {https://doi.org/10.1103/jsww-v4y9} {\bibfield  {journal} {\bibinfo
  {journal} {Phys. Rev. B}\ }\textbf {\bibinfo {volume} {112}},\ \bibinfo
  {pages} {085423} (\bibinfo {year} {2025})}\BibitemShut {NoStop}%
\bibitem [{\citenamefont {Bhattacharyya}\ \emph {et~al.}(2019)\citenamefont
  {Bhattacharyya}, \citenamefont {Adroja}, \citenamefont {Panda}, \citenamefont
  {Saha}, \citenamefont {Das}, \citenamefont {Machado}, \citenamefont
  {Cigarroa}, \citenamefont {Grant}, \citenamefont {Fisk}, \citenamefont
  {Hillier},\ and\ \citenamefont {Manfrinetti}}]{exp-NL-2019-prl}%
  \BibitemOpen
  \bibfield  {author} {\bibinfo {author} {\bibfnamefont {A.}~\bibnamefont
  {Bhattacharyya}}, \bibinfo {author} {\bibfnamefont {D.~T.}\ \bibnamefont
  {Adroja}}, \bibinfo {author} {\bibfnamefont {K.}~\bibnamefont {Panda}},
  \bibinfo {author} {\bibfnamefont {S.}~\bibnamefont {Saha}}, \bibinfo {author}
  {\bibfnamefont {T.}~\bibnamefont {Das}}, \bibinfo {author} {\bibfnamefont
  {A.~J.~S.}\ \bibnamefont {Machado}}, \bibinfo {author} {\bibfnamefont
  {O.~V.}\ \bibnamefont {Cigarroa}}, \bibinfo {author} {\bibfnamefont {T.~W.}\
  \bibnamefont {Grant}}, \bibinfo {author} {\bibfnamefont {Z.}~\bibnamefont
  {Fisk}}, \bibinfo {author} {\bibfnamefont {A.~D.}\ \bibnamefont {Hillier}},\
  and\ \bibinfo {author} {\bibfnamefont {P.}~\bibnamefont {Manfrinetti}},\
  }\bibfield  {title} {\bibinfo {title} {{Evidence of a Nodal Line in the
  Superconducting Gap Symmetry of Noncentrosymmetric ${\mathrm{ThCoC}}_{2}$}},\
  }\href {https://doi.org/10.1103/PhysRevLett.122.147001} {\bibfield  {journal}
  {\bibinfo  {journal} {Phys. Rev. Lett.}\ }\textbf {\bibinfo {volume} {122}},\
  \bibinfo {pages} {147001} (\bibinfo {year} {2019})}\BibitemShut {NoStop}%
\bibitem [{\citenamefont {Nayak}\ \emph {et~al.}(2021)\citenamefont {Nayak},
  \citenamefont {Steinbok}, \citenamefont {Roet}, \citenamefont {Koo},
  \citenamefont {Margalit}, \citenamefont {Feldman}, \citenamefont {Almoalem},
  \citenamefont {Kanigel}, \citenamefont {Fiete}, \citenamefont {Yan},
  \citenamefont {Oreg}, \citenamefont {Avraham},\ and\ \citenamefont
  {Beidenkopf}}]{exp-NSC-2021-np}%
  \BibitemOpen
  \bibfield  {author} {\bibinfo {author} {\bibfnamefont {A.~K.}\ \bibnamefont
  {Nayak}}, \bibinfo {author} {\bibfnamefont {A.}~\bibnamefont {Steinbok}},
  \bibinfo {author} {\bibfnamefont {Y.}~\bibnamefont {Roet}}, \bibinfo {author}
  {\bibfnamefont {J.}~\bibnamefont {Koo}}, \bibinfo {author} {\bibfnamefont
  {G.}~\bibnamefont {Margalit}}, \bibinfo {author} {\bibfnamefont
  {I.}~\bibnamefont {Feldman}}, \bibinfo {author} {\bibfnamefont
  {A.}~\bibnamefont {Almoalem}}, \bibinfo {author} {\bibfnamefont
  {A.}~\bibnamefont {Kanigel}}, \bibinfo {author} {\bibfnamefont {G.~A.}\
  \bibnamefont {Fiete}}, \bibinfo {author} {\bibfnamefont {B.}~\bibnamefont
  {Yan}}, \bibinfo {author} {\bibfnamefont {Y.}~\bibnamefont {Oreg}}, \bibinfo
  {author} {\bibfnamefont {N.}~\bibnamefont {Avraham}},\ and\ \bibinfo {author}
  {\bibfnamefont {H.}~\bibnamefont {Beidenkopf}},\ }\bibfield  {title}
  {\bibinfo {title} {{Evidence of topological boundary modes with topological
  nodal-point superconductivity}},\ }\href
  {https://doi.org/10.1038/s41567-021-01376-z} {\bibfield  {journal} {\bibinfo
  {journal} {Nat. Phys.}\ }\textbf {\bibinfo {volume} {17}},\ \bibinfo {pages}
  {1413} (\bibinfo {year} {2021})}\BibitemShut {NoStop}%
\bibitem [{\citenamefont {Yang}\ \emph {et~al.}(2023)\citenamefont {Yang},
  \citenamefont {Zhong}, \citenamefont {Mardanya}, \citenamefont {Cochran},
  \citenamefont {Chapai}, \citenamefont {Mine}, \citenamefont {Zhang},
  \citenamefont {S\'anchez-Barriga}, \citenamefont {Cheng}, \citenamefont
  {Clark}, \citenamefont {Yin}, \citenamefont {Blawat}, \citenamefont {Cheng},
  \citenamefont {Belopolski}, \citenamefont {Nagashima}, \citenamefont
  {Najafzadeh}, \citenamefont {Gao}, \citenamefont {Yao}, \citenamefont
  {Bansil}, \citenamefont {Jin}, \citenamefont {Chang}, \citenamefont {Shin},
  \citenamefont {Okazaki},\ and\ \citenamefont {Hasan}}]{exp-2023-prl}%
  \BibitemOpen
  \bibfield  {author} {\bibinfo {author} {\bibfnamefont {X.~P.}\ \bibnamefont
  {Yang}}, \bibinfo {author} {\bibfnamefont {Y.}~\bibnamefont {Zhong}},
  \bibinfo {author} {\bibfnamefont {S.}~\bibnamefont {Mardanya}}, \bibinfo
  {author} {\bibfnamefont {T.~A.}\ \bibnamefont {Cochran}}, \bibinfo {author}
  {\bibfnamefont {R.}~\bibnamefont {Chapai}}, \bibinfo {author} {\bibfnamefont
  {A.}~\bibnamefont {Mine}}, \bibinfo {author} {\bibfnamefont {J.}~\bibnamefont
  {Zhang}}, \bibinfo {author} {\bibfnamefont {J.}~\bibnamefont
  {S\'anchez-Barriga}}, \bibinfo {author} {\bibfnamefont {Z.-J.}\ \bibnamefont
  {Cheng}}, \bibinfo {author} {\bibfnamefont {O.~J.}\ \bibnamefont {Clark}},
  \bibinfo {author} {\bibfnamefont {J.-X.}\ \bibnamefont {Yin}}, \bibinfo
  {author} {\bibfnamefont {J.}~\bibnamefont {Blawat}}, \bibinfo {author}
  {\bibfnamefont {G.}~\bibnamefont {Cheng}}, \bibinfo {author} {\bibfnamefont
  {I.}~\bibnamefont {Belopolski}}, \bibinfo {author} {\bibfnamefont
  {T.}~\bibnamefont {Nagashima}}, \bibinfo {author} {\bibfnamefont
  {S.}~\bibnamefont {Najafzadeh}}, \bibinfo {author} {\bibfnamefont
  {S.}~\bibnamefont {Gao}}, \bibinfo {author} {\bibfnamefont {N.}~\bibnamefont
  {Yao}}, \bibinfo {author} {\bibfnamefont {A.}~\bibnamefont {Bansil}},
  \bibinfo {author} {\bibfnamefont {R.}~\bibnamefont {Jin}}, \bibinfo {author}
  {\bibfnamefont {T.-R.}\ \bibnamefont {Chang}}, \bibinfo {author}
  {\bibfnamefont {S.}~\bibnamefont {Shin}}, \bibinfo {author} {\bibfnamefont
  {K.}~\bibnamefont {Okazaki}},\ and\ \bibinfo {author} {\bibfnamefont {M.~Z.}\
  \bibnamefont {Hasan}},\ }\bibfield  {title} {\bibinfo {title} {Coexistence of
  bulk-nodal and surface-nodeless cooper pairings in a superconducting dirac
  semimetal},\ }\href {https://doi.org/10.1103/PhysRevLett.130.046402}
  {\bibfield  {journal} {\bibinfo  {journal} {Phys. Rev. Lett.}\ }\textbf
  {\bibinfo {volume} {130}},\ \bibinfo {pages} {046402} (\bibinfo {year}
  {2023})}\BibitemShut {NoStop}%
\bibitem [{\citenamefont {Yadav}\ \emph {et~al.}(2024)\citenamefont {Yadav},
  \citenamefont {Ghosh}, \citenamefont {Kumar}, \citenamefont {Thamizhavel},\
  and\ \citenamefont {Paulose}}]{exp-2024-prb}%
  \BibitemOpen
  \bibfield  {author} {\bibinfo {author} {\bibfnamefont {C.~S.}\ \bibnamefont
  {Yadav}}, \bibinfo {author} {\bibfnamefont {S.~K.}\ \bibnamefont {Ghosh}},
  \bibinfo {author} {\bibfnamefont {P.}~\bibnamefont {Kumar}}, \bibinfo
  {author} {\bibfnamefont {A.}~\bibnamefont {Thamizhavel}},\ and\ \bibinfo
  {author} {\bibfnamefont {P.~L.}\ \bibnamefont {Paulose}},\ }\bibfield
  {title} {\bibinfo {title} {{Signature of point nodal superconductivity in the
  Dirac semimetal PdTe}},\ }\href {https://doi.org/10.1103/PhysRevB.110.054515}
  {\bibfield  {journal} {\bibinfo  {journal} {Phys. Rev. B}\ }\textbf {\bibinfo
  {volume} {110}},\ \bibinfo {pages} {054515} (\bibinfo {year}
  {2024})}\BibitemShut {NoStop}%
\bibitem [{\citenamefont {Shang}\ \emph {et~al.}(2025)\citenamefont {Shang},
  \citenamefont {Zhao}, \citenamefont {Hu}, \citenamefont {Wu}, \citenamefont
  {Xia}, \citenamefont {Ajeesh}, \citenamefont {Nicklas}, \citenamefont {Xu},
  \citenamefont {Zhan}, \citenamefont {Gawryluk}, \citenamefont {Shi},\ and\
  \citenamefont {Shiroka}}]{NLSC-2025-Ad}%
  \BibitemOpen
  \bibfield  {author} {\bibinfo {author} {\bibfnamefont {T.}~\bibnamefont
  {Shang}}, \bibinfo {author} {\bibfnamefont {J.}~\bibnamefont {Zhao}},
  \bibinfo {author} {\bibfnamefont {L.-H.}\ \bibnamefont {Hu}}, \bibinfo
  {author} {\bibfnamefont {W.}~\bibnamefont {Wu}}, \bibinfo {author}
  {\bibfnamefont {K.}~\bibnamefont {Xia}}, \bibinfo {author} {\bibfnamefont
  {M.~O.}\ \bibnamefont {Ajeesh}}, \bibinfo {author} {\bibfnamefont
  {M.}~\bibnamefont {Nicklas}}, \bibinfo {author} {\bibfnamefont
  {Y.}~\bibnamefont {Xu}}, \bibinfo {author} {\bibfnamefont {Q.}~\bibnamefont
  {Zhan}}, \bibinfo {author} {\bibfnamefont {D.~J.}\ \bibnamefont {Gawryluk}},
  \bibinfo {author} {\bibfnamefont {M.}~\bibnamefont {Shi}},\ and\ \bibinfo
  {author} {\bibfnamefont {T.}~\bibnamefont {Shiroka}},\ }\bibfield  {title}
  {\bibinfo {title} {{Discovery of Nodal-Line Superconductivity in Chiral
  Crystals}},\ }\href {https://doi.org/https://doi.org/10.1002/adma.202511385}
  {\bibfield  {journal} {\bibinfo  {journal} {Advanced Materials}\ ,\ \bibinfo
  {pages} {e11385}} (\bibinfo {year} {2025})}\BibitemShut {NoStop}%
\bibitem [{\citenamefont {Yip}(2013)}]{eff_theoty_Yip_tw}%
  \BibitemOpen
  \bibfield  {author} {\bibinfo {author} {\bibfnamefont {S.-K.}\ \bibnamefont
  {Yip}},\ }\bibfield  {title} {\bibinfo {title} {{Models of superconducting
  Cu:Bi${}_{2}$Se${}_{3}$: Single- versus two-band description}},\ }\href
  {https://doi.org/10.1103/PhysRevB.87.104505} {\bibfield  {journal} {\bibinfo
  {journal} {Phys. Rev. B}\ }\textbf {\bibinfo {volume} {87}},\ \bibinfo
  {pages} {104505} (\bibinfo {year} {2013})}\BibitemShut {NoStop}%
\bibitem [{\citenamefont {Fu}\ and\ \citenamefont
  {Kane}(2007)}]{TRS_2007_Fu_inversion}%
  \BibitemOpen
  \bibfield  {author} {\bibinfo {author} {\bibfnamefont {L.}~\bibnamefont
  {Fu}}\ and\ \bibinfo {author} {\bibfnamefont {C.~L.}\ \bibnamefont {Kane}},\
  }\bibfield  {title} {\bibinfo {title} {Topological insulators with inversion
  symmetry},\ }\href {https://doi.org/10.1103/PhysRevB.76.045302} {\bibfield
  {journal} {\bibinfo  {journal} {Phys. Rev. B}\ }\textbf {\bibinfo {volume}
  {76}},\ \bibinfo {pages} {045302} (\bibinfo {year} {2007})}\BibitemShut
  {NoStop}%
\bibitem [{\citenamefont {Sergienko}(2004)}]{symP-2004-prb}%
  \BibitemOpen
  \bibfield  {author} {\bibinfo {author} {\bibfnamefont {I.~A.}\ \bibnamefont
  {Sergienko}},\ }\bibfield  {title} {\bibinfo {title} {{Mixed-parity
  superconductivity in centrosymmetric crystals}},\ }\href
  {https://doi.org/10.1103/PhysRevB.69.174502} {\bibfield  {journal} {\bibinfo
  {journal} {Phys. Rev. B}\ }\textbf {\bibinfo {volume} {69}},\ \bibinfo
  {pages} {174502} (\bibinfo {year} {2004})}\BibitemShut {NoStop}%
\bibitem [{\citenamefont {Kozii}\ and\ \citenamefont
  {Fu}(2015)}]{symP-2015-prl}%
  \BibitemOpen
  \bibfield  {author} {\bibinfo {author} {\bibfnamefont {V.}~\bibnamefont
  {Kozii}}\ and\ \bibinfo {author} {\bibfnamefont {L.}~\bibnamefont {Fu}},\
  }\bibfield  {title} {\bibinfo {title} {{Odd-Parity Superconductivity in the
  Vicinity of Inversion Symmetry Breaking in Spin-Orbit-Coupled Systems}},\
  }\href {https://doi.org/10.1103/PhysRevLett.115.207002} {\bibfield  {journal}
  {\bibinfo  {journal} {Phys. Rev. Lett.}\ }\textbf {\bibinfo {volume} {115}},\
  \bibinfo {pages} {207002} (\bibinfo {year} {2015})}\BibitemShut {NoStop}%
\bibitem [{\citenamefont {Luke}\ \emph {et~al.}(1993)\citenamefont {Luke},
  \citenamefont {Keren}, \citenamefont {Le}, \citenamefont {Wu}, \citenamefont
  {Uemura}, \citenamefont {Bonn}, \citenamefont {Taillefer},\ and\
  \citenamefont {Garrett}}]{symp-1993-prl}%
  \BibitemOpen
  \bibfield  {author} {\bibinfo {author} {\bibfnamefont {G.~M.}\ \bibnamefont
  {Luke}}, \bibinfo {author} {\bibfnamefont {A.}~\bibnamefont {Keren}},
  \bibinfo {author} {\bibfnamefont {L.~P.}\ \bibnamefont {Le}}, \bibinfo
  {author} {\bibfnamefont {W.~D.}\ \bibnamefont {Wu}}, \bibinfo {author}
  {\bibfnamefont {Y.~J.}\ \bibnamefont {Uemura}}, \bibinfo {author}
  {\bibfnamefont {D.~A.}\ \bibnamefont {Bonn}}, \bibinfo {author}
  {\bibfnamefont {L.}~\bibnamefont {Taillefer}},\ and\ \bibinfo {author}
  {\bibfnamefont {J.~D.}\ \bibnamefont {Garrett}},\ }\bibfield  {title}
  {\bibinfo {title} {{Muon spin relaxation in ${\mathrm{UPt}}_{3}$}},\ }\href
  {https://doi.org/10.1103/PhysRevLett.71.1466} {\bibfield  {journal} {\bibinfo
   {journal} {Phys. Rev. Lett.}\ }\textbf {\bibinfo {volume} {71}},\ \bibinfo
  {pages} {1466} (\bibinfo {year} {1993})}\BibitemShut {NoStop}%
\bibitem [{\citenamefont {Luke}\ \emph {et~al.}(1998)\citenamefont {Luke},
  \citenamefont {Fudamoto}, \citenamefont {Kojima}, \citenamefont {Larkin},
  \citenamefont {Merrin}, \citenamefont {Nachumi}, \citenamefont {Uemura},
  \citenamefont {Maeno}, \citenamefont {Mao}, \citenamefont {Mori},
  \citenamefont {Nakamura},\ and\ \citenamefont {Sigrist}}]{symp-1998-nat}%
  \BibitemOpen
  \bibfield  {author} {\bibinfo {author} {\bibfnamefont {G.~M.}\ \bibnamefont
  {Luke}}, \bibinfo {author} {\bibfnamefont {Y.}~\bibnamefont {Fudamoto}},
  \bibinfo {author} {\bibfnamefont {K.~M.}\ \bibnamefont {Kojima}}, \bibinfo
  {author} {\bibfnamefont {M.~I.}\ \bibnamefont {Larkin}}, \bibinfo {author}
  {\bibfnamefont {J.}~\bibnamefont {Merrin}}, \bibinfo {author} {\bibfnamefont
  {B.}~\bibnamefont {Nachumi}}, \bibinfo {author} {\bibfnamefont {Y.~J.}\
  \bibnamefont {Uemura}}, \bibinfo {author} {\bibfnamefont {Y.}~\bibnamefont
  {Maeno}}, \bibinfo {author} {\bibfnamefont {Z.~Q.}\ \bibnamefont {Mao}},
  \bibinfo {author} {\bibfnamefont {Y.}~\bibnamefont {Mori}}, \bibinfo {author}
  {\bibfnamefont {H.}~\bibnamefont {Nakamura}},\ and\ \bibinfo {author}
  {\bibfnamefont {M.}~\bibnamefont {Sigrist}},\ }\bibfield  {title} {\bibinfo
  {title} {{Time-reversal symmetry-breaking superconductivity in
  Sr$_{2}$RuO$_{4}$}},\ }\href {https://doi.org/10.1038/29038} {\bibfield
  {journal} {\bibinfo  {journal} {Nature}\ }\textbf {\bibinfo {volume} {394}},\
  \bibinfo {pages} {558} (\bibinfo {year} {1998})}\BibitemShut {NoStop}%
\bibitem [{\citenamefont {Lee}\ \emph {et~al.}(2009)\citenamefont {Lee},
  \citenamefont {Zhang},\ and\ \citenamefont {Wu}}]{symp-2009-prl}%
  \BibitemOpen
  \bibfield  {author} {\bibinfo {author} {\bibfnamefont {W.-C.}\ \bibnamefont
  {Lee}}, \bibinfo {author} {\bibfnamefont {S.-C.}\ \bibnamefont {Zhang}},\
  and\ \bibinfo {author} {\bibfnamefont {C.}~\bibnamefont {Wu}},\ }\bibfield
  {title} {\bibinfo {title} {Pairing state with a time-reversal symmetry
  breaking in feas-based superconductors},\ }\href
  {https://doi.org/10.1103/PhysRevLett.102.217002} {\bibfield  {journal}
  {\bibinfo  {journal} {Phys. Rev. Lett.}\ }\textbf {\bibinfo {volume} {102}},\
  \bibinfo {pages} {217002} (\bibinfo {year} {2009})}\BibitemShut {NoStop}%
\bibitem [{\citenamefont {Schemm}\ \emph {et~al.}(2014)\citenamefont {Schemm},
  \citenamefont {Gannon}, \citenamefont {Wishne}, \citenamefont {Halperin},\
  and\ \citenamefont {Kapitulnik}}]{symp-2014-sci}%
  \BibitemOpen
  \bibfield  {author} {\bibinfo {author} {\bibfnamefont {E.~R.}\ \bibnamefont
  {Schemm}}, \bibinfo {author} {\bibfnamefont {W.~J.}\ \bibnamefont {Gannon}},
  \bibinfo {author} {\bibfnamefont {C.~M.}\ \bibnamefont {Wishne}}, \bibinfo
  {author} {\bibfnamefont {W.~P.}\ \bibnamefont {Halperin}},\ and\ \bibinfo
  {author} {\bibfnamefont {A.}~\bibnamefont {Kapitulnik}},\ }\bibfield  {title}
  {\bibinfo {title} {{Observation of broken time-reversal symmetry in the
  heavy-fermion superconductor {UPt}$_{\textrm{3}}$}},\ }\href
  {https://doi.org/10.1126/science.1248552} {\bibfield  {journal} {\bibinfo
  {journal} {Science}\ }\textbf {\bibinfo {volume} {345}},\ \bibinfo {pages}
  {190} (\bibinfo {year} {2014})}\BibitemShut {NoStop}%
\bibitem [{\citenamefont {Ghosh}\ \emph {et~al.}(2021)\citenamefont {Ghosh},
  \citenamefont {Smidman}, \citenamefont {Shang}, \citenamefont {Annett},
  \citenamefont {Hillier}, \citenamefont {Quintanilla},\ and\ \citenamefont
  {Yuan}}]{symp-2021-jpcs}%
  \BibitemOpen
  \bibfield  {author} {\bibinfo {author} {\bibfnamefont {S.~K.}\ \bibnamefont
  {Ghosh}}, \bibinfo {author} {\bibfnamefont {M.}~\bibnamefont {Smidman}},
  \bibinfo {author} {\bibfnamefont {T.}~\bibnamefont {Shang}}, \bibinfo
  {author} {\bibfnamefont {J.~F.}\ \bibnamefont {Annett}}, \bibinfo {author}
  {\bibfnamefont {A.~D.}\ \bibnamefont {Hillier}}, \bibinfo {author}
  {\bibfnamefont {J.}~\bibnamefont {Quintanilla}},\ and\ \bibinfo {author}
  {\bibfnamefont {H.}~\bibnamefont {Yuan}},\ }\bibfield  {title} {\bibinfo
  {title} {Recent progress on superconductors with time-reversal symmetry
  breaking},\ }\href {https://doi.org/10.1088/1361-648X/abaa06} {\bibfield
  {journal} {\bibinfo  {journal} {J. Phys.: Condens. Matter}\ }\textbf
  {\bibinfo {volume} {33}},\ \bibinfo {pages} {033001} (\bibinfo {year}
  {2021})}\BibitemShut {NoStop}%
\bibitem [{\citenamefont {Guguchia}\ \emph {et~al.}(2023)\citenamefont
  {Guguchia}, \citenamefont {Mielke}, \citenamefont {Das}, \citenamefont
  {Gupta}, \citenamefont {Yin}, \citenamefont {Liu}, \citenamefont {Yin},
  \citenamefont {Christensen}, \citenamefont {Tu}, \citenamefont {Gong},
  \citenamefont {Shumiya}, \citenamefont {Hossain}, \citenamefont
  {Gamsakhurdashvili}, \citenamefont {Elender}, \citenamefont {Dai},
  \citenamefont {Amato}, \citenamefont {Shi}, \citenamefont {Lei},
  \citenamefont {Fernandes}, \citenamefont {Hasan}, \citenamefont {Luetkens},\
  and\ \citenamefont {Khasanov}}]{symp-2023-np}%
  \BibitemOpen
  \bibfield  {author} {\bibinfo {author} {\bibfnamefont {Z.}~\bibnamefont
  {Guguchia}}, \bibinfo {author} {\bibfnamefont {C.}~\bibnamefont {Mielke}},
  \bibinfo {author} {\bibfnamefont {D.}~\bibnamefont {Das}}, \bibinfo {author}
  {\bibfnamefont {R.}~\bibnamefont {Gupta}}, \bibinfo {author} {\bibfnamefont
  {J.-X.}\ \bibnamefont {Yin}}, \bibinfo {author} {\bibfnamefont
  {H.}~\bibnamefont {Liu}}, \bibinfo {author} {\bibfnamefont {Q.}~\bibnamefont
  {Yin}}, \bibinfo {author} {\bibfnamefont {M.~H.}\ \bibnamefont
  {Christensen}}, \bibinfo {author} {\bibfnamefont {Z.}~\bibnamefont {Tu}},
  \bibinfo {author} {\bibfnamefont {C.}~\bibnamefont {Gong}}, \bibinfo {author}
  {\bibfnamefont {N.}~\bibnamefont {Shumiya}}, \bibinfo {author} {\bibfnamefont
  {M.~S.}\ \bibnamefont {Hossain}}, \bibinfo {author} {\bibfnamefont
  {T.}~\bibnamefont {Gamsakhurdashvili}}, \bibinfo {author} {\bibfnamefont
  {M.}~\bibnamefont {Elender}}, \bibinfo {author} {\bibfnamefont
  {P.}~\bibnamefont {Dai}}, \bibinfo {author} {\bibfnamefont {A.}~\bibnamefont
  {Amato}}, \bibinfo {author} {\bibfnamefont {Y.}~\bibnamefont {Shi}}, \bibinfo
  {author} {\bibfnamefont {H.~C.}\ \bibnamefont {Lei}}, \bibinfo {author}
  {\bibfnamefont {R.~M.}\ \bibnamefont {Fernandes}}, \bibinfo {author}
  {\bibfnamefont {M.~Z.}\ \bibnamefont {Hasan}}, \bibinfo {author}
  {\bibfnamefont {H.}~\bibnamefont {Luetkens}},\ and\ \bibinfo {author}
  {\bibfnamefont {R.}~\bibnamefont {Khasanov}},\ }\bibfield  {title} {\bibinfo
  {title} {{Tunable unconventional kagome superconductivity in charge ordered
  RbV$_{3}$Sb$_{5}$ and KV$_{3}$Sb$_{5}$ }},\ }\href
  {https://doi.org/10.1038/s41467-022-35718-z} {\bibfield  {journal} {\bibinfo
  {journal} {Nat. Commun.}\ }\textbf {\bibinfo {volume} {14}},\ \bibinfo
  {pages} {153} (\bibinfo {year} {2023})}\BibitemShut {NoStop}%
\bibitem [{\citenamefont {Deng}\ \emph {et~al.}(2024)\citenamefont {Deng},
  \citenamefont {Liu}, \citenamefont {Guguchia}, \citenamefont {Yang},
  \citenamefont {Liu}, \citenamefont {Wang}, \citenamefont {Xie}, \citenamefont
  {Shao}, \citenamefont {Ma}, \citenamefont {Liège}, \citenamefont
  {Bourdarot}, \citenamefont {Yan}, \citenamefont {Qin}, \citenamefont
  {Mielke}, \citenamefont {Khasanov}, \citenamefont {Luetkens}, \citenamefont
  {Wu}, \citenamefont {Chang}, \citenamefont {Liu}, \citenamefont
  {Christensen}, \citenamefont {Kreisel}, \citenamefont {Andersen},
  \citenamefont {Huang}, \citenamefont {Zhao}, \citenamefont {Bourges},
  \citenamefont {Yao}, \citenamefont {Dai},\ and\ \citenamefont
  {Yin}}]{symp-2024-nm}%
  \BibitemOpen
  \bibfield  {author} {\bibinfo {author} {\bibfnamefont {H.}~\bibnamefont
  {Deng}}, \bibinfo {author} {\bibfnamefont {G.}~\bibnamefont {Liu}}, \bibinfo
  {author} {\bibfnamefont {Z.}~\bibnamefont {Guguchia}}, \bibinfo {author}
  {\bibfnamefont {T.}~\bibnamefont {Yang}}, \bibinfo {author} {\bibfnamefont
  {J.}~\bibnamefont {Liu}}, \bibinfo {author} {\bibfnamefont {Z.}~\bibnamefont
  {Wang}}, \bibinfo {author} {\bibfnamefont {Y.}~\bibnamefont {Xie}}, \bibinfo
  {author} {\bibfnamefont {S.}~\bibnamefont {Shao}}, \bibinfo {author}
  {\bibfnamefont {H.}~\bibnamefont {Ma}}, \bibinfo {author} {\bibfnamefont
  {W.}~\bibnamefont {Liège}}, \bibinfo {author} {\bibfnamefont
  {F.}~\bibnamefont {Bourdarot}}, \bibinfo {author} {\bibfnamefont {X.-Y.}\
  \bibnamefont {Yan}}, \bibinfo {author} {\bibfnamefont {H.}~\bibnamefont
  {Qin}}, \bibinfo {author} {\bibfnamefont {C.}~\bibnamefont {Mielke}},
  \bibinfo {author} {\bibfnamefont {R.}~\bibnamefont {Khasanov}}, \bibinfo
  {author} {\bibfnamefont {H.}~\bibnamefont {Luetkens}}, \bibinfo {author}
  {\bibfnamefont {X.}~\bibnamefont {Wu}}, \bibinfo {author} {\bibfnamefont
  {G.}~\bibnamefont {Chang}}, \bibinfo {author} {\bibfnamefont
  {J.}~\bibnamefont {Liu}}, \bibinfo {author} {\bibfnamefont {M.~H.}\
  \bibnamefont {Christensen}}, \bibinfo {author} {\bibfnamefont
  {A.}~\bibnamefont {Kreisel}}, \bibinfo {author} {\bibfnamefont {B.~M.}\
  \bibnamefont {Andersen}}, \bibinfo {author} {\bibfnamefont {W.}~\bibnamefont
  {Huang}}, \bibinfo {author} {\bibfnamefont {Y.}~\bibnamefont {Zhao}},
  \bibinfo {author} {\bibfnamefont {P.}~\bibnamefont {Bourges}}, \bibinfo
  {author} {\bibfnamefont {Y.}~\bibnamefont {Yao}}, \bibinfo {author}
  {\bibfnamefont {P.}~\bibnamefont {Dai}},\ and\ \bibinfo {author}
  {\bibfnamefont {J.-X.}\ \bibnamefont {Yin}},\ }\bibfield  {title} {\bibinfo
  {title} {Evidence for time-reversal symmetry-breaking kagome
  superconductivity},\ }\href {https://doi.org/10.1038/s41563-024-01995-w}
  {\bibfield  {journal} {\bibinfo  {journal} {Nat. Mater.}\ }\textbf {\bibinfo
  {volume} {23}},\ \bibinfo {pages} {1639} (\bibinfo {year}
  {2024})}\BibitemShut {NoStop}%
\bibitem [{\citenamefont {Yoshida}\ \emph {et~al.}(2025)\citenamefont
  {Yoshida}, \citenamefont {Takeda}, \citenamefont {Yan}, \citenamefont
  {Kanemori}, \citenamefont {Ortiz}, \citenamefont {Oey}, \citenamefont
  {Wilson}, \citenamefont {Konczykowski}, \citenamefont {Ishihara},
  \citenamefont {Shibauchi},\ and\ \citenamefont {Yamashita}}]{symp-2025-sci}%
  \BibitemOpen
  \bibfield  {author} {\bibinfo {author} {\bibfnamefont {H.}~\bibnamefont
  {Yoshida}}, \bibinfo {author} {\bibfnamefont {H.}~\bibnamefont {Takeda}},
  \bibinfo {author} {\bibfnamefont {J.}~\bibnamefont {Yan}}, \bibinfo {author}
  {\bibfnamefont {Y.}~\bibnamefont {Kanemori}}, \bibinfo {author}
  {\bibfnamefont {B.~R.}\ \bibnamefont {Ortiz}}, \bibinfo {author}
  {\bibfnamefont {Y.~M.}\ \bibnamefont {Oey}}, \bibinfo {author} {\bibfnamefont
  {S.~D.}\ \bibnamefont {Wilson}}, \bibinfo {author} {\bibfnamefont
  {M.}~\bibnamefont {Konczykowski}}, \bibinfo {author} {\bibfnamefont
  {K.}~\bibnamefont {Ishihara}}, \bibinfo {author} {\bibfnamefont
  {T.}~\bibnamefont {Shibauchi}},\ and\ \bibinfo {author} {\bibfnamefont
  {M.}~\bibnamefont {Yamashita}},\ }\bibfield  {title} {\bibinfo {title}
  {Observation of anomalous thermal hall effect in a kagome superconductor},\
  }\href {https://doi.org/10.1126/sciadv.adu2973} {\bibfield  {journal}
  {\bibinfo  {journal} {Sci. Adv.}\ }\textbf {\bibinfo {volume} {11}},\
  \bibinfo {pages} {eadu2973} (\bibinfo {year} {2025})}\BibitemShut {NoStop}%
\bibitem [{\citenamefont {Wang}\ and\ \citenamefont
  {Fu}(2017)}]{symp-2017-prl}%
  \BibitemOpen
  \bibfield  {author} {\bibinfo {author} {\bibfnamefont {Y.}~\bibnamefont
  {Wang}}\ and\ \bibinfo {author} {\bibfnamefont {L.}~\bibnamefont {Fu}},\
  }\bibfield  {title} {\bibinfo {title} {Topological phase transitions in
  multicomponent superconductors},\ }\href
  {https://doi.org/10.1103/PhysRevLett.119.187003} {\bibfield  {journal}
  {\bibinfo  {journal} {Phys. Rev. Lett.}\ }\textbf {\bibinfo {volume} {119}},\
  \bibinfo {pages} {187003} (\bibinfo {year} {2017})}\BibitemShut {NoStop}%
\bibitem [{\citenamefont {Kanasugi}\ and\ \citenamefont
  {Yanase}(2022)}]{symp-2022-cn}%
  \BibitemOpen
  \bibfield  {author} {\bibinfo {author} {\bibfnamefont {S.}~\bibnamefont
  {Kanasugi}}\ and\ \bibinfo {author} {\bibfnamefont {Y.}~\bibnamefont
  {Yanase}},\ }\bibfield  {title} {\bibinfo {title} {{Anapole superconductivity
  from $\mathcal{PT}$-symmetric mixed-parity interband pairing}},\ }\href
  {https://doi.org/10.1038/s42005-022-00804-7} {\bibfield  {journal} {\bibinfo
  {journal} {Commun. Phys.}\ }\textbf {\bibinfo {volume} {5}},\ \bibinfo
  {pages} {39} (\bibinfo {year} {2022})}\BibitemShut {NoStop}%
\bibitem [{\citenamefont {Jackiw}\ and\ \citenamefont
  {Rebbi}(1976)}]{corner_1976_prd}%
  \BibitemOpen
  \bibfield  {author} {\bibinfo {author} {\bibfnamefont {R.}~\bibnamefont
  {Jackiw}}\ and\ \bibinfo {author} {\bibfnamefont {C.}~\bibnamefont {Rebbi}},\
  }\bibfield  {title} {\bibinfo {title} {Solitons with fermion number
  \textonehalf{}},\ }\href {https://doi.org/10.1103/PhysRevD.13.3398}
  {\bibfield  {journal} {\bibinfo  {journal} {Phys. Rev. D}\ }\textbf {\bibinfo
  {volume} {13}},\ \bibinfo {pages} {3398} (\bibinfo {year}
  {1976})}\BibitemShut {NoStop}%
\bibitem [{\citenamefont {Lee}\ \emph {et~al.}(2007)\citenamefont {Lee},
  \citenamefont {Zhang},\ and\ \citenamefont {Xiang}}]{corner_2007_prl}%
  \BibitemOpen
  \bibfield  {author} {\bibinfo {author} {\bibfnamefont {D.-H.}\ \bibnamefont
  {Lee}}, \bibinfo {author} {\bibfnamefont {G.-M.}\ \bibnamefont {Zhang}},\
  and\ \bibinfo {author} {\bibfnamefont {T.}~\bibnamefont {Xiang}},\ }\bibfield
   {title} {\bibinfo {title} {Edge solitons of topological insulators and
  fractionalized quasiparticles in two dimensions},\ }\href
  {https://doi.org/10.1103/PhysRevLett.99.196805} {\bibfield  {journal}
  {\bibinfo  {journal} {Phys. Rev. Lett.}\ }\textbf {\bibinfo {volume} {99}},\
  \bibinfo {pages} {196805} (\bibinfo {year} {2007})}\BibitemShut {NoStop}%
\bibitem [{\citenamefont {Wan}\ \emph {et~al.}(2011)\citenamefont {Wan},
  \citenamefont {Turner}, \citenamefont {Vishwanath},\ and\ \citenamefont
  {Savrasov}}]{WSM-2011-prb-Wan}%
  \BibitemOpen
  \bibfield  {author} {\bibinfo {author} {\bibfnamefont {X.}~\bibnamefont
  {Wan}}, \bibinfo {author} {\bibfnamefont {A.~M.}\ \bibnamefont {Turner}},
  \bibinfo {author} {\bibfnamefont {A.}~\bibnamefont {Vishwanath}},\ and\
  \bibinfo {author} {\bibfnamefont {S.~Y.}\ \bibnamefont {Savrasov}},\
  }\bibfield  {title} {\bibinfo {title} {{Topological semimetal and Fermi-arc
  surface states in the electronic structure of pyrochlore iridates}},\ }\href
  {https://doi.org/10.1103/PhysRevB.83.205101} {\bibfield  {journal} {\bibinfo
  {journal} {Phys. Rev. B}\ }\textbf {\bibinfo {volume} {83}},\ \bibinfo
  {pages} {205101} (\bibinfo {year} {2011})}\BibitemShut {NoStop}%
\bibitem [{\citenamefont {Xu}\ \emph {et~al.}(2015)\citenamefont {Xu},
  \citenamefont {Belopolski}, \citenamefont {Alidoust}, \citenamefont
  {Neupane}, \citenamefont {Bian}, \citenamefont {Zhang}, \citenamefont
  {Sankar}, \citenamefont {Chang}, \citenamefont {Yuan}, \citenamefont {Lee},
  \citenamefont {Huang}, \citenamefont {Zheng}, \citenamefont {Ma},
  \citenamefont {Sanchez}, \citenamefont {Wang}, \citenamefont {Bansil},
  \citenamefont {Chou}, \citenamefont {Shibayev}, \citenamefont {Lin},
  \citenamefont {Jia},\ and\ \citenamefont {Hasan}}]{WSM-2015-sci}%
  \BibitemOpen
  \bibfield  {author} {\bibinfo {author} {\bibfnamefont {S.-Y.}\ \bibnamefont
  {Xu}}, \bibinfo {author} {\bibfnamefont {I.}~\bibnamefont {Belopolski}},
  \bibinfo {author} {\bibfnamefont {N.}~\bibnamefont {Alidoust}}, \bibinfo
  {author} {\bibfnamefont {M.}~\bibnamefont {Neupane}}, \bibinfo {author}
  {\bibfnamefont {G.}~\bibnamefont {Bian}}, \bibinfo {author} {\bibfnamefont
  {C.}~\bibnamefont {Zhang}}, \bibinfo {author} {\bibfnamefont
  {R.}~\bibnamefont {Sankar}}, \bibinfo {author} {\bibfnamefont
  {G.}~\bibnamefont {Chang}}, \bibinfo {author} {\bibfnamefont
  {Z.}~\bibnamefont {Yuan}}, \bibinfo {author} {\bibfnamefont {C.-C.}\
  \bibnamefont {Lee}}, \bibinfo {author} {\bibfnamefont {S.-M.}\ \bibnamefont
  {Huang}}, \bibinfo {author} {\bibfnamefont {H.}~\bibnamefont {Zheng}},
  \bibinfo {author} {\bibfnamefont {J.}~\bibnamefont {Ma}}, \bibinfo {author}
  {\bibfnamefont {D.~S.}\ \bibnamefont {Sanchez}}, \bibinfo {author}
  {\bibfnamefont {B.}~\bibnamefont {Wang}}, \bibinfo {author} {\bibfnamefont
  {A.}~\bibnamefont {Bansil}}, \bibinfo {author} {\bibfnamefont
  {F.}~\bibnamefont {Chou}}, \bibinfo {author} {\bibfnamefont {P.~P.}\
  \bibnamefont {Shibayev}}, \bibinfo {author} {\bibfnamefont {H.}~\bibnamefont
  {Lin}}, \bibinfo {author} {\bibfnamefont {S.}~\bibnamefont {Jia}},\ and\
  \bibinfo {author} {\bibfnamefont {M.~Z.}\ \bibnamefont {Hasan}},\ }\bibfield
  {title} {\bibinfo {title} {{Discovery of a Weyl fermion semimetal and
  topological Fermi arcs}},\ }\href {https://doi.org/10.1126/science.aaa9297}
  {\bibfield  {journal} {\bibinfo  {journal} {Science}\ }\textbf {\bibinfo
  {volume} {349}},\ \bibinfo {pages} {613} (\bibinfo {year}
  {2015})}\BibitemShut {NoStop}%
\bibitem [{\citenamefont {Armitage}\ \emph {et~al.}(2018)\citenamefont
  {Armitage}, \citenamefont {Mele},\ and\ \citenamefont
  {Vishwanath}}]{WSM-2018-RMP}%
  \BibitemOpen
  \bibfield  {author} {\bibinfo {author} {\bibfnamefont {N.~P.}\ \bibnamefont
  {Armitage}}, \bibinfo {author} {\bibfnamefont {E.~J.}\ \bibnamefont {Mele}},\
  and\ \bibinfo {author} {\bibfnamefont {A.}~\bibnamefont {Vishwanath}},\
  }\bibfield  {title} {\bibinfo {title} {{Weyl and Dirac semimetals in
  three-dimensional solids}},\ }\href
  {https://doi.org/10.1103/RevModPhys.90.015001} {\bibfield  {journal}
  {\bibinfo  {journal} {Rev. Mod. Phys.}\ }\textbf {\bibinfo {volume} {90}},\
  \bibinfo {pages} {015001} (\bibinfo {year} {2018})}\BibitemShut {NoStop}%
\bibitem [{\citenamefont {Kane}\ and\ \citenamefont
  {Mele}(2005)}]{KMmodel_2005_prl}%
  \BibitemOpen
  \bibfield  {author} {\bibinfo {author} {\bibfnamefont {C.~L.}\ \bibnamefont
  {Kane}}\ and\ \bibinfo {author} {\bibfnamefont {E.~J.}\ \bibnamefont
  {Mele}},\ }\bibfield  {title} {\bibinfo {title} {${Z}_{2}$ topological order
  and the quantum spin hall effect},\ }\href
  {https://doi.org/10.1103/PhysRevLett.95.146802} {\bibfield  {journal}
  {\bibinfo  {journal} {Phys. Rev. Lett.}\ }\textbf {\bibinfo {volume} {95}},\
  \bibinfo {pages} {146802} (\bibinfo {year} {2005})}\BibitemShut {NoStop}%
\end{thebibliography}%
\end{document}